\documentclass[a4paper,11pt]{article}
\pdfoutput=1 

\usepackage{jheppub} 

\usepackage[T1]{fontenc} 
\usepackage{relsize}
\usepackage{graphics}      
\usepackage{graphicx}      
\usepackage{longtable}     
\usepackage{url}           
\usepackage{bm,dsfont}            
\usepackage{amsmath}
\usepackage{amssymb}
\usepackage{float}
\usepackage{physics}
\usepackage{tensor}
\usepackage{slashed}
\usepackage{bbold,xcolor}
\usepackage{bm}
\usepackage[utf8]{inputenc}
\usepackage{mathtools}
\usepackage{soul}
\usepackage{lipsum}
\usepackage{enumitem}

\definecolor{lcolor}{rgb}{0.5,0,0}
\definecolor{citcolor}{rgb}{0,0.3,0.0}

\newcommand{\Hcal}{\mathcal{H}}
\newcommand{\Mcal}{\mathcal{M}}
\newcommand{\Ncal}{\mathcal{N}}
\newcommand{\Tcal}{\mathcal{T}}
\newcommand{\Scal}{\mathcal{S}}

\newcommand{\Rcal}{\mathcal{R}}
\newcommand{\Ical}{\mathcal{I}}
\newcommand{\Jcal}{\mathcal{J}}
\newcommand{\Pcal}{\mathcal{P}}
\newcommand{\vect}[1]{\boldsymbol{#1}_{\perp}}
\newcommand{\kt}{\vect{k}}
\newcommand{\Acalts}{\boldsymbol{\mathcal{\tilde{A}}}_{\perp}} 
 
\newcommand{\btone}{\boldsymbol{b}_{1\perp}}
\newcommand{\bttwo}{\boldsymbol{b}_{2\perp}}
\newcommand{\ktone}{\boldsymbol{k}_{1\perp}}
\newcommand{\kttwo}{\boldsymbol{k}_{2\perp}} 
\newcommand{\ltone}{\boldsymbol{l}_{1\perp}}
\newcommand{\lttwo}{\boldsymbol{l}_{2\perp}} 
\newcommand{\ztone}{\boldsymbol{z}_{1\perp}}
\newcommand{\zttwo}{\boldsymbol{z}_{2\perp}}
\newcommand{\vtone}{\boldsymbol{v}_{1\perp}}
\newcommand{\vttwo}{\boldsymbol{v}_{2\perp}}
\newcommand{\At}{\boldsymbol{A}_{\perp}} 
\newcommand{\Ats}{\tilde{\boldsymbol{A}}_{\perp}} 

\newcommand{\pt}{\vect{p}}
\newcommand{\Pt}{\vect{P}}

\newcommand{\lt}{\vect{l}}
\newcommand{\at}{\vect{a}}
\newcommand{\bt}{\vect{b}}
\newcommand{\vt}{\vect{v}}
\newcommand{\ellt}{\vect{\ell}}

\newcommand{\xt}{\vect{x}}
\newcommand{\yt}{\vect{y}}
\newcommand{\zt}{\vect{z}}

\newcommand{\rt}{\vect{r}}
\newcommand{\Rt}{\vect{R}}

\newcommand{\et}{\vect{\epsilon}}

\newcommand{\der}{\mathrm{d}}

\title{The importance of kinematic twists and genuine saturation effects in dijet production at the Electron-Ion Collider}

\author[a]{Renaud Boussarie}
\author[b,c]{Heikki Mäntysaari}
\author[d,e,f]{Farid Salazar}
\author[e]{Bj\"orn Schenke}

\affiliation[a]{CPHT, CNRS, Ecole Polytechnique, Institut Polytechnique de Paris, 91128 Palaiseau, France}

\affiliation[b]{
Department of Physics, University of Jyväskylä,  P.O. Box 35, 40014 University of Jyväskylä, Finland
}
\affiliation[c]{
Helsinki Institute of Physics, P.O. Box 64, 00014 University of Helsinki, Finland
}

\affiliation[d]{Department of Physics and Astronomy, Stony Brook University,\\ Stony Brook, NY 11794, USA}
\affiliation[e]{Physics Department, Brookhaven National Laboratory, \\ Bldg. 510A, Upton, NY 11973, USA}

\affiliation[f]{Center for Frontiers in Nuclear Science (CFNS), Stony Brook University,
Stony Brook, NY 11794, USA}

\emailAdd{renaud.boussarie@polytechnique.edu}
\emailAdd{heikki.mantysaari@jyu.fi}
\emailAdd{farid.salazarwong@stonybrook.edu}
\emailAdd{bschenke@bnl.gov}

\abstract{ We compute the differential yield for quark anti-quark dijet production in high-energy electron-proton and electron-nucleus collisions at small $x$ as a function of the relative momentum $\Pt$ and momentum imbalance $\kt$ of the dijet system for different photon virtualities $Q^2$, and study the elliptic and quadrangular anisotropies in the relative angle between $\Pt$ and $\kt$. We review and extend the analysis in \cite{Mantysaari:2019hkq}, which compared the results of the Color Glass Condensate (CGC) with those obtained using the transverse momentum dependent (TMD) framework. In particular, we include in our comparison the improved TMD (ITMD) framework, which resums kinematic power corrections of the ratio $k_\perp$ over the hard scale $Q_\perp$. By comparing ITMD and CGC results we are able to isolate genuine higher saturation contributions in the ratio $Q_s/Q_\perp$ which are resummed only in the CGC. These saturation 
contributions are in addition to those in the Weizsäcker-Williams gluon TMD that appear in powers of $Q_s/k_\perp$. We provide numerical estimates of these contributions for inclusive dijet production at the future Electron-Ion Collider, and identify kinematic windows where they can become relevant in the measurement of dijet and dihadron azimuthal correlations. 
We argue that such measurements will allow the detailed experimental study of both kinematic power corrections and genuine gluon saturation effects.
}

\begin{document}
\maketitle
\flushbottom
\newpage 
\section{Introduction}
In the past decades, high energy collider experiments have successfully verified that quantum chromodynamics (QCD) is the theory of strong interactions of quarks and gluons (partons) inside hadrons and nuclei. A systematic program to uncover the structure of hadrons began with the introduction of parton distributions functions (PDFs) which describe the parton densities as functions of the longitudinal momentum fraction $x$. Beyond this one-dimensional picture, transverse momentum dependent (TMD) PDFs have been introduced to characterize the three-momenta of partons inside hadrons \cite{Collins:1981uw,Mulders:2000sh,Meissner:2007rx}. Complementarily, generalized parton distributions (GPDs) have been defined by furnishing PDFs with the two dimensional transverse spatial distribution of partons resulting in a tomographic picture of hadrons and nuclei \cite{Ji:1996ek,Radyushkin:1997ki,Mueller:1998fv}.

While a partonic description is appropriate at moderate values of $x$, at sufficiently high energies (or small $x$), gluon densities grow quickly resulting in a large parton occupation number. At sufficiently small $x$ this growth is expected to eventually be tamed by non-linear QCD effects \cite{Gribov:1984tu,Mueller:1985wy} (see also \cite{Balitsky:2015qba,Balitsky:2016dgz} for a recent discussion on the transition between non-linear evolution at small-$x$ and linear evolution moderate-$x$ for gluon distributions). An appropriate description of the fundamental degrees of freedom of hadrons and nuclei in this regime is in terms of classical strong gluon fields. The Color Glass Condensate (CGC) is a semi-classical effective field theory (EFT) for small-$x$ gluons in this regime \cite{McLerran:1993ni,McLerran:1993ka,McLerran:1994vd,Iancu:2000hn,Iancu:2001ad,Ferreiro:2001qy,Iancu:2003xm,Gelis:2010nm,Kovchegov:2012mbw,Albacete:2014fwa,Blaizot:2016qgz}.  The CGC has been applied for a variety of processes in proton-nucleus collisions as well as DIS: structure functions (inclusive \cite{Albacete:2010sy,Beuf:2020dxl} and diffractive \cite{Kowalski:2008sa}), semi-inclusive production (photon \cite{JalilianMarian:2012bd,Roy:2018jxq,Ducloue:2017kkq,Kolbe:2020tlq}, inclusive single hadron \cite{Tribedy:2011aa,Albacete:2012xq,Lappi:2013zma,Marquet:2009ca,Iancu:2020jch}, dihadron/dijet \cite{Lappi:2012nh,Albacete:2010pg,Zheng:2014vka,Dumitru:2018kuw}, quarkonia \cite{Fujii:2013yja,Ma:2018bax,Kang:2013hta,Ducloue:2016pqr}), and exclusive processes (deeply virtual Compton scattering and vector meson \cite{Kowalski:2006hc,Rezaeian:2013tka,Lappi:2010dd,Lappi:2013am,Mantysaari:2020lhf,Mantysaari:2019jhh,Mantysaari:2017dwh,Mantysaari:2016ykx,Mantysaari:2017slo}, dijet \cite{Mantysaari:2019csc,Altinoluk:2015dpi,Salazar:2019ncp,Boussarie:2019ero}, trijet \cite{Kovner:2021lty,Ayala:2016lhd,Boussarie:2016ogo} production) to name a few (for a recent review see \cite{Morreale:2021pnn}).

Among these various processes, forward particle azimuthal angle correlations are powerful observables to access the small-$x$ structure of hadrons and nuclei at current and future collider experiments \cite{Kharzeev:2004bw}. A paradigmatic example is that of inclusive dihadron production in $\rm{d} + \rm{Au}$ collisions at RHIC, where a suppression in the back-to-back peak relative to $p+p$  collisions \cite{Braidot:2010zh,Adare:2011sc} might signal the presence of gluon saturation \cite{Marquet:2007vb,Lappi:2012nh,Albacete:2018ruq}\footnote{Another mechanism that could provide a suppression of the back-to-back peak is the momentum broadening due to cold nuclear matter energy loss and coherent power corrections, which has been studied for dihadron production in $\mathrm{p+p}$ and $\mathrm{d+Au}$ in \cite{Kang:2011bp}.}. At the future Electron-Ion Collider (EIC)~\cite{AbdulKhalek:2021gbh,Aschenauer:2017jsk,Accardi:2012qut} and at the LHeC/FCC-he~\cite{Agostini:2020fmq} a similar measurement has been proposed where a depletion of the back-to-back peak in forward (electron-going) dihadron production is expected to be observed when going from proton to nuclear deeply inelastic scattering (DIS)~\cite{Zheng:2014vka,Dominguez:2011wm}. This measurement at the EIC offers two advantages: i) there is no need to assume a hybrid dilute-dense factorization in which one convolutes with PDFs of the proton since in DIS the projectile is a virtual photon whose kinematics can be reconstructed by measuring the scattered electron, ii) in the correlation (back-to-back kinematics) limit, it is only sensitive to one type of gluon TMD distribution\footnote{Unlike $\rm{p/d} + \rm{p/Au}$, which is sensitive to eight different types of gluon TMDs \cite{Dominguez:2011wm,Petreska:2018cbf,Albacete:2018ruq}.}: the Weizs\"acker-Williams (WW) type. Furthermore, measurements of the azimuthal distribution of the momentum imbalance $\kt$ of the dihadron/dijet system provide the opportunity to access the linearly polarized WW gluon TMD. In particular, the elliptic anisotropy in the angle between the momentum imbalance $\kt$ and the relative dijet momentum $\Pt$ is proportional to the ratio of linearly polarized to unpolarized gluon pairs \cite{Dominguez:2011br,Metz:2011wb,Dumitru:2015gaa}. Both unpolarized and linearly polarized WW gluon TMD are expected to be sensitive to saturation effects at small $x$ as they resum contributions in powers of $Q_s/k_\perp$, making the studies of dihadron and dijet anisotropies promising channels to investigate gluon saturation \cite{Zheng:2014vka,Dumitru:2018kuw}.

While the TMD framework for inclusive dijet production is expected to hold near back-to-back kinematics, there are important higher twist corrections that must be resummed for more controlled phenomenological predictions. At small $x$, the CGC EFT approach to high-energy QCD~\cite{Blaizot:2016qgz,Iancu:2003xm,Gelis:2010nm,Kovchegov:2012mbw} encodes these higher order corrections in the dipole and quadrupole correlators of light-like Wilson lines\footnote{Note that only the higher order corrections that are leading in powers of the center-of-mass energy $W$ are resummed in the CGC EFT.}. In \cite{Mantysaari:2019hkq}, we computed the inclusive dijet production differential cross section in the CGC formalism employing the Gaussian approximation \cite{Marquet:2010cf,Dominguez:2011wm,Fujii:2006ab} and at realistic energies for the future EIC, and observed deviations from the TMD framework at large momentum imbalance $k_\perp \sim P_\perp$. In addition, we also found differences at small momentum imbalance if also $Q_s \sim P_\perp$, which were enhanced in nuclear DIS due to the enhanced nuclear saturation scale $Q_s^2\propto A^{1/3}$. These genuine saturation contributions are different than those existing in the WW gluon TMD at small $x$ which scale as $Q_s/k_\perp$ \cite{Dominguez:2011wm,Dumitru:2015gaa}.

When studying power corrections to semi-inclusive processes with several hard and semi-hard scales involved, it is necessary to distinguish terms according to the ratios they resum. Indeed, the presence of a semi-hard $k_\perp$ and a saturation scale $Q_s$ in addition to the hard scale $Q_\perp$ implies that power-suppressed terms can scale as $k_\perp/Q_\perp$ or $Q_s/Q_\perp$, which can be significantly enhanced when compared to the simpler $\Lambda_{\mathrm{QCD}}/Q_\perp$ twist corrections. This approach has been recently applied to the production of quark anti-quark dijets in $p+p$ and $p+A$ collisions in \cite{Fujii:2020bkl}. The present manuscript follows the same spirit of \cite{Fujii:2020bkl} combined with the results in \cite{Altinoluk:2019fui,Altinoluk:2019wyu}, which allow us to make the distinction of higher twists into kinematic ($k_\perp/Q_\perp$)  and  genuine ($Q_s/Q_\perp$) saturation contributions for the electroproduction of a quark anti-quark dijet. On the level of the differential cross section they can be schematically identified as follows:
\begin{align}
    \der \sigma_{\rm{CGC}} = \underbrace{\der \sigma_{\rm{TMD}} + \overbrace{\mathcal{O}\left(\frac{k_\perp}{Q_\perp} \right)}^{\mathrm{kinematic}} }_{ \mathlarger{\der \sigma_{\mathrm{ITMD}} } } + \overbrace{\mathcal{O}\left(\frac{Q_s}{Q_\perp} \right)}^{\mathrm{genuine}} \,,
\end{align}
where the hard scale $Q_\perp$ of the process is built from the relative momentum $P_\perp$ and the virtuality of the photon $Q$, both controlling the dipole size $r_\perp$. For simplicity, one can define $Q_\perp = \mathrm{max}(Q,P_\perp)$.

In practice, the so-called genuine saturation terms correspond to the contribution of genuine higher twist operators. At small $x$, they correspond to operators with higher powers of the gluon field strength tensor in the form of $gF^{\mu\nu}$ insertions along with gauge links that maintain gauge invariance. As we will prove in this article, these corrections yield powers of $Q_s/Q_\perp$ in the CGC because of semi-perturbative gluon loops whose internal transverse momenta peak around $Q_s$.
The ITMD framework resums only the kinematic power corrections (kinematic twists) leaving aside the genuine saturation contributions. This framework was originally introduced to study dijet production in proton-nucleus collisions \cite{Kotko:2015ura,vanHameren:2016ftb,Petreska:2018cbf}, and recently extended for DIS and photo-production (\cite{Altinoluk:2019fui,Altinoluk:2019wyu}). By comparing TMD, ITMD and CGC frameworks, we are able to identify the separate roles of kinematic and genuine higher twists in the production of quark anti-quark dijets at the EIC, which is the goal of the present manuscript.

This manuscript is organized as follows. In Sec.\,\ref{sec:field_stregnth_WilsonLines} we begin by briefly reviewing the results in \cite{Boussarie:2020vzf} relating the product of two light-like Wilson lines to a transverse gauge link. This allow us to generalize the small dipole size expansion in \cite{Dominguez:2011wm}, and isolate kinematic and genuine higher twists.  In Sec.\,\ref{sec:inclusive_dijets_highenergy}, we discuss the inclusive production of dijets in high energy (small-$x$) DIS. We start with the computation in the CGC formalism in momentum space, where we also take the opportunity to set up the notations. We choose to express the differential cross-section in terms of the momentum imbalance $\kt$ and the relative momentum $\Pt$ of the dijet pair. Then, we review the derivation of the TMD limit for back-to-back dijets ($k_\perp,Q_s \ll Q_\perp$) and introduce the WW gluon TMD. Next, we study the ITMD framework which resums the kinematic twists (in powers of $k_\perp/Q_\perp$), and obtain analytic expressions for the hard factors. We briefly discuss genuine higher twists and argue that these are parametrically of the order $Q_s/Q_\perp$. In Sec.\,\ref{sec:set_up_dijets}, we discuss the setup for our numerical computation: the initial conditions, the small-$x$ evolution, the Gaussian approximation for high energy CGC correlators and the WW gluon TMD, and the computation of the harmonics of the differential cross-section in the angle $\phi = \phi_{\Pt} - \phi_{\kt}$. Our numerical results for the production of quark anti-quark dijets at realistic EIC kinematics are presented in Sec.\,\ref{sec:numerical_results_dijets}. We focus on the hadronic sub-amplitudes $\gamma^*_{\mathrm{L}/\mathrm{T}} + \mathrm{p/Au} \rightarrow q \bar{q}+X$, and compare the results in the TMD, ITMD and CGC formalisms. We first present the angle averaged yield (over $\phi_{\Pt}$ and $\phi_{\kt}$) as a function of momentum imbalance $k_\perp$ at different values of $P_\perp$ and virtuality $Q$. Then, we focus on the production close to back-to-back kinematics at various values of virtuality and relative jet momentum. Finally, we present our results for the elliptic and quadrangular azimuthal anisotropies in the angle $\phi$.  In Sec.\,\ref{sec:conclusions_dijets} we summarize our main results and comment about their implications to the measurements of azimuthal dijets and dihadron correlations at the EIC.

\section{Kinematic and genuine higher twists in the Wilson line pairs}
\label{sec:field_stregnth_WilsonLines}
In the CGC, the scattering of a colored particle moving along the minus light-cone direction through the background field $A^\mu$ of a nucleus propagating in the plus light-cone direction can be characterized (in the eikonal approximation) as a color rotation of the particle given by the light-like Wilson line:
\begin{align}
    V(\vect{z}) = \Pcal \exp( ig \int_{-\infty}^\infty \der z^- A^{+,a}  (z^-,\vect{z}) t^a )  
    \label{eq:WilsonLine} \,,
\end{align}
where $t^a$ are the generators of $SU(3)$ in the fundamental representation, and $A^{+,a}$ is in  Lorenz gauge $\partial_\mu A^\mu =0$. Here $\Pcal$ stands for path ordering such that the operator at $z=-\infty$ is in the rightmost position, while that at $z=+\infty$ is in the leftmost position\footnote{Other authors may use the opposite convention for the path ordering. In that case, an additional $-$ sign would appear in the exponent of the Wilson line.}. For the sake of clarity, we will decorate the gauge field with a \textit{tilde} $\tilde{A}$ when working in $\tilde{A}^+ =0$ light-cone gauge as opposed to Lorenz gauge $\partial_\mu A^\mu=0$.

At the level of the amplitude, the inclusive production of a quark anti-quark dijet in the CGC will contain the product of two Wilson lines $ V(\vect{x}) V^\dagger(\yt)$ corresponding to the color rotation of the quark and the anti-quark, respectively. The seminal work in \cite{Dominguez:2011wm} established the connection between CGC and TMD amplitudes (cross-sections) by noting that the first term in the expansion in powers of $\rt=\xt-\yt$ of this product produces the transverse gauge field $\Ats^i$ in light-cone gauge $\tilde{A}^+ = 0$:
\begin{align}
    \mathbb{1} - V(\vect{x}) V^\dagger(\yt) = \underbrace{ig \rt^i \Ats^i}_{\mathrm{TMD}} + \mathcal{O}(\rt^2) \,,
    \label{eq:dipole_expansion_generic}
\end{align}
where $\Ats^i = \frac{i}{g} V\partial^i V^\dagger$ (see Secs.\,\ref{sec:inclusive_dijets_CGC} and \ref{sec:inclusive_dijets_TMD}). Such transverse gauge fields can be written in terms of the field strength tensor $F^{i+}$, which constitute the building blocks of small-$x$ gluon TMDs (see Appendix\,\ref{sec:Appendix_WW}).

It seems natural to wonder what is the behavior of the higher terms in powers of $\rt$ in the expansion in Eq.\,\eqref{eq:dipole_expansion_generic}. This is the approach followed in \cite{Dumitru:2016jku}, where part of the quadratic term in $\rt$ has been isolated. However, one quickly realizes that order by order in the $\rt$ expansion, the resulting terms are cumbersome to organize in a way which preserves explicit QCD gauge invariance for the involved operators. Recent developments on the CGC/TMD correspondence rely on the possibility to express the product of two Wilson lines as a transverse gauge link at $x^-=-\infty$ \cite{Boussarie:2020vzf}:
\begin{align}
    V(\xt)V^\dagger(\yt) = \mathcal{P} \exp\left[-ig \int_{\yt}^{\xt} \der \zt^i  \Ats^i(\zt) \right] \,.
    \label{eq:dipole_transversegaugelink}
\end{align}
One can then show (see Appendix\,\ref{sec:Wilson_and_gaugelink} for details) that as a consequence of recursive relations satisfied by the transverse gauge link, it is possible to re-organize all terms beyond the linear expansion in $g\Ats^i$ as follows:
\begin{align}
    \mathbb{1}-V(\xt)V^\dagger(\yt) &= \underbrace{ig \int_{\yt}^{\xt} \der \zt^i  \Ats^i(\zt)}_{\mathrm{ITMD}} \nonumber \\ &+ \underbrace{g^2 \int_{\yt}^{\xt} \der \ztone^i  \int_{\yt}^{\ztone} \der \zttwo^j  \Ats^i(\ztone) V(\ztone) V^\dagger(\zttwo) \Ats^j(\zttwo)}_{\mathrm{genuine \ higher \ twists \ (g.h.t.)}} \,.
    \label{eq:dipole_itmd_twist}
\end{align}
We note that this is an exact relation, and that it is organized in a gauge invariant way circumventing the difficulty introduced by the naive expansion in the dipole size $\rt$.

The infinite series obtained via iteration of the right-hand side of Eq.\,\eqref{eq:dipole_itmd_twist} re-expresses the multiple scattering of a quark anti-quark pair as a combination of gluon saturation (embodied in the strong transverse gauge field $\Ats^i(\zt) \sim 1/g$) and multiple scattering in light-cone gauge\footnote{We note that the physical picture of gluon saturation is gauge dependent. In Lorenz gauge $\partial_\mu A^\mu = 0$, high gluon-density manifests as multiple scattering encoded in the Wilson lines. On the other hand, in light-cone gauge $\tilde{A}^+ =0$ the phenomenon of gluon saturation becomes manifest in the WW gluon distribution, which represents the gluon number density (see e.g. Sects. 2.5 and 2.6 in the review article \cite{Iancu:2003xm}).}. In Sec.\,\ref{sec:inclusive_dijets_iTMD} we will show explicitly that when the linear term in Eq.~\eqref{eq:dipole_itmd_twist} is used to calculate the amplitude for quark anti-quark dijet production, it will only involve the WW gluon TMD (single scattering) but with a modified hard factor that resums all kinematic power corrections in $k_\perp/Q_\perp$ (ITMD). On the other hand, the higher order terms in the quadratic term $(g\Ats)^2$ (higher order operators in $g F^{i+}$) resum contributions in powers of $Q_s/Q_\perp$, corresponding to double scattering (and beyond through iteration), which will be discussed in Sec.\,\ref{sec:inclusive_dijets_ght}.

\section{Inclusive dijet production in high energy DIS}
\label{sec:inclusive_dijets_highenergy}

In this section, we review the computation of the differential cross section for the production of a forward quark anti-quark dijet in DIS within the CGC EFT
\begin{align}
    e(k_e)+A(P_A) \rightarrow e(k_e')+q(k_1)+\bar{q}(k_2)+X,
\end{align}
where $A$ can be proton or a nucleus. The kinematic variables are detailed in Table\,\ref{tab:kinematics_dijet}.

The leptonic and hadronic parts of this process can be separated by expressing the production cross-section as follows: 
\begin{align}
\label{eq:eA_xs}
    \frac{\der \sigma^{e+A\rightarrow e' + q\bar{q}+X}}{\der W^2 \der Q^2 \der^2 \ktone \der^2 \kttwo \der \eta_1 \der \eta_2} = \sum_{\lambda=\mathrm{L},\mathrm{T}} f_{\lambda}(W^2,Q^2) \frac{\der \sigma^{\gamma_{\lambda}^*+A\rightarrow q\bar{q}+X}}{ \der^2 \ktone \der^2 \kttwo \der \eta_1 \der \eta_2} \,,
\end{align}
where $\lambda$ is the polarization of the exchanged virtual photon, $Q^2$ is its virtuality, and $W^2$ is the center of mass energy per nucleon of the $\gamma^*-A$ system. The leptonic part is contained in the photon fluxes:
\begin{align}
    f_{\mathrm{L}}(W^2,Q^2) &= \frac{\alpha_{\rm{em}}}{\pi Q^2 sy} (1-y) \,, \\
    f_{\mathrm{T}}(W^2,Q^2)  &= \frac{\alpha_{\rm{em}}}{2\pi Q^2 sy} \left[1 +(1-y)^2 \right] \,,
\end{align}
where the inelasticity is given by $y=\frac{W^2+Q^2-m_n^2}{s-m_n^2}$ and $\sqrt{s}$ is the center of mass energy per nucleon in the $e+A$ collision.

\begin{table}[H]
\centering
\caption{ Kinematic variables} \label{tab:kinematics_dijet} 
\begin{tabular}{ll}
\\ \hline \hline
$P_A $ & nucleus four-momentum \\
$P_n $ & nucleon four-momentum \\
$k_e \ (k_e')$ & incoming (outgoing) electron four-momentum \\ 
$q=k_e-k_e'$ & virtual photon four-momentum \\
$k_{1,2}$ & quark (anti-quark) four-momentum \\ 
$z_{1,2}$  & quark (anti-quark) longitudinal momentum fraction \\
$\eta_{1,2}$ & quark (anti-quark) rapidity \\
$\mathbf{k_{1,2\perp}}$ &  quark (anti-quark) transverse momentum \\
$\kt = \ktone + \kttwo$ &  quark anti-quark dijet transverse momentum imbalance \\
$\Pt = z_2 \ktone - z_1 \kttwo$ &  quark anti-quark dijet relative transverse momentum  \\
$s=(P_n+k_e)^2$ & nucleon-electron system center of momentum energy squared  \\
$W^2=(P_n+q)^2$ & nucleon-virtual photon system center of momentum energy squared\\
$m^2_n = P_n^2$ & nucleon invariant mass squared\\
$M^2_{q\bar{q}} = (k_1 + k_2)^2$ & invariant mass squared of the dijet system.\\
$Q^2=-q^2$ & virtuality squared of the exchanged photon \\ \hline \hline
\end{tabular}
\end{table}

In the present work, we focus on the hadronic content of the scattering encoded in the sub-process:
\begin{align}
    \gamma_{\lambda}^*(q)+A(P_A) \rightarrow q(k_1)+\bar{q}(k_2)+X.
\end{align}
We will work in light-cone coordinates\footnote{We follow the convention $a^{\pm} = \frac{1}{\sqrt{2}}(a^0 \pm a^3)$, such that $a^\mu b_\mu = a^+ b^- + a^- b^+ - \at \cdot \bt$, where $\at$ and $\bt$ are the 2-dimensional transverse vectors. We will denote the magnitude of transverse vectors as $a_\perp = (\at \cdot \at )^{1/2}$ .}, and in a frame where the proton or nucleus propagates in the plus light-cone direction and the virtual photon in the minus light-cone direction:
\begin{align}
    P_A &= \left(P_A^+, \frac{m_A^2}{2P_A^+},\vect{0} \right) \,,  \\
    q & = \left(-\frac{Q^2}{2q^-} ,q^-, \vect{0} \right) \,,
\end{align}
with $m_A$ the proton (nucleus) mass.
The momenta of the produced $q$ and $\bar{q}$ are given by
\begin{align}
    k_1 &= \left(\frac{\ktone^2}{2k_1^-}, k_1^- ,\ktone \right) \,, \\
    k_2 &= \left(\frac{\kttwo^2}{2k_2^-}, k_2^- ,\kttwo \right) \,,
\end{align}
respectively. We will denote the longitudinal momentum fractions $z_{1,2} = k_{1,2}^-/q^-$, and we will neglect the quark masses.

\subsection{Inclusive quark anti-quark production in the CGC}
\label{sec:inclusive_dijets_CGC}

In the CGC EFT, the large-$x$ degrees of freedom in the nucleus are treated as stochastic color sources $\rho^a_A$ which generate the strong classical gluon field $A^{\mu}$, characterizing the small-$x$ gluon content of the nucleus.

For a fast moving nucleus along the plus light-cone direction, the color sources generate a current of the form
\begin{align}
    J^{\mu}(x^-,\xt) = \delta^{\mu+} \rho_A(x^-,\xt) \,,
\end{align}
which is independent of light-cone time $x^+$.

It can be easily verified that the Yang-Mills equations $\left[D_{\mu}, F^{\mu\nu} \right] = J^\nu $ are satisfied by the gauge field
\begin{align}
    A^{+}(x) =  \alpha(x^-,\xt), \quad A^-=\At^i =0 \,,
    \label{eq:wrongLCgaugeA}
\end{align}
where $\alpha(x^-,\xt)$ solves the Poisson equation
\begin{align}
    \nabla_\perp^2 \alpha(x^-,\xt) = - \rho_A(x^-,\xt) \,.
\end{align}
The solution in Eq.\,\eqref{eq:wrongLCgaugeA} is in Lorenz gauge $\partial_\mu A^\mu =0$.

An infinite set of solutions $\tilde{A}$ can be found by application of gauge transformations
\begin{align}
    \tilde{A}^\mu(x) = \Omega(x) A^\mu(x) \Omega^{-1}(x) -\frac{1}{ig} \Omega(x) \partial^\mu \Omega^{-1}(x) \,.
\end{align}
One notable choice is the light-cone gauge $\tilde{A}^+ = 0$ , which can be obtained from the gauge transformation
\begin{align}
    \Omega_{\xt}(x^-) = \Pcal \exp( ig \int_{x^-}^{+\infty} \der z^- A^{+}  (z^-,\vect{x})  ) \,.
    \label{eq:gauge_trans_A+_Ai}
\end{align}
In such a case one obtains a solution in terms of a purely transverse gauge field: 
\begin{align}
    \Ats^i(x^-,\xt) = \frac{i}{g} \Omega_{\xt}(x^-) \partial^i \Omega^{-1}_{\xt}(x^-), \quad \tilde{A}^{+}=\tilde{A}^{-} = 0 \,. 
    \label{eq:transversegauge_gaugerotation}
\end{align}
Note that this choice relies implicitly on the boundary condition $\Ats^i(+\infty,\xt) = 0$ which is allowed by the freedom of the residual subgauge condition in light-cone gauges. It will be useful to denote the transverse gauge field at $x^- =-\infty$ by simply
suppressing the light-cone argument:
\begin{align}
    \Ats^i(\xt) \equiv \Ats^i(-\infty,\xt) = \frac{i}{g} V(\xt) \partial^i V^\dagger(\xt) \,,
\end{align}
where we used $\Omega_{\xt}(x^-=-\infty) = V(\xt)$. 

The expectation value of any observable in the CGC is then computed in perturbation theory in the presence of the background field $A^{\mu}$ for a given configuration of sources $\rho_A$, and then averaging over all possible configurations:
\begin{align}
    \left \langle \mathcal{O} [\rho_A]\right \rangle_Y = \int \left[\mathcal{D} \rho_A \right] W_Y[\rho_A] \mathcal{O}[\rho_A] \,,
\end{align}
where $W_Y[\rho_A]$ is a gauge invariant weight functional for the configuration of $\rho_A$, at the rapidity scale $Y=\log(1/x)$ (see also Ref.~\cite{Ducloue:2019ezk} for a discussion of different evolution rapidity variables). Here $x$ is the typical longitudinal momentum fraction probed by the observable.

Following this prescription, the differential cross-section reads:
\begin{align}
    \frac{\der \sigma^{\gamma_{\lambda}^*+A\rightarrow q\bar{q}+X}}{ \der^2 \ktone \der^2 \kttwo \der \eta_1 \der \eta_2}  =  \frac{1}{4 (2\pi)^6} \frac{1}{2q^-} (2\pi) \delta(k_1^- + k_2^- - q^-) \left \langle \Mcal^{\lambda\dagger}_{\mathrm{CGC}}[\rho_A] \Mcal^{\lambda}_{\mathrm{CGC}} [\rho_A] \right \rangle_Y \,,
\end{align}
where $\Mcal^{\lambda}_{\mathrm{CGC}} [\rho_A]$ is the CGC amplitude for the production of quark anti-quark in the collision of a virtual photon $\gamma^*$ with the external classical background field (shock-wave) produced by the source configuration $\rho_A$ characterizing the nucleus $A$. 

\subsubsection{Amplitude}

\begin{figure}[h]
\centering
\includegraphics[width = 3.00in]{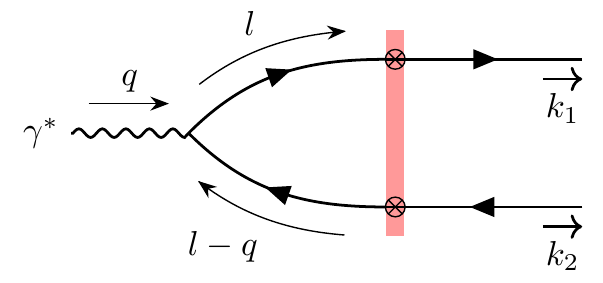}
\caption{CGC EFT diagram for the production of a quark anti-quark pair in DIS. The red rectangle and crosses represent the effective interaction of the quark and the anti-quark with the background field representing the small-$x$ content of the nucleus.}
\label{CGC_diagram}
\end{figure}

At leading order in the CGC, the forward production of a dijet pair proceeds by the splitting of the virtual photon $\gamma^*$ into a quark anti-quark dipole that subsequently scatters off the field $A^\mu$, and then fragments into two jets (see Fig.\,\ref{CGC_diagram})\footnote{ We expect that the contribution from the quark initiated channel $\gamma^* +q \rightarrow q +g$ to be negligible in low-$x$ kinematics because of the dominance of gluon exchanges. For a recent study of this process and the impact of gluon saturation and multiple parton scattering see \cite{Zhang:2021tcc}.}

The result for the computation of quark anti-quark production in DIS within the CGC is well-known and has been obtained in \cite{Dominguez:2011wm,Dominguez:2011br}. Here we briefly review the computation of the amplitude for this process following momentum space Feynman rules \cite{Gelis:2002nn}. The expression for the amplitude is
\begin{align}
    \Scal^{\lambda} [\rho_A]=  \int \frac{\der^4 l}{(2\pi)^4} \bar{u}_{\sigma}(k_1) \Tcal^q(k_1,l) S^0(l) \left(-i e e_f \slashed{\epsilon}(q,\lambda) \right) S^0(l-q) \Tcal^q(l-q,-k_2) v_{\sigma'}(k_2) \,,
\end{align}
where we followed the standard momentum space Feynman rules of QCD+QED in $\partial_\mu A^\mu = 0$ gauge for QCD, supplemented by the effective CGC vertex for quark interaction with the back-ground field. The Feynman propagator for a free massless quark reads
\begin{align}
    S^0(l) = \frac{i \slashed{l}}{l^2 + i\epsilon} \,,
\end{align}
and the effective CGC vertex for the eikonal multiple scattering of a quark (anti-quark) off the color field $A^{\mu}$ \cite{McLerran:1998nk,Balitsky:2001mr}:
\begin{align}
    \mathcal{T}^q_{ij}(l,l') =  (2\pi) \delta(l^--l'^-) \gamma^- \mathrm{sgn}(l^-) \int \der^2\vect{z} e^{-i(\lt-\lt')\cdot \vect{z}} V_{ij}^{\mathrm{sgn}(l^-)}(\vect{z}) \,, \label{eq:CGCEFT_qvertex}
\end{align}
where $i,j$ represent the color indices, and $l$ and $l'$ are the outgoing and incoming momenta of the quark, and the light-like Wilson line in the fundamental representation was introduced in Eq.\,\eqref{eq:WilsonLine}. The
superscript $\mathrm{sgn}(p)$ in Eq.\,\eqref{eq:CGCEFT_qvertex} denotes matrix exponentiation: $V
^{+1}(\zt) = V(\zt)$ and $V^{-1}
(\zt) = V^\dagger
(\zt)$,
where the latter follows from the unitarity of $V(\zt)$.

It is convenient to work in $A_{\mathrm{QED}}^-=0$ light-cone gauge for the QED part of the amplitude and in a frame where the transverse momentum of the photon is $\vect{0}$. The polarization vectors for the virtual photon read\footnote{In the Lorenz gauge $\partial_\mu A^\mu = 0$ for QED, the longitudinally photon polarization takes the form $\epsilon^{\lambda= 0,\mu} =\left(\frac{Q}{2q^-}, \frac{q^-}{Q} ,\vect{0} \right)$.}
\begin{align}
    \epsilon^{\lambda= 0,\mu} &=\left(\frac{Q}{q^-}, 0 ,\vect{0} \right) \,, \\
    \epsilon^{\lambda = \pm 1,\mu} &= \left( 0, 0, \et^{\lambda} \right)\,,
\end{align}
where $\et^{\pm 1} = \frac{1}{\sqrt{2}}(1,\pm i)$, $\lambda = 0$ denotes the longitudinal polarization, and $\lambda = \pm 1$ the two transverse polarizations.

Subtracting the non-interacting piece and factoring out an overall delta function $2\pi \delta(k_1^- + k_2^- - q^-)$, the amplitude can be organized as follows
\begin{align}
    \Mcal^\lambda_{\mathrm{CGC}}[\rho_A] = \frac{ee_f q^-}{\pi}  \int \der^2 \xt \der^2 \yt e^{-i \ktone\cdot \xt } e^{-i \kttwo\cdot \yt } \Ncal^\lambda_{\sigma\sigma'}(\xt-\yt)  \left[\mathbb{1} - V(\xt) V^\dagger(\yt) \right] \label{eq:M-matrix} \,,
\end{align}
with the perturbative factor: 
\begin{align}
    \Ncal^\lambda_{\sigma\sigma'} (\rt) &= -i (2q^-)  \int \frac{\der^4 l}{(2\pi)^2} \frac{ N^{\lambda}_{\sigma\sigma'}(l) e^{i \lt\cdot \rt} \delta(k_1^- - l^-) }{(l^2 + i \epsilon)((q-l)^2+i \epsilon)}  \label{eq:dijet-LO-pert} \,, \\
    N^{\lambda}_{\sigma\sigma'}(l)  &= \frac{1}{(2q^-)^2}\left[ \bar{u}_{\sigma}(k_1) \gamma^- \slashed{l} \slashed{\epsilon}(q,\lambda)  (\slashed{q}-\slashed{l})\gamma^- v_{\sigma'}(k_2) \right] \,,
\end{align}
where $\sigma$ and $\sigma'$ are the helicities of the quark and anti-quark respectively. 

The computation of the perturbative factors is straightforward, we briefly outline it in Appendices\,\ref{sec:explicit_spinor_rep} and \ref{sec:computing_N_CGC_dijets}. We find
\begin{align}
    \Ncal^{\lambda=0}_{\sigma\sigma'}(\rt) &= 2   (z_1z_2)^{3/2} Q K_0(\varepsilon r_\perp) \delta_{\sigma,- \sigma'} \,, 
    \label{eq:dijet-LO-pert_L}\\
    \Ncal^{\lambda= \pm 1}_{\sigma\sigma'}(\rt) &=  (z_1z_2)^{1/2} \left[  (z_1-z_2) - \sigma \lambda \right] \frac{i \varepsilon \rt \cdot \et^{\lambda} }{r_\perp} K_1(\varepsilon r_\perp) \delta_{\sigma,- \sigma'} \,,
    \label{eq:dijet-LO-pert_T}
\end{align}
where $\varepsilon^2 = z_1 z_2 Q^2$.
\subsubsection{Differential Cross-section}
The color structure obtained by squaring the amplitude, summing over the colors of the quark and the anti-quark in the final state, and averaging over the different color charge configurations $\rho_A$ is given by
\begin{align}
    \Xi_Y(\xt,\yt;\yt',\xt') = 1-S^{(2)}(\xt,\yt)-S^{(2)}(\yt',\xt') + S^{(4)}(\xt,\yt;\yt',\xt') \,,
    \label{eq:LO_colorstructure}
\end{align}
where we define the CGC average of the dipole and quadrupole operators respectively as
\begin{align}
    \label{eq:dipole} S^{(2)}(\xt,\yt) &= \frac{1}{N_c} \left \langle \Tr[V(\xt)V^\dagger(\yt)] \right \rangle_Y \,, \\
    \label{eq:quadrupole} S^{(4)}(\xt,\yt;\yt',\xt') &= \frac{1}{N_c} \left \langle \Tr[V(\xt)V^\dagger(\yt)V(\yt')V^\dagger(\xt')] \right \rangle_Y \,.
\end{align}
The impact factor is obtained by squaring the perturbative factor in Eq.\,\eqref{eq:dijet-LO-pert} and summing over the helicities of the quark and the anti-quark in the final state:
\begin{align}
    \Rcal^{\lambda}(\rt,\rt') &= \sum_{\sigma,\sigma'}\Ncal^{\dagger \lambda}_{\sigma\sigma'} (\rt') \Ncal^{\lambda}_{\sigma\sigma'} (\rt)\,.
    \label{eq:impact-factor-LO}
\end{align}

The differential cross-section for quark anti-quark production in the CGC is then
\begin{multline}
\label{eq:cgc_xs}
    \frac{\der \sigma_{\mathrm{CGC}}^{\gamma_{\lambda}^*+A\rightarrow q\bar{q}+X}}{ \der^2 \ktone \der^2 \kttwo \der \eta_1 \der \eta_2} =  \frac{\alpha_\mathrm{em} e_f^2 N_c \delta_z}{ (2\pi)^6}   \int \der^8 \Pi \  e^{-i \ktone\cdot (\xt -\xt' )} e^{-i \kttwo\cdot (\yt-\yt') } \\
    \times \Xi_Y(\xt,\yt;\yt',\xt') \Rcal^{\lambda}(\xt-\yt,\xt'-\yt') \,,
\end{multline}
where $\alpha_\mathrm{em} = e^2/(4\pi)$ is the electromagnetic coupling constant, $\delta_z = \delta(1-z_1-z_2)$ is an overall minus light-cone momentum conserving delta function, and for convenience we defined the measure: 
\begin{align}
    \der^8 \Pi = \der^2 \xt \der^2 \yt \der^2 \yt' \der^2 \xt' \,.
\end{align}
The impact factors corresponding to longitudinally and transversely polarized photons can be easily obtained by inserting Eqs.\,\eqref{eq:dijet-LO-pert_L} and \eqref{eq:dijet-LO-pert_T} in Eq.\,\eqref{eq:impact-factor-LO}:
\begin{align}
    \Rcal^{L}(\rt,\rt') &=  8 (z_1 z_2)^3 Q K_0 (\varepsilon r_\perp) Q K_0 (\varepsilon r'_\perp) \,, 
    \label{eq:HardTMDL}\\
    \Rcal^{T}(\rt,\rt') &= 2(z_1z_2) \left[z_1^2 + z_2^2 \right] \frac{\rt \cdot \rt'}{r_\perp r'_\perp} \varepsilon K_1(\varepsilon r_\perp ) \varepsilon K_1(\varepsilon r'_\perp ) \,.
    \label{eq:HardTMDT}
\end{align}
For transversely polarized photon contributions we averaged over both polarizations $\lambda=\pm 1$. The obtained differential cross section agrees with the previous calculations presented in Ref.~\cite{Dominguez:2011wm,Dominguez:2011br,Metz:2011wb}.

A convenient choice of momentum variables to characterize the production of dijets is
\begin{align}
    \kt &= \ktone + \kttwo \,, \label{eq:momentum_imbalance}\\
    \Pt &= z_2 \ktone -z_1 \kttwo \,,
    \label{eq:relative_momentum}
\end{align}
which are conjugate to the coordinate variables:
\begin{align}
    \bt & = z_1 \xt+z_2\yt \,,\\
    \rt &= \xt-\yt \,,
\end{align}
respectively the dipole impact parameter and the dipole relative vector. 

The momentum imbalance $\kt$ measures the deviations of the dijet system from being back-to-back in the transverse plane, while the relative momentum $\Pt$ is closely related to the invariant mass of the dijet pair
\begin{align}
    M^2_{q\bar{q}} = (k_1 + k_2)^2 = \frac{P_\perp^2}{z_1z_2} \,.
\end{align}
This pair of momentum variables is also useful to establish the relation between the CGC EFT and the TMD framework as will be shown in the next section.

With this alternative choice of coordinates, the differential cross-section is written as
\begin{align}
    \frac{\der \sigma_{\mathrm{CGC}}^{\gamma_{\lambda}^*+A\rightarrow q\bar{q}+X}}{ \der^2 \Pt \der^2 \kt \der \eta_1 \der \eta_2} =  \frac{\alpha_\mathrm{em} e_f^2 N_c \delta_z}{ (2\pi)^6}   &\int \der^8 \tilde{\Pi} \  e^{-i \Pt \cdot (\rt -\rt' )} e^{-i \kt \cdot (\bt-\bt') } \nonumber \\
    &\quad \quad \tilde{\Xi}_Y(\rt,\bt,\rt',\bt') \Rcal^{\lambda}(\rt,\rt') \,,
    \label{eq:cgc_xs2}
\end{align}
where the measure is defined as
\begin{align}
    \der^8 \tilde{\Pi} = \der^2 \bt \der^2 \rt \der^2 \bt' \der^2 \rt' \,,
\end{align}
and
\begin{align}
    \tilde{\Xi}_Y(\rt,\bt,\rt',\bt') = \Xi_Y(\bt + z_2 \rt, \bt - z_1 \rt;\bt' - z_1 \rt', \bt' + z_2 \rt') \,,
\end{align}
with the right-hand side defined in Eq.\,\eqref{eq:LO_colorstructure}.

\subsection{The TMD limit}
\label{sec:inclusive_dijets_TMD}

The transverse momentum dependent (TMD) framework for the inclusive production of a quark anti-quark pair is expected to hold when the momentum imbalance is small relative to the typical transverse momenta of the $q$ and $\bar{q}$, i.e., $k_\perp \ll P_\perp$. In coordinate space, this condition is expected to be equivalent to the small dipole expansion $r_\perp \ll b_\perp$ (see \cite{Dominguez:2011wm}), although this assertion will be corrected in Section~\ref{sec:inclusive_dijets_ght}. Such an expansion applied to the Wilson line correlators in the CGC amplitude (Eq.\,\eqref{eq:M-matrix}) results in:
\begin{align}
    \left[\mathbb{1} - V\left(\xt \right) V^\dagger\left(\yt \right) \right] = -\rt^j \left[ V(\bt) \partial^j V^\dagger(\bt) \right] + \mathcal{O}(\rt^2) \,.
    \label{eq:dipole_smallr_expansion}
\end{align}
The correlator appearing on the right hand side is the pure gauge transverse gauge field one would obtain if working in the $\tilde{A}^+ = 0$ gauge \cite{JalilianMarian:1996xn}:
\begin{align}
    \Ats^i(\bt) = \frac{i}{g} \left[V(\bt) \partial^i V^\dagger(\bt) \right] \,,
    \label{eq:A_coordinate}
\end{align}
as introduced in Sec.\,\ref{sec:inclusive_dijets_CGC}.

Inserting the result of Eq.\,\eqref{eq:dipole_smallr_expansion} in Eq.\,\eqref{eq:M-matrix} and ignoring higher order terms in $r_\perp$ (or equivalently in $1/Q_\perp$), we obtain the amplitude in the TMD framework:
\begin{align}
    \Mcal^\lambda_{\mathrm{TMD}} = 2 g ee_f q^-\  \Ical^{\lambda,i}(\Pt) \Acalts^i(\kt) \,,
    \label{eq:M_TMD}
\end{align}
where
\begin{align}
    \Ical_{\sigma\sigma'}^{\lambda,i}(\Pt) &= \int \frac{\der^2 \rt}{2\pi}  e^{-i \Pt \cdot \rt }  \Ncal^\lambda_{\sigma\sigma'}(\rt) i \rt^i \,, 
    \label{eq:I_TMD} \\
    \Acalts^i(\kt) &=  \int \der^2 \bt  e^{-i \kt \cdot \bt } \Ats^i(\bt) \label{eq:Amomentum} \,.
\end{align}
The perturbative factors in Eq.\eqref{eq:I_TMD} for longitudinally and transversely polarized photons are given respectively by:
\begin{align}
    \Ical_{\sigma\sigma'}^{\lambda=0,i}(\Pt) &=    4 (z_1z_2)^{3/2} Q \delta_{\sigma,- \sigma'} \frac{\Pt^i}{(P_\perp^2 + \varepsilon^2)^2} \,,  \\
    \Ical_{\sigma\sigma'}^{\lambda=\pm 1,i}(\Pt) &=  (z_1z_2)^{1/2} \left[  (z_2-z_1) + \sigma \lambda \right] \delta_{\sigma,- \sigma'} \frac{1}{P_\perp^2 + \varepsilon^2} \left(\delta^{ij} - \frac{2\Pt^i\Pt^j}{P_\perp^2 + \varepsilon^2} \right) \et^{\lambda,j} \,.
\end{align}
Then the differential cross-section reads:
\begin{align}
    \frac{\der \sigma_{\mathrm{TMD}}^{\gamma_{\lambda}^*+A\rightarrow q\bar{q}+X}}{ \der^2 \Pt \der^2 \kt \der \eta_1 \der \eta_2} =  \alpha_s \alpha_\mathrm{em} e_f^2 \delta_z \Hcal_{\rm{TMD}}^{ij,\lambda}(\Pt)  x G^{ij}(x,\kt) \,,
\end{align}
where the hard factor is
\begin{align}
    \Hcal_{\rm{TMD}}^{ij,\lambda}(\Pt) = \frac{1}{2} \sum_{\sigma,\sigma'}\Ical^{\dagger \lambda,i}_{\sigma\sigma'} (\Pt) \Ical^{\lambda,j}_{\sigma\sigma'} (\Pt) \,,
\end{align}
and the soft factor is the Weizsäcker-Williams \cite{Kharzeev:2003wz,Dominguez:2011wm} gluon TMD defined as (see Appendix\,\ref{sec:Appendix_WW} for more details):
\begin{align}
    xG^{ij} (x,\kt) = \frac{4}{(2\pi)^3} \Big\langle \Tr[ \Acalts^{\dagger,i}(\kt) \Acalts^{j}(\kt) ]  \Big\rangle_{x}
    \label{eq:WW_gluon_1} \,.
\end{align}
While at large $k_\perp$, one expects a power tail $xG^{ii}(x,\kt)  \sim 1/k_\perp^2$ from perturbative QCD, at smaller momentum imbalance $k_\perp \lesssim Q_s(x)$ the saturation framework predicts logarithmic behavior $xG^{ii}(x,\kt)  \sim \log(Q_s^2(x)/k_\perp^2)$ \cite{Dominguez:2011br}. This behavior at low momentum imbalance results in a nuclear suppression of back-to-back dihadron production in DIS \cite{Zheng:2014vka}. 

The differential cross-section for quark anti-quark production in DIS within the TMD limit is given by  \cite{Dominguez:2011wm,Dominguez:2011br,Metz:2011wb}:
\begin{align}
    \frac{\der \sigma_{\mathrm{TMD}}^{\gamma_{\mathrm{L}}^*+A\rightarrow q\bar{q}+X}}{ \der^2 \Pt \der^2 \kt \der \eta_1 \der \eta_2} &=    \alpha_s \alpha_\mathrm{em} e_f^2 \delta_z  (z_1 z_2)^3  \frac{8Q^2 P_\perp^2}{(P_\perp^2 + \varepsilon^2)^4}\nonumber \\
    & \quad \times \left[xG^0(x,k_\perp) + xh^0(x,k_\perp) \cos 2\phi  \right] \,,
    \label{eq:dijetTMD_L}\\
    \frac{\der \sigma_{\mathrm{TMD}}^{\gamma_{\mathrm{T}}^*+A\rightarrow q\bar{q}+X}}{ \der^2 \Pt \der^2 \kt \der \eta_1 \der \eta_2} &=   \alpha_s \alpha_\mathrm{em} e_f^2 \delta_z  (z_1 z_2)  \left[z_1^2 + z_2^2 \right]  \frac{P_\perp^4 + \varepsilon^4}{(P_\perp^2 + \varepsilon^2)^4}\nonumber \\
    & \quad \times \left[xG^0(x,k_\perp) - \frac{2\varepsilon^2 P_\perp^2}{(P_\perp^4 + \varepsilon^4)}xh^0(x,k_\perp) \cos 2\phi  \right] \,,
    \label{eq:dijetTMD_T}
\end{align}
where $\phi$ is the angle between $\Pt$ and $\kt$, and the WW correlator has been decomposed into trace and traceless parts:
\begin{align}
    xG^{ij} (x,\kt) = \frac{1}{2}\delta^{ij}xG^0 (x,k_\perp) + \frac{1}{2}\left(\frac{2\kt^i\kt^j}{k_\perp^2} - \delta^{ij} \right)xh^0(x,k_\perp) \,.
    \label{eq:ww_decomposition}
\end{align}
The trace and traceless parts are known as unpolarized and linearly polarized gluon WW TMD distributions, respectively. They obey the inequality \begin{align}
    xh^{0} (x,k_\perp) \leq xG^{0} (x,k_\perp) \label{eq:TMD_bound}\,,
\end{align}
which renders the cross-sections in Eqs.\,\eqref{eq:dijetTMD_L} and \eqref{eq:dijetTMD_T} positive \cite{Mulders:2000sh}.

Note that in the TMD framework the elliptic anisotropy defined as
\begin{align}
    v_{2,\lambda}  = \int \der \phi_{\Pt}
    \der \phi_{\kt} e^{2i(\phi_{\Pt}-\phi_{\kt})}\frac{\der \sigma^{\gamma_{\lambda}^*+A\rightarrow q\bar{q}+X}}{ \der^2 \Pt \der^2 \kt \der \eta_1 \der \eta_2} \Bigg/ \int \der \phi_{\Pt}
    \der \phi_{\kt} \frac{\der \sigma^{\gamma_{\lambda}^*+A\rightarrow q\bar{q}+X}}{ \der^2 \Pt \der^2 \kt \der \eta_1 \der \eta_2} \,,
\end{align}
is proportional to the ratio of the linearly polarized to the unpolarized distribution \cite{Dominguez:2011br,Dumitru:2015gaa}:
\begin{align}
    v_{2,\rm{L}} &= \frac{1}{2}\frac{xh^{0} (x,k_\perp)}{xG^{0} (x,k_\perp)} \,,
    \label{eq:v2LTMD}\\
    v_{2,\rm{T}} &= - \frac{\varepsilon^2 P_\perp^2}{P_\perp^4 + \varepsilon^4} \frac{xh^{0} (x,k_\perp)}{xG^{0} (x,k_\perp)} 
     \label{eq:v2TTMD}\,.
\end{align}

\subsection{The improved TMD framework: resumming kinematic twists}
\label{sec:inclusive_dijets_iTMD}

While the CGC EFT takes into account all possible twist corrections that are not suppressed by a factor of $1/W$, the so-called small-$x$ improved TMD framework (ITMD for short) \cite{Kotko:2015ura} only resums specific power corrections which correspond to kinematic corrections in the hard sub-amplitude: while the TMD framework expresses the observable as
\begin{align}
    \der \sigma \sim {\cal H}^{ij,\lambda}_{\mathrm{TMD}}(\Pt) xG^{ij}(x,\kt) \,,
\end{align}
the ITMD framework rewrites it as 
\begin{align}
    \der \sigma \sim {\cal H}^{ij,\lambda}_{\mathrm{ITMD}}(\Pt,\kt) xG^{ij}(x,\kt) \,,
\end{align} with ${\cal H}^{ij,\lambda}_{\mathrm{ITMD}}(\Pt,\vect{0}) = {\cal H}^{ij,\lambda}_{\mathrm{TMD}}(\Pt)$. It is crucial that the distribution $xG^{ij}$ is the same for both schemes. Indeed, the ITMD framework neglects higher order corrections $g \Ats^i$ in the operators that appear in the CGC EFT. Unlike the TMD approximation, where one expands in $\rt$, our starting point is the truncation of the expansion introduced in Sec.\,\ref{sec:field_stregnth_WilsonLines} (see Eq.\eqref{eq:dipole_itmd_twist}):
\begin{align}
    \left[ \mathbb{1}-V(\xt)V^\dagger(\yt) \right]_{\mathrm{ITMD}} &= ig \int_{\yt}^{\xt} \der \zt^i  \Ats^i(\zt) \,.
\end{align}
This expression can be cast into the following form (see Appendix\,\ref{sec:Wilson_and_gaugelink}):
\begin{equation}
    \left[ \mathbb{1}-V(\xt)V^{\dagger}(\yt) \right]_{\mathrm{ITMD}} =i \rt^{i} \int \der^{2}\vt \int\frac{\der^{2} \ellt }{(2\pi)^{2}}e^{-i \ellt \cdot \vt } \ g \Ats^{i}(\vt) \left(\frac{e^{i\ellt\cdot\xt}-e^{i\ellt\cdot\yt}}{i\ellt\cdot\rt}\right)
    \label{eq:1-body-operator} \,.
\end{equation}
When plugged into the amplitude from Eq.~(\ref{eq:M-matrix}), it yields:
\begin{align}
    \Mcal_{\rm{ITMD}}^{\lambda} & = 2 g e e_{f}q^{-} \Jcal^{\lambda,i}(\Pt,\kt) \Acalts^i(\kt)
    \label{eq:M-matrix-ITMD} \,,
\end{align}
where $\Acalts^i(\kt)$ has been defined in Eq.\,\eqref{eq:Amomentum} and the hard factor is defined as
\begin{align}
    \Jcal_{\sigma\sigma'}^{\lambda,i}(\Pt,\kt) = \int \frac{\der^2 \rt}{2\pi} e^{-i\Pt\cdot\rt}   \Ncal_{\sigma\sigma^{\prime}}^{\lambda}(\rt) \ i \rt^{i} \left(\frac{e^{iz_{2}(\kt\cdot\rt)}-e^{-iz_{1}(\kt\cdot\rt)}}{i\kt\cdot\rt}\right) \,.
    \label{eq:ITMD_J}
\end{align}
Note that by neglecting powers of $\kt \cdot \rt$ (or equivalently powers in $k_\perp / Q_\perp$) one recovers Eq.\,\eqref{eq:I_TMD} and thus Eq.\,\eqref{eq:M-matrix-ITMD} reduces to the TMD limit in Eq.\,\eqref{eq:M_TMD}.

The differential cross-section for the production of an quark anti-quark pair in the ITMD framework reads:
\begin{align}
    \frac{\der \sigma_{\mathrm{ITMD}}^{\gamma_{\lambda}^*+A\rightarrow q\bar{q}+X}}{ \der^2 \Pt \der^2 \kt \der \eta_1 \der \eta_2} =  \alpha_s \alpha_\mathrm{em} e_f^2 \delta_z \Hcal_{\rm{ITMD}}^{ij,\lambda}(\Pt,\kt)  x G^{ij}(x,\kt) \,,
    \label{eq:ITMD_xsec}
\end{align}
where
\begin{align}
    \Hcal_{\rm{ITMD}}^{ij,\lambda}(\Pt,\kt) = \frac{1}{2} \sum_{\sigma,\sigma'}\Jcal^{\dagger \lambda,i}_{\sigma\sigma'} (\Pt,\kt) \Jcal^{\lambda,j}_{\sigma\sigma'} (\Pt,\kt) \,.
    \label{eq:ITMD_hard}
\end{align}
The computation is almost identical to that in the TMD framework, except for the hard factor in Eq.\,\eqref{eq:ITMD_hard}, whose dependence on $\kt$ resums kinematic power corrections in $k_\perp/Q_\perp$. 
The hard factor in the ITMD framework (Eq.\,\eqref{eq:ITMD_J}) can be computed fully analytically (see Appendix \ref{sec:computing_J_hardfactor_ITMD_dijets} for details). When the virtual photon is longitudinally polarized, we find
\begin{align}
    \Jcal^{\lambda=0,i}_{\sigma,\sigma'}(\Pt,\kt) &= 2 (z_1 z_2)^{3/2} Q \delta_{\sigma,-\sigma'} \left \{ \frac{\ktone^{i}-\kttwo^{i}}{(\ktone^{2}+\varepsilon^{2})(\kttwo^{2}+\varepsilon^{2})} \right. \nonumber \\
    &+\frac{ \left[\kt^{2}\Pt^{i}-\left(\Pt\cdot\kt\right)\kt^{i}\right]}{\mathcal{X}^{3}} \left[ \arctan\left(\frac{\kt \cdot \ktone}{\mathcal{X}} \right) + \arctan\left(\frac{\kt \cdot \kttwo}{\mathcal{X}} \right)\right] \nonumber \\
    & \left. - \frac{\left[\kt^{2}\Pt^{i}-\left(\Pt\cdot\kt\right)\kt^{i}\right]}{\mathcal{X}^2}\left(\frac{-\ktone\cdot\kttwo + \varepsilon^2}{(\ktone^{2}+\varepsilon^{2})(\kttwo^{2}+\varepsilon^{2})}\right) \right \} \,,
    \label{eq:ITMD_hardL_analytic}
\end{align}
and for the transversely polarized photon, we have
\begin{align}
    \Jcal^{\lambda=\pm 1,i}_{\sigma,\sigma'}(\Pt,\kt) &= (z_1 z_2)^{1/2} \left[(z_2 -z_1) + \sigma \lambda \right] \delta_{\sigma,-\sigma'} \et^{\lambda = \pm 1,j} \nonumber \\
    & \times \left\{ \frac{\varepsilon^{2}(\delta^{ij}\kt^{2}-\kt^{i}\kt^{j})}{\mathcal{X}^{3}}\left[ \arctan\left(\frac{\kt \cdot \ktone}{\mathcal{X}} \right) + \arctan\left(\frac{\kt \cdot \kttwo}{\mathcal{X}} \right)\right] \right. \nonumber \\
    & +\frac{1}{\kt^{2}}\frac{\varepsilon^{2}(\delta^{ij}\kt^{2}-\kt^{i}\kt^{j})\kt^{k}}{\mathcal{X}^2}\left(\frac{\ktone^{k}}{\ktone^{2}+\varepsilon^{2}}+\frac{\kttwo^{k}}{\kttwo^{2}+\varepsilon^{2}}\right)\nonumber \\
    & \left. +\frac{1}{\kt^{2}}(\kt^{i}\delta^{jk}+\kt^{j}\delta^{ik}-\kt^{k}\delta^{ij})\left(\frac{\ktone^{k}}{\ktone^{2}+\varepsilon^{2}}+\frac{\kttwo^{k}}{\kttwo^{2}+\varepsilon^{2}}\right) \right\} \,,
    \label{eq:ITMD_hardT_analytic}
\end{align}
where we used $\ktone = \Pt + z_1 \kt$,\  $ \kttwo = -\Pt + z_2 \kt$ (see Eqs.\,\eqref{eq:momentum_imbalance} and \eqref{eq:relative_momentum}), and we defined:
\begin{align}
    \mathcal{X}^2 &= \Pt^{2}\kt^{2}-(\Pt\cdot\kt)^{2}+\varepsilon^{2}\kt^{2} \,.
\end{align}
These expressions will be the ones used in our numerical evaluation of the ITMD differential cross-section\footnote{While this work was in progress, the hard ITMD factors were computed independently in \cite{Altinoluk:2021ygv}. We have verified that both results are equivalent.}. Some interesting limits are discussed in Appendix\,\ref{sec:computing_J_hardfactor_ITMD_dijets}.

\subsection{Genuine higher twists}\label{sec:inclusive_dijets_ght}

In the landmark article on the equivalence between CGC observables and TMD-factorized observables~\cite{Dominguez:2011wm}, the notion of the correlation limit was introduced as a means to justify taking the Taylor expansion from which TMD distributions emerge as correlators of Wilson lines and derivatives of Wilson lines. While this limit does correspond to the TMD limit, there is an additional subtlety worth discussing. Indeed, it is common to identify the correlation limit with the back-to-back kinematic limit. Given that $\rt$ and $\bt$ are respectively Fourier conjugated to $\Pt$ and $\kt$, in the $k_\perp \ll P_\perp$ limit one can justify assuming $r_\perp \ll b_\perp$.
More recent insights on the CGC/TMD correspondence~\cite{Altinoluk:2019wyu} have revealed the limit of this hypothesis. Because in CGC observables dipole sizes appear both in the non-perturbative correlators and in the hard subamplitudes, powers of $\rt$ can actually be enhanced by powers of the saturation scale $Q_s$ via non-perturbative effects in the target. Assuming that the $k_\perp \ll P_\perp$ limit (or more precisely $k_\perp \ll Q_\perp$) and the $r_\perp\ll b_\perp$ limit are indistinguishable is tantamount to neglecting all powers of $Q_s/Q_\perp$. As we will show in this section, there is actually a non-zero contribution from genuine higher twists even in the $k_\perp \rightarrow 0$ limit. These $Q_s/Q_\perp$ corrections are taken into account neither in the TMD nor  ITMD framework.\footnote{Recall the saturation scale $Q_s$ appears in the TMD (and ITMD) framework only as powers $Q_s/k_\perp$ in the WW gluon TMD.}

The quadratic term in Eq.\,\eqref{eq:dipole_itmd_twist} can be cast into the following form (see Appendix\,\ref{sec:Wilson_and_gaugelink}):
\begin{align}
    \left[\mathbb{1}-V(\xt)V^{\dagger}(\yt)\right]&_{\mathrm{g.h.t}} = \int\frac{{\rm d}^{2}\ltone}{(2\pi)^{2}}\frac{{\rm d}^{2}\lttwo}{(2\pi)^{2}}\int{\rm d}^{2}\vtone{\rm d}^{2}\vttwo{\rm e}^{-i(\ltone\cdot\vtone)-i(\lttwo\cdot\vttwo)}\frac{\rt^{i}\rt^{j}}{i(\lttwo\cdot\rt)}\nonumber \\
    & \times\left(\frac{{\rm e}^{i(\ltone+\lttwo)\cdot\boldsymbol{x}_{\perp}}-{\rm e}^{i(\ltone+\lttwo)\cdot\boldsymbol{y}_{\perp}}}{i(\ltone+\lttwo)\cdot\rt}-{\rm e}^{i(\lttwo\cdot\boldsymbol{y}_{\perp})}\frac{{\rm e}^{i(\ltone\cdot\boldsymbol{x}_{\perp})}-{\rm e}^{i(\ltone\cdot\boldsymbol{y}_{\perp})}}{i(\ltone\cdot\rt)}\right)\nonumber \\
    & \times g^2 A^{i}(\vtone)V(\vtone)V^{\dagger}(\vttwo)A^{j}(\vttwo)\,. \label{eq:ght}
\end{align}
The last line in the r.h.s. of this equation is a building block to construct genuine higher twist TMD distributions, e.g. gauge invariant distributions with 4 field strength tensors carrying non-zero transverse momenta, but it is worth noting that the $V(\vtone)V^{\dagger}(\vttwo)$ operator could also be recursively expanded in powers of $g\Ats^i$ \textit{ad nauseam}.
When plugged into the CGC amplitude from Eq.~(\ref{eq:M-matrix}), we obtain the following  $(g^2\Ats^2)$-suppressed contribution:
\begin{align}
{\cal M}_{\rm g.h.t.}^{\lambda} & =\frac{ee_{f}q^{-}g^{2}}{\pi}\int\frac{{\rm d}^{2}\boldsymbol{\ell}_{\perp}}{(2\pi)^{2}}\int{\rm d}^{2}\vtone{\rm d}^{2}\vttwo{\rm e}^{-i(\boldsymbol{\ell}_{\perp}\cdot\vtone)-i(\kt-\boldsymbol{\ell}_{\perp})\cdot\vttwo}\nonumber \\
 & \times A^{i}(\vtone)V(\vtone)V^{\dagger}(\vttwo)A^{j}(\vttwo)\label{eq:M2}\\
 & \times\int{\rm d}^{2}\rt{\rm e}^{-i(\ktone\cdot\rt)}{\cal N}_{\sigma\sigma^{\prime}}^{\lambda}(\rt)\nonumber \\
 & \times\frac{\rt^{i}\rt^{j}}{i(\kt-\boldsymbol{\ell}_{\perp})\cdot\rt}\left(\frac{{\rm e}^{i(\kt\cdot\rt)}-1}{i(\kt\cdot\rt)}-\frac{{\rm e}^{i(\boldsymbol{\ell}_{\perp}\cdot\rt)}-1}{i(\boldsymbol{\ell}_{\perp}\cdot\rt)}\right)\nonumber .
\end{align}
We can now see the mechanism through which a $Q_s/P_\perp$ correction emerges from such a contribution. The 2 gluons in the amplitude both carry transverse momenta, whose sum is $\kt$. The momentum difference $\ellt$, which is Fourier conjugated to the difference in positions $\vtone$ and $\vttwo$, provides an additional loop variable. In CGC descriptions for the target, the momentum integral will typically peak around $|\ellt| \sim Q_s$, while due to the hard factors the dipole sizes peak around $r_\perp \sim 1/Q_\perp$. Overall, every power of $(\ellt\cdot\rt)$ yields a $Q_s/Q_\perp$ correction, regardless of the magnitude of $\kt$.
Taking the limit $\kt \rightarrow \boldsymbol{0}_\perp$ (back-to-back) in the hard subfactor, we get the non-zero correction
\begin{align}
\left.{\cal M}_{\rm g.h.t.}^{\lambda}\right|_{\kt\rightarrow \boldsymbol{0}_{\perp}} & =\frac{ee_{f}q^{-}g^{2}}{\pi}\int\frac{{\rm d}^{2}\boldsymbol{\ell}_{\perp}}{(2\pi)^{2}}\int{\rm d}^{2}\vtone{\rm d}^{2}\vttwo{\rm e}^{-i(\boldsymbol{\ell}_{\perp}\cdot\vtone)-i(\kt-\boldsymbol{\ell}_{\perp})\cdot\vttwo}\nonumber \\
 & \times A^{i}(\vtone)V(\vtone)V^{\dagger}(\vttwo)A^{j}(\vttwo)\nonumber\\
 & \times\int{\rm d}^{2}\rt{\rm e}^{-i(\boldsymbol{P}_{\perp}\cdot\rt)}{\cal N}_{\sigma\sigma^{\prime}}^{\lambda}(\rt)\nonumber \\
 & \times\frac{\rt^{i}\rt^{j}}{i(\boldsymbol{\ell}_{\perp}\cdot\rt)}\left(\frac{{\rm e}^{i(\boldsymbol{\ell}_{\perp}\cdot\rt)}-1-i(\boldsymbol{\ell}_{\perp}\cdot\rt)}{i(\boldsymbol{\ell}_{\perp}\cdot\rt)}\right) \label{eq:M2-b2b}\,.
\end{align}
We finally see the difference between the so-called correlation (small dipole) limit and the kinematic back-to-back limit: genuine higher twists contribute in the latter but not in the former.
Let us illustrate this more concretely by taking the leading power of $Q_\perp$ in Eq.~(\ref{eq:M2-b2b}). From the previous discussion, we know that each power of $(\ellt\cdot\rt)$ yields a power correction. From the leading term of such an expansion, we get
\begin{align}
\left.{\cal M}_{{\rm g.h.t.}}^{\lambda}\right|_{\kt \rightarrow \boldsymbol{0}_{\perp}} & \!\!\!\!\!\! \simeq\frac{ee_{f}q^{-}g^{2}}{2\pi}\int{\rm d}^{2}\bt{\rm e}^{-i(\kt\cdot\bt)}A^{i}(\bt)A^{j}(\bt) \int{\rm d}^{2}\rt{\rm e}^{-i(\boldsymbol{P}_{\perp}\cdot\rt)}\rt^{i}\rt^{j}{\cal N}_{\sigma\sigma^{\prime}}^{\lambda}(\rt)\label{eq:M2-b2b-LT}. 
\end{align}
It is particularly revealing to compare this to the TMD amplitude, given by
\begin{align}
{\cal M}_{{\rm TMD}}^{\lambda} & =\frac{ee_{f}q^{-}g}{\pi}\int{\rm d}^{2}\bt{\rm e}^{-i(\kt\cdot\bt)}A^{i}(\bt) \int{\rm d}^{2}\rt{\rm e}^{-i(\boldsymbol{P}_{\perp}\cdot\rt)}i\rt^{i}{\cal N}_{\sigma\sigma^{\prime}}^{\lambda}(\rt)\label{eq:M-TMD}.
\end{align}
We immediately see that in ${\cal M}_{{\rm g.h.t.}}^{\lambda}$, the hard factor is enhanced by a power of $r_\perp$ hence suppressed by a power of the hard scale after the integral is taken. This additional power is compensated by the fact that the operator now has its dimension (hence its twist) increased by 1 due to the presence of an additional $gA^j$ insertion. The action of a higher twist operator on target states will be one higher power of the typical target scale: $Q_s$ in the CGC or $\Lambda_{\rm QCD}$ in dilute schemes.

To sum up this section: genuine higher twist corrections are suppressed by a power of the ratio between a target scale and a hard scale, and they contribute to the observable regardless of how $k_\perp$ compares to $P_\perp$ or $Q$. In the CGC, we expect the ratio to be linear in $Q_s$, which means comparing CGC predictions to TMD predictions for dense targets in the exact back-to-back limit would provide a probe for genuine saturation effects.

\section{Numerical setup}
\label{sec:set_up_dijets}

\subsection{Initial conditions and small-$x$ evolution}
\label{sec:set_up_dijets_IC}

The CGC target average of the dipole and quadrupole operators, Eqs.~\eqref{eq:dipole} and~\eqref{eq:quadrupole}, are necessary ingredients to evaluate the dijet production cross section \eqref{eq:cgc_xs}, and are defined as correlators of Wilson lines $V(\xt)$. The Weizsäcker-Williams gluon distribution, Eq.~\eqref{eq:WW_gluon_1}, can also be written in terms of the dipole operator (and its derivatives) as we will show explicitly below.
The energy (or momentum fraction $x$) dependence of the Wilson lines is given by the JIMWLK evolution equation~\cite{JalilianMarian:1996xn,JalilianMarian:1997jx, JalilianMarian:1997gr,Iancu:2001md, Ferreiro:2001qy, Iancu:2001ad, Iancu:2000hn}. In the so-called planar limit where the number of colors $N_c$ is considered large enough to neglect $1/(N_c^2-1)$ corrections, it reduces to the Balitsky-Kovchegov (BK) evolution equation~\cite{Balitsky:1995ub,Kovchegov:1999yj}, which is a closed equation to describe the energy evolution of the dipole operator.

The JIMWLK evolution of Wilson lines on a transverse lattice can be solved numerically~\cite{Blaizot:2002np,Rummukainen:2003ns,Mantysaari:2018zdd,Cali:2021tsh}, which would allow one to directly evaluate the 2- and 4-point correlators, Eqs.~\eqref{eq:dipole} and~\eqref{eq:quadrupole}. However, this is computationally demanding, and instead we will employ the Gaussian approximation discussed in Sec.~\ref{sec:gaussian_approx} which allows one to express higher point correlators in terms of the dipole correlator only\footnote{The calculation of WW gluon distribution has been done beyond the Gaussian approximation in \cite{Dumitru:2015gaa}.
}.

We use the dipole-target scattering amplitude obtained in  Ref.~\cite{Lappi:2013zma} (see also Ref.~\cite{Albacete:2010sy}). Here, at the initial $x=0.01$ the functional form of the dipole operator (in case of proton targets) is obtained from a McLerran-Venugopalan model~\cite{McLerran:1993ni} based parametrization and written as (see also Refs.~\cite{Dumitru:2020gla,Dumitru:2021tvw}) 
\begin{equation}
    S^{(2)}(\xt,\yt)_{x=0.01} = \exp\left[ -\frac{\rt^2 Q_{s,0}^2}{4} \ln \left( \frac{1}{r_\perp \Lambda_\text{QCD}} + e\right) \right].
\end{equation}
The dipole amplitude at $x<0.01$ is obtained by solving the BK equation over $Y=\ln\frac{0.01}{x}$ units of rapidity, including running coupling corrections~\cite{Balitsky:2006wa}.
The free parameters ($Q_{s,0}^2$, the scale of the running coupling in coordinate space and the proton transverse area $S_\perp$) are determined in Ref.~\cite{Lappi:2013zma} by performing a fit to the proton structure function data from HERA~\cite{Aaron:2009aa}. We note that there has recently been progress to promote the structure function calculations to next-to-leading order accuracy~\cite{Beuf:2021qqa,Ducloue:2017ftk,Beuf:2017bpd,Beuf:2016wdz,Lappi:2016oup,Lappi:2016fmu,Hanninen:2017ddy,Beuf:2020dxl}, but for consistency we use the leading order result from Ref.~\cite{Lappi:2013zma} in our leading order calculations.

To generalize the dipole-proton operator $S^{(2)}$ to describe the interaction with a heavy nucleus, we again follow Ref.~\cite{Lappi:2013zma}. At the initial condition, the dipole operator is written as 
\begin{equation}
    S^{(2)}(\xt,\yt)_{x=0.01} = \exp\left[ -S_\perp A T_A(\bt)  \frac{\rt^2 Q_{s,0}^2}{4} \ln \left( \frac{1}{r_\perp \Lambda_\text{QCD}} +  e\right) \right] \,,
\end{equation}
with $S_\perp=18.81$ mb
, and $T_A$ is the spatial density profile obtained from the Woods-Saxon distribution.
To obtain the dipole amplitude at smaller $x$, the BK equation is solved at fixed impact parameter. This approach results in vanishing nuclear effects in the dilute region and successful phenomenology, see e.g. Refs.~\cite{Mantysaari:2019nnt,Ducloue:2016pqr,Ducloue:2015gfa}. Since the determination of the impact parameter is subtle in electron-nucleus collisions (see however Refs.~\cite{Zheng:2014cha,Lappi:2014foa}), in this work (as in Ref.~\cite{Mantysaari:2019hkq}) we use an impact parameter independent  dipole amplitude, obtained by evaluating the dipole at the median impact parameter $\langle b_\perp\rangle$ defined as  $\langle b_\perp \rangle = \int \der^2\bt b_\perp T_A(\bt)$ with the normalization $\int \der^2 \bt T_A(\bt)=1$, as a proxy for minimum bias collisions. This results in a nuclear \textit{oohmp} factor of the effective initial nuclear saturation scale 
\begin{align}
    Q^2_{s,\rm{A}0} = S_\perp A T_A(\langle b_\perp \rangle) \ Q^2_{s,0} \,,
\end{align}
relative to the proton saturation scale. For gold nuclei, we find $S_\perp A T_A(\langle b_\perp \rangle) \approx 3.1$ .

\subsection{Gaussian approximation for high energy correlators}
\label{sec:gaussian_approx}

Assuming that the color charges in the target are Gaussian distributed both at the initial condition and after the small-$x$ evolution, it becomes possible to express all higher point correlators in terms of the two-point correlator only. This is referred to as a Gaussian approximation or Gaussian truncation~\cite{Marquet:2010cf,Dominguez:2011wm,Fujii:2006ab}, and is used in this work to express the quadrupole operator~\eqref{eq:quadrupole} in terms of the dipole operator~\eqref{eq:dipole} satisfying the BK evolution as discussed in the previous Section. The validity of the Gaussian approximation has been numerically confirmed in Ref.~\cite{Dumitru:2011vk} by comparing the approximatively calculated quadrupole operator to the one obtained from JIMWLK-evolved Wilson lines, and analytically justified in \cite{Iancu:2011nj}. It has also been used for example in phenomenological analyses of multi particle production~\cite{Lappi:2012nh,Dusling:2017aot,Dusling:2017dqg} and to evaluate the higher point correlators in the next-to-leading order BK evolution equation~\cite{Balitsky:2008zza,Lappi:2016fmu} in Ref.~\cite{Lappi:2020srm}. Following Ref.~\cite{Dominguez:2011wm}, the quadrupole operator in the Gaussian approximation can be written as
\begin{align}
    S^{(4)}_Y&(\xt,\yt;\yt',\xt') \nonumber \\
    =&\  \mathrm{exp}\left(-\Gamma_Y(\xt-\yt)-\Gamma_Y(\yt'-\xt')\right) \nonumber \\
    &\left[ \left( \frac{\sqrt{\Delta_Y} + F_Y(\xt,\yt';\yt,\xt')}{2\sqrt{\Delta_Y}} -\frac{ F_Y(\xt,\yt;\yt',\xt')}{\sqrt{\Delta_Y}} \right)\mathrm{exp}\left(\frac{N_c}{4}\sqrt{\Delta_Y}\right) \right.  \nonumber \\
    &\left. + \left( \frac{\sqrt{\Delta_Y} - F_Y(\xt,\yt';\yt,\xt')}{2\sqrt{\Delta_Y}} +\frac{ F_Y(\xt,\yt;\yt',\xt')}{\sqrt{\Delta_Y}} \right)\mathrm{exp}\left(-\frac{N_c}{4}\sqrt{\Delta_Y}\right) \right] \nonumber \\
    &\times \mathrm{exp}\left(-\frac{N_c}{4}F_Y(\xt,\yt';\yt,\xt') + \frac{1}{2N_c}F_Y(\xt,\yt;\yt',\xt')\right).
\end{align}
Here we used the following definitions:
\begin{align}
\Gamma_Y(\xt-\yt) &= - \ln S^{(2)}(\xt,\yt) \\
\Delta_Y &= F_Y^2(\xt,\yt';\yt,\xt')+ \frac{4}{N_c^2} F_Y(\xt,\yt;\yt',\xt') F_Y(\xt,\xt';\yt',\yt), \\
F_Y(\xt,\yt;\yt',\xt')& = \frac{1}{C_F} \ln\left[\frac{S^{(2)}(\xt-\yt')   S^{(2)}(\yt-\xt')}{S^{(2)}(\xt-\xt') S^{(2)}(\yt-\yt')}\right],
\end{align}
where $C_F = (N_c^2-1)/(2N_c)=4/3$ is the fundamental Casimir. Note that we assumed that the dipole only depends on its relative vector and not the impact parameter.

The Weizsäcker-Williams distribution from Eq.~\eqref{eq:WW_gluon_1} can be conveniently written in terms of $\Gamma$ and its derivatives as \cite{Dominguez:2011wm,Dumitru:2016jku,Lappi:2017skr} 
\begin{align}
    xG^{ij} (x,\kt) = \frac{S_\perp}{\alpha_s} \frac{2 C_F}{C_A} \int \frac{\der^2 \Rt}{(2\pi)^4}  e^{-i \kt \cdot \Rt} \frac{\partial^i \partial^j \Gamma(\Rt) }{\Gamma(\Rt) } \left[1- \mathrm{exp}\left(-\frac{C_A}{C_F} \Gamma_Y(\Rt)  \right)\right].
\end{align}
where $C_A=N_c =3$ is the adjoint Casimir, and $S_\perp$ represents the transverse area of the target, which factors out when the dipole is translationally invariant (impact parameter independent). The trace and traceless components $xG^0$ and $xh^0$ defined in Eq.~\eqref{eq:ww_decomposition} can now be written as
\begin{align}
    xG^{0} (x,\kt) = \frac{S_\perp (N_c^2 -1)}{(2\pi)^3\alpha_s N_c} \int R_\perp \der R_\perp &J_0(R_\perp k_\perp)   \left[1- \mathrm{exp}\left(-\frac{C_A}{C_F} \Gamma_Y(R_\perp)  \right)\right] \nonumber \\
    & \times \frac{1}{\Gamma_Y(R_\perp)} \left[\frac{\der^2}{\der R_\perp^2} + \frac{1}{R_\perp}\frac{\der}{\der R_\perp} \right] \Gamma_Y(R_\perp) \,,
\end{align}
\begin{align}
    xh^{0} (x,\kt) = \frac{S_\perp (N_c^2 -1)}{(2\pi)^3\alpha_s N_c} \int R_\perp \der R_\perp &J_2(R_\perp k_\perp)  \left[1- \mathrm{exp}\left(-\frac{C_A}{C_F} \Gamma_Y(R_\perp)  \right)\right] \nonumber \\
    & \times \frac{1}{\Gamma_Y(R_\perp)} \left[\frac{1}{R_\perp}\frac{\der}{\der R_\perp} - \frac{\der^2}{\der R_\perp^2}  \right] \Gamma_Y(R_\perp).
\end{align}
With these results, it becomes possible to evaluate all cross sections in terms of the BK evolved dipole operator $S^{(2)}$ only.

\subsection{Computing harmonics}
\label{sec:harmonics}
Before we proceed, we note that due to the translational invariance of the dipole, the differential cross-section is proportional to the overall area of the target (proton/nucleus) $S_\perp$. Thus, we shall study the differential yield
\begin{align}
    \frac{\der N^{\gamma_{\lambda}^*+A\rightarrow q\bar{q}+X}}{ \der^2 \Pt \der^2 \kt \der \eta_1 \der \eta_2} = \frac{1}{S_\perp}\frac{\der \sigma^{\gamma_{\lambda}^*+A\rightarrow q\bar{q}+X}}{ \der^2 \Pt \der^2 \kt \der \eta_1 \der \eta_2} \,.
\end{align}
Due to overall rotational invariance, the differential yield is independent of the angle $\Phi = \phi_{\Pt} + \phi_{\kt}$; thus it is sufficient to characterize the transverse momenta of the jets with $P_\perp$, $k_\perp$ and the relative angle $\phi = \phi_{\Pt} - \phi_{\kt}$. It is then convenient to decompose the differential yield in angular modes with respect to $\phi$ :
\begin{align}
    \frac{\der N^{\gamma_{\lambda}^*+A\rightarrow q\bar{q}+X}}{ \der^2 \Pt \der^2 \kt \der \eta_1 \der \eta_2} =  N^\lambda_0(P_\perp,k_\perp) + 2 \sum_{n=1}^\infty N^\lambda_n(P_\perp,k_\perp) \cos(n\phi)  \,,
    \label{eq:mode_decomposition}
\end{align}
where the modes are given by
\begin{align}
    N^\lambda_n(P_\perp,k_\perp) = \frac{1}{S_\perp}\int \frac{\der \phi_{\Pt}}{2\pi}
    \frac{\der \phi_{\kt}}{2\pi} e^{in(\phi_{\Pt}-\phi_{\kt})}\frac{\der \sigma^{\gamma_{\lambda}^*+A\rightarrow q\bar{q}+X}}{ \der^2 \Pt \der^2 \kt \der \eta_1 \der \eta_2} \label{eq:modes_cross-section}\,.
\end{align}
From these quantities, we can then compute the elliptic and quadrangular anisotropies:
\begin{align}
    v_{2,\lambda} = \langle \cos 2 \phi \rangle = N^\lambda_2 / N^\lambda_0 \,, \\
    v_{4,\lambda} = \langle \cos 4 \phi \rangle = N^\lambda_4 / N^\lambda_0 \,.
\end{align}
In the TMD limit the only non-vanishing mode is $v_{2,\lambda}$ for which there are explicit expressions in terms of the kinematic variables and the WW gluon TMD (see Eqs.\,\eqref{eq:v2LTMD} and \eqref{eq:v2TTMD}). In the ITMD and CGC there are no simple expressions for the anisotropies, thus they have to be computed numerically by first evaluating the differential yield as a function of relative angle $\phi$, and subsequent integration. While this is the approach we follow to compute these modes in the ITMD framework, for the computation in the CGC it is advantageous to use the following identity:
\begin{align}
    e^{iA\cos \alpha} = \sum_{n=-\infty}^{\infty} (-i)^n J_n(A) e^{-in\alpha} 
    \label{eq:identity}\,.
\end{align}
Inserting the CGC differential cross-section Eq.\,\eqref{eq:cgc_xs2} into Eq.\eqref{eq:modes_cross-section} and using the identity above, we can perform analytically the integrals over $\phi_{\Pt}$ and $\phi_{\kt}$ resulting in
\begin{align}
    N^{\lambda,\mathrm{CGC}}_n(P_\perp,k_\perp) =   \frac{\alpha_\mathrm{em} e_f^2 N_c S_\perp \delta_z}{ (2\pi)^6}  (-1)^n \int & \der^2 \Rt \der^2 \rt \der^2 \rt' J_n(R_\perp k_\perp) J_n(|\rt-\rt'| P_\perp) \nonumber  \\
    & \tilde{\Xi}_Y(\rt,\bt,\rt',\bt') \Rcal^{\lambda}(\rt,\rt') \,.
\end{align}
The computation of the modes and anisotropies for dijet production in the CGC reduces to the  6 dimensional integration above. Our results are shown in the next section.

\section{Numerical results}
\label{sec:numerical_results_dijets}

In this section we numerically evaluate the differential yield for the inclusive quark anti-quark production in proton and nuclear DIS. As argued in Sec.\,\ref{sec:harmonics} (see Eq.\,\eqref{eq:mode_decomposition}) we can write the differential yield as
\begin{align}
    \frac{\der N^{\gamma_{\lambda}^*+A\rightarrow q\bar{q}+X}}{ \der^2 \Pt \der^2 \kt \der \eta_1 \der \eta_2} =  N^\lambda_0(P_\perp,k_\perp)\left[1  + 2 \sum_{n=1}^\infty v_{n,\lambda}(P_\perp,k_\perp) \cos(n\phi) \right]  \,.
\end{align}
We will focus our study on the averaged angle differential yield $N^\lambda_0(P_\perp,k_\perp)$, the elliptic anisotropy $v_{2,\lambda}(P_\perp,k_\perp)$ and the quadrangular anisotropy $v_{4,\lambda}(P_\perp,k_\perp)$. The small-$x$ BK evolution will be carried up to $Y= \log(0.01/x_g)$, where 
\begin{align}
    x_g = \frac{Q^2 + M^2_{q\bar{q}} + k_\perp^2}{W^2 + Q^2 - m_n^2}  
\end{align}
is the fraction of the target plus momentum transferred to the dijet system in the $t$-channel exchange.

\begin{figure}[h!]
\centering
    \includegraphics[width = 2.8in]{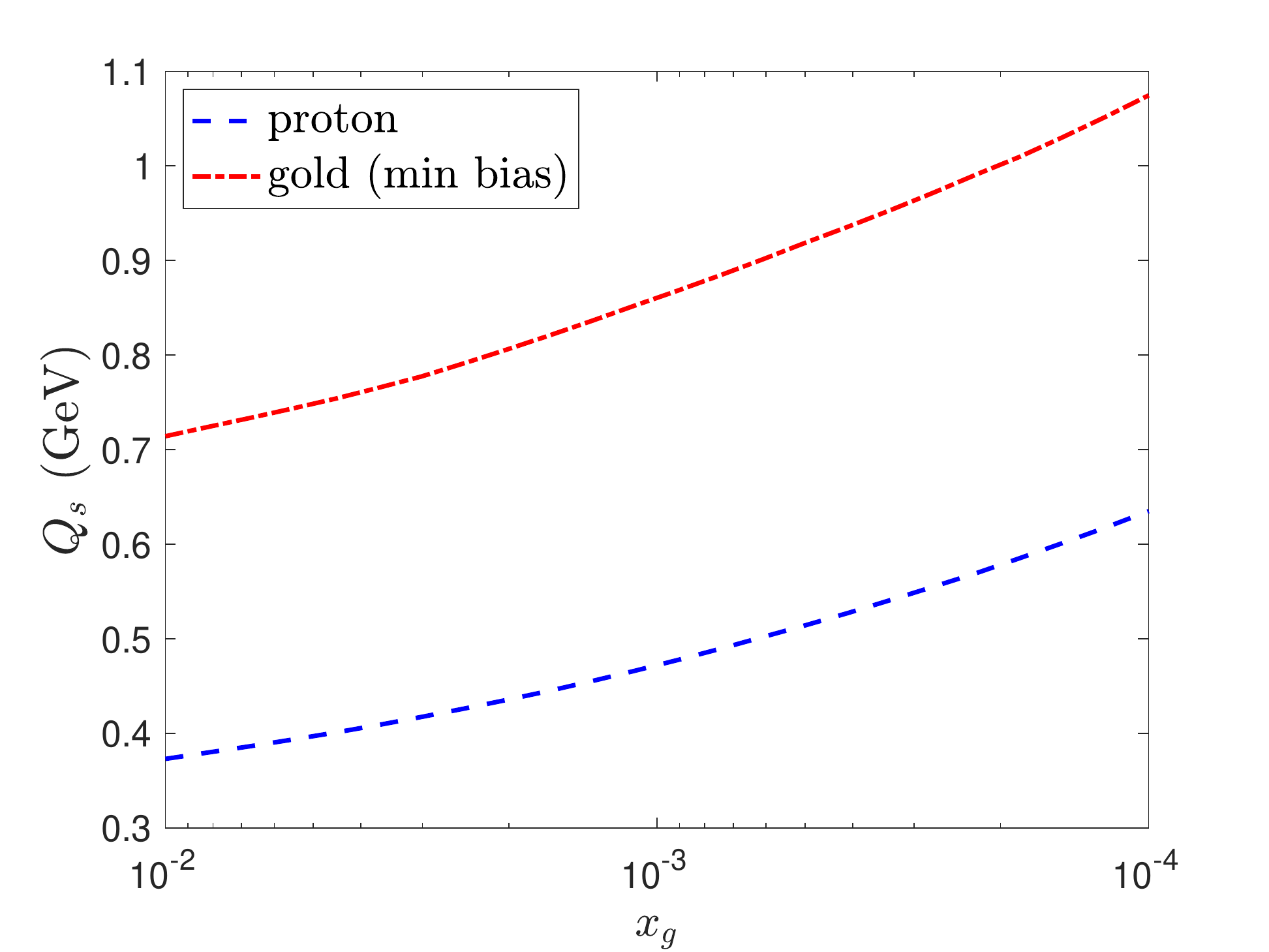}
    \includegraphics[width = 2.8in]{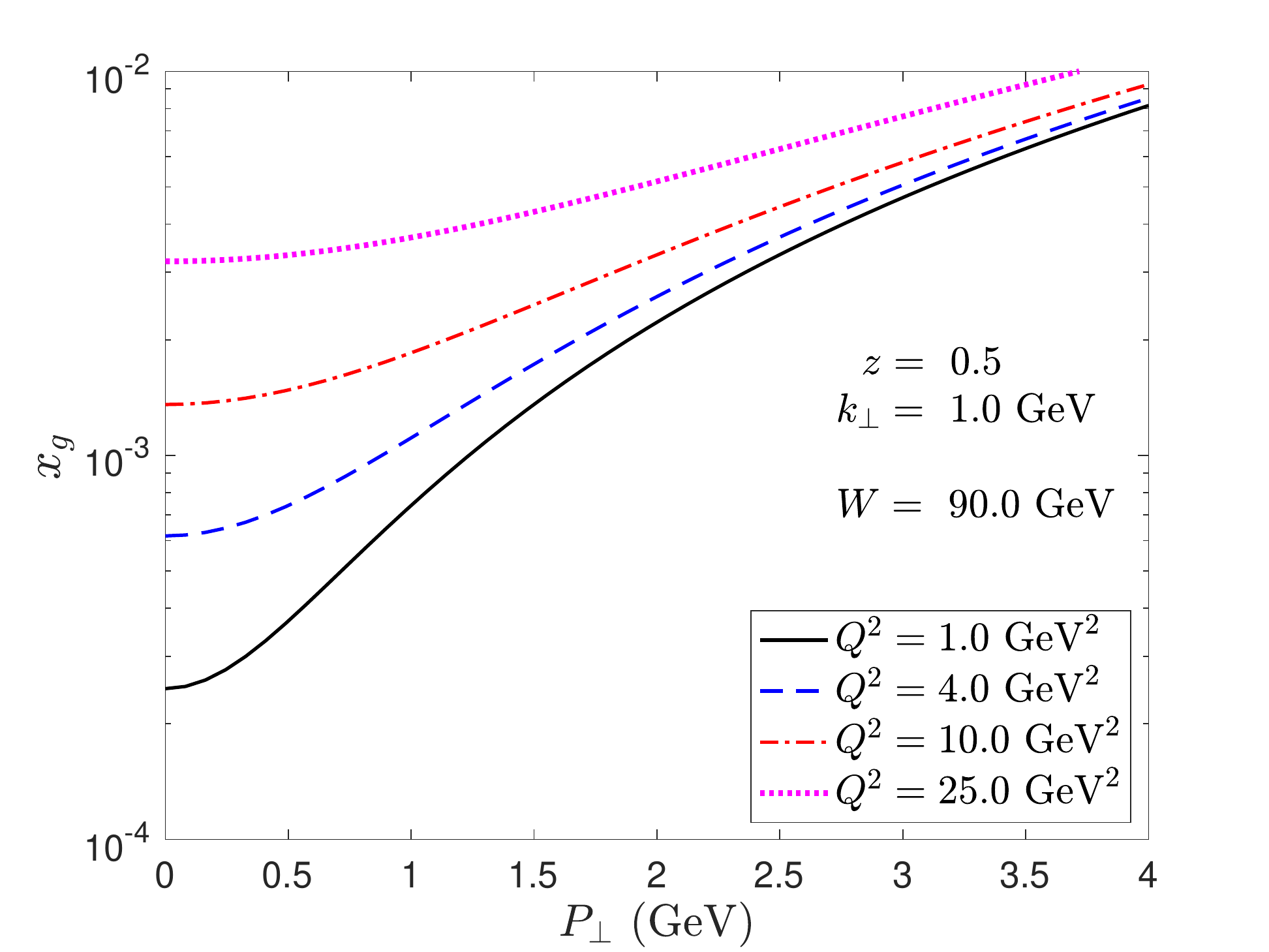}
\caption{Left: Saturation scale $Q_s$ as a function of $x_g$ for proton, and gold at median impact parameter (proxy for minimum bias collision). Right: values of $x_g$ as a function of the kinematic variables.}
\label{Fig_Qs_xg_dep}
\end{figure}

The left panel in Fig.\,\ref{Fig_Qs_xg_dep} shows the saturation scale $Q_s$ as a function of $x_g$ for proton, and for gold at median impact parameter (as a proxy of minimum bias collisions) following our initial conditions (see Sec.\,\ref{sec:set_up_dijets_IC})\footnote{The modest values of the saturation scale displayed in Fig.\,\ref{Fig_Qs_xg_dep} are a result of the parametrization of the dipole amplitude with MV initial conditions in \cite{Lappi:2013zma}. Other parametrizations can result in larger values of the saturation scale.}. The saturation scale has been defined as
\begin{align}
    Q_s = \frac{\sqrt{2}}{r_s},\quad \quad S_{Y}^{(2)}(r_s) = \exp(-1/2) \,,
\end{align}
where $Y = \log(\frac{x_0}{x_g})$. The right panel in Fig.\,\ref{Fig_Qs_xg_dep} displays the value of $x_g$ as a function of the kinematic variables. At the projected top EIC energies, requiring $x_g \leq 10^{-2}$ constrains the transverse momenta and virtualities of the dijet system as shown.

Our results will be shown separately for transversely and longitudinally polarized photons in collisions off protons and gold nuclei at a center of mass energy $W=90 \ \rm{GeV}$. We choose a configuration for the quark and the anti-quark in which they have identical longitudinal momenta $z_1=z_2=\frac{1}{2}$ (note that the momentum fractions are related to pseudorapidities as $z_i = 2 E_n |\boldsymbol{k}_{i\perp}| e^{-\eta_i}/W^2$ where $E_n$ is the nucleon energy). Experimentally it is not directly possible to determine the photon polarization in dijet production, and as such the cross section can not be directly measured for the longitudinal and transverse polarization states separately. As the photon flux factors $f_\lambda$ in the electron-nucleus cross section (Eq.\,\eqref{eq:eA_xs}) depend on inelasticity, some insight can in principle be obtained by doing the same measurement at different $\sqrt{s}$ resulting in different inelasticities $y$. Additionally, as discussed in Ref.~\cite{Dumitru:2018kuw}, the polarization dependent cross sections and elliptic modulations can be extracted from the data assuming that the functional form of the $v_{2,\lambda}$ coefficients is know from theory.

We will perform our computations in the TMD framework (Sec.\,\ref{sec:inclusive_dijets_TMD}), the improved TMD framework (Sec.\,\ref{sec:inclusive_dijets_iTMD}), and the CGC EFT (Sec.\,\ref{sec:inclusive_dijets_CGC}). We remind the reader that by comparing the TMD and the improved TMD framework we gain access to kinematic power corrections:
\begin{align}
    \der \sigma_{\mathrm{ITMD}} - \der \sigma_{\mathrm{TMD}} = \mathcal{O}\left(\frac{k_\perp}{Q_\perp}\right) \,,
\end{align}
while the comparison of the CGC with the ITMD will help us assess the role of genuine saturation contributions:
\begin{align}
    \der \sigma_{\mathrm{CGC}} - \der \sigma_{\mathrm{ITMD}} = \mathcal{O}\left(\frac{Q_s}{Q_\perp}\right) \,.
\end{align}
These saturation contributions are present in addition to those appearing in the WW gluon TMD that resum powers of $Q_s/k_\perp$.

\subsection{Angle averaged differential yield}
\label{sec:numerical_results_dijets_yield_angleaveraged}

\begin{figure}[h!]
\begin{minipage}{1.0\textwidth}
    \centering
    \includegraphics[width = 5.0in]{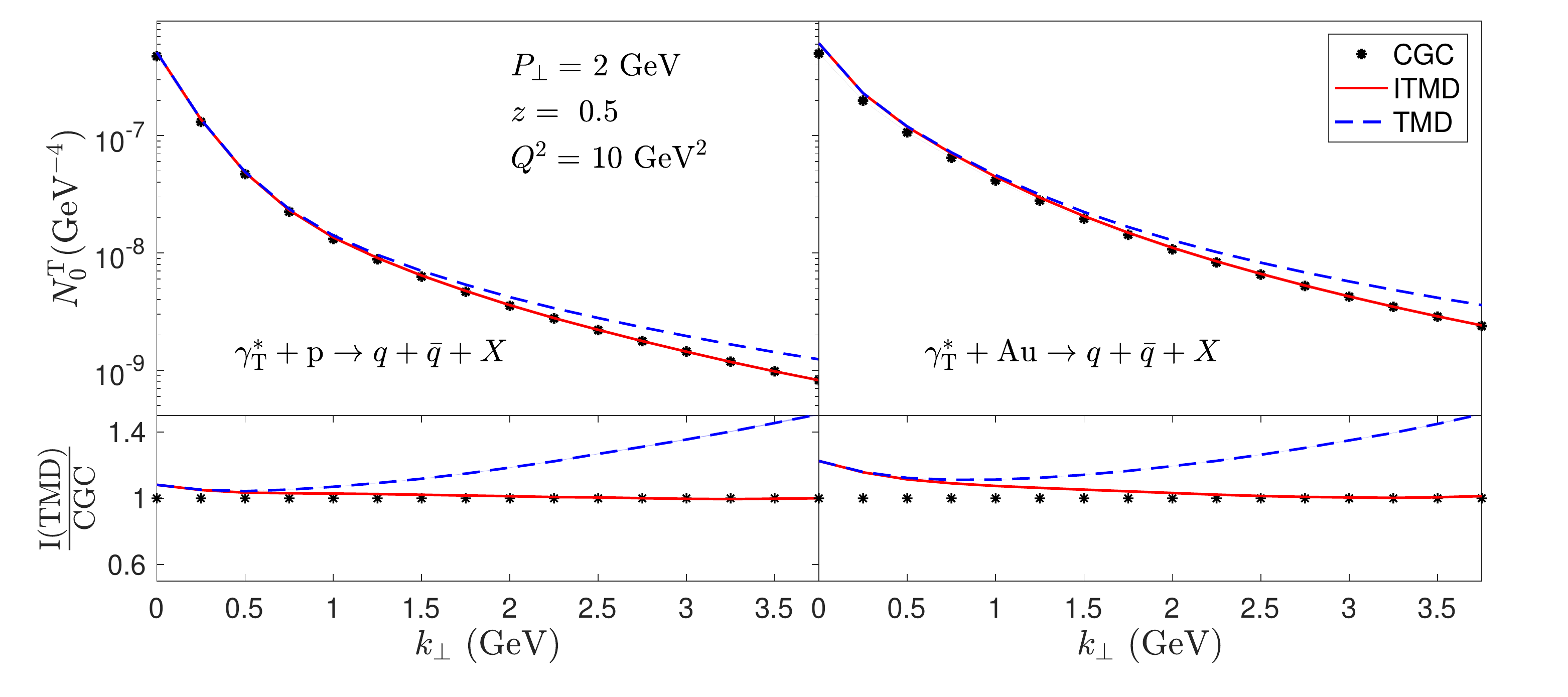}
\caption{Top: Differential yield (averaged over azimuthal angles $\phi_{\Pt}$ and $\phi_{\kt}$) for production of quark anti-quark pairs in $\gamma^*_{\mathrm{T}} +  p$ (left) and $\gamma^*_{\mathrm{T}} + \mathrm{Au}$ (right) scattering. Bottom: Ratio of the differential yield in the (I)TMD to the CGC.}
\label{Fig_N0T}
\end{minipage}
\begin{minipage}{1.0\textwidth}
    \centering
    \includegraphics[width = 5.7in]{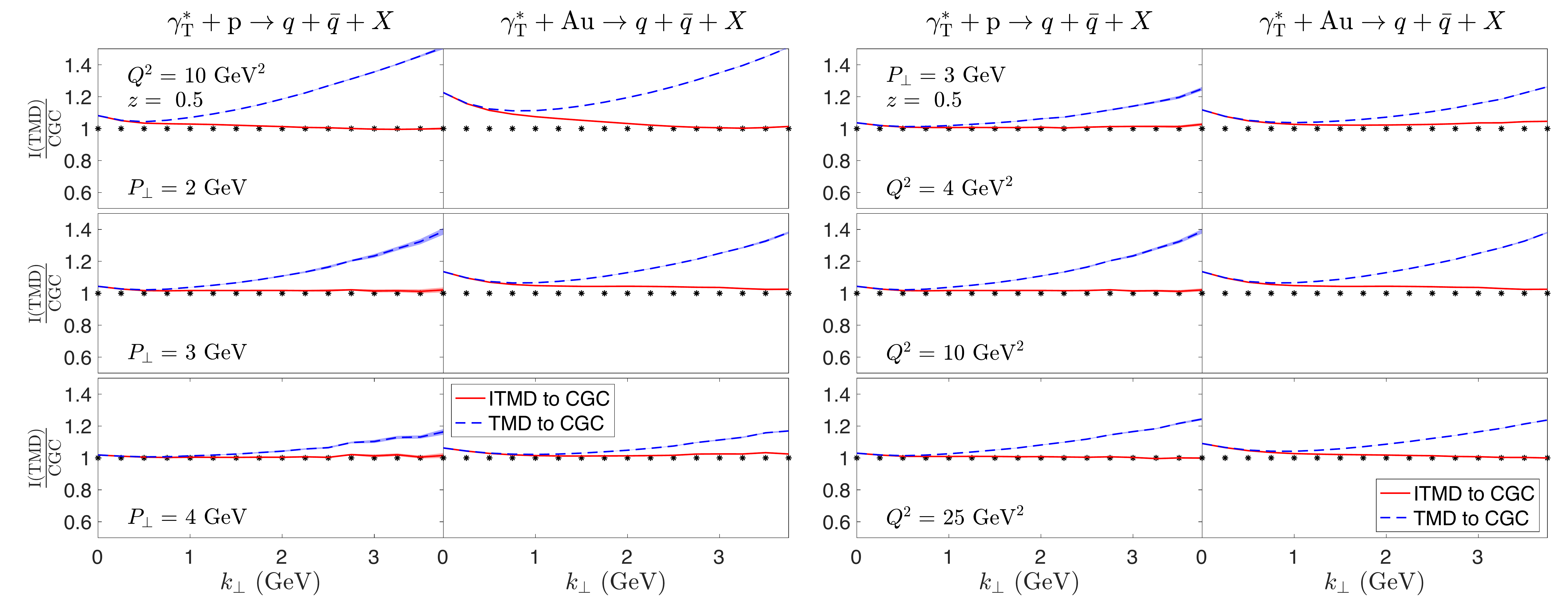}
\caption{Ratio of the averaged angle differential yield  (I)TMD to CGC as a function of $k_\perp$ at different values of  $Q$ (left), and at different values of $P_\perp$ (right).}
\label{Fig_ratio_N0T}
\end{minipage}
\end{figure}

We start by presenting our results for the angle averaged differential yield. In Fig.\,\ref{Fig_N0T} we show our results for $N^{\rm{T}}_0(P_\perp,k_\perp)$ as a function of momentum imbalance $k_\perp$, and at fixed virtuality $Q^2=10 \ \rm{GeV}^2$ and relative momentum $P_\perp = 2 \ \rm{GeV}$.

In both proton and gold collisions, the CGC results in a suppression of $\sim 20\%$ relative to the TMD when $P_\perp \sim k_\perp$. This suppression is enhanced as the momentum imbalance $k_\perp$ is increased (away from the quark anti-quark back-to-back configuration) signaling the breakdown of the TMD framework. When kinematic power corrections in $k_\perp/P_\perp$ are resummed into the hard factors of the improved TMD framework, we observe an excellent agreement between the CGC and ITMD results for  $k_\perp > P_\perp$ and up to large values of $k_\perp$ relative to $P_\perp$, suggesting that genuine higher twist effects are suppressed for non-back-to-back configurations and the suppression in the yield observed in the CGC relative to the TMD is driven by kinematic power corrections. Interestingly, the genuine saturation effects become visible in the back-to-back regime where deviations between TMD (or ITMD\footnote{Recall that TMD and ITMD match in the back-to-back limit $k_\perp \ll Q_\perp$.}) and the CGC
are observed, and which are enhanced from $~7\%$ in $e+p$ to about $\sim 20\%$ in $e+\rm{Au}$ collisions.

In Fig.\,\ref{Fig_ratio_N0T}, we present the ratios ITMD/CGC and TMD/CGC, and study their dependence as a function of $k_\perp$ at different virtualities $Q$ and relative momenta $P_\perp$. As expected, increasing either $Q$ or $P_\perp$ eventually reduces the kinematic twists resulting in smaller differences between the CGC and TMD at a given $k_\perp$. We also observe a systematic reduction of the genuine higher twist contributions at moderate $P_\perp$ and $Q$ values. Thus we expect that the improved TMD framework will provide good agreement with the CGC in the production of quark anti-quark pairs in DIS for kinematics in which $P_\perp$ or $Q$ are significantly larger than the saturation scale. On the other hand, genuine twist corrections have a significant impact on the measurement of back-to-back dihadrons/dijets at low virtualities and transverse momenta in nuclear DIS where the saturation scale is enhanced. 

\begin{figure}[h!]
\begin{minipage}{1.0\textwidth}
    \centering
\includegraphics[width = 5.0in]{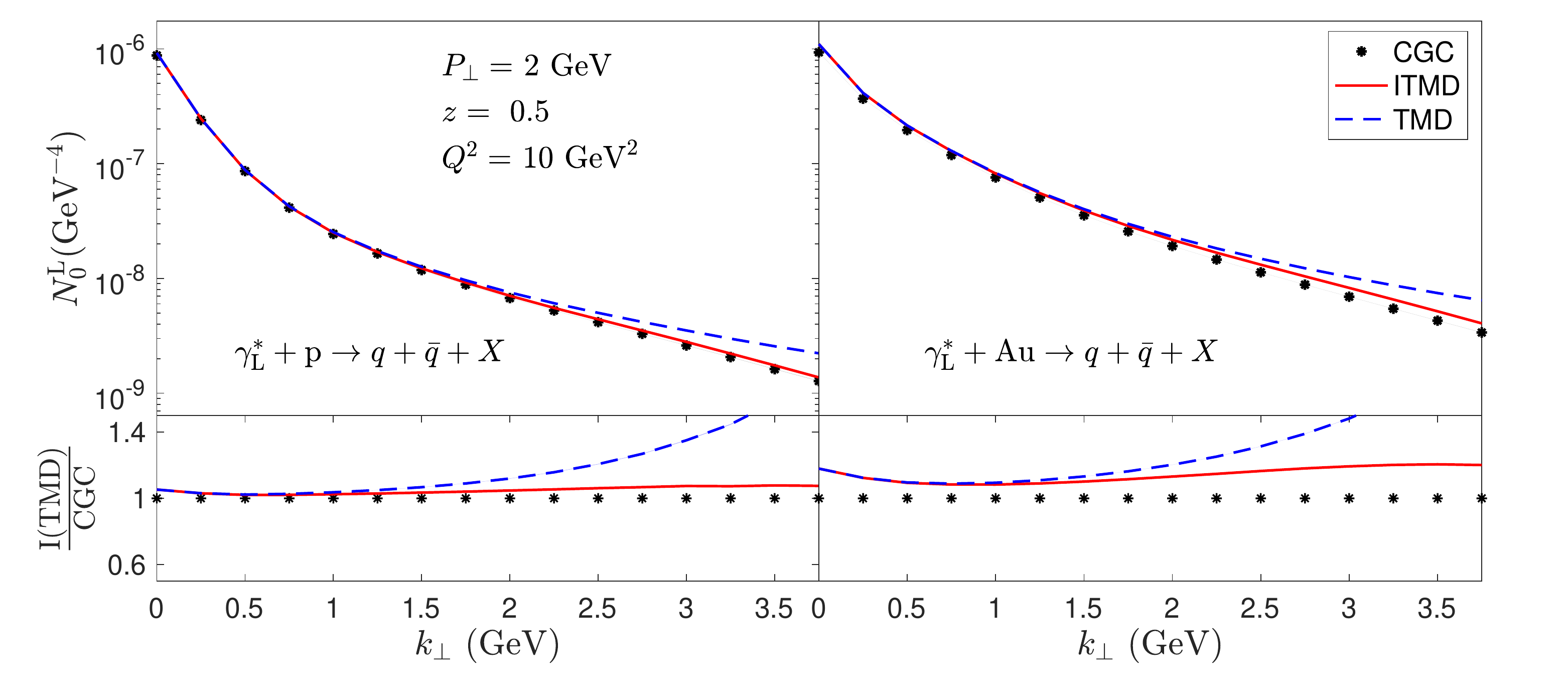}
\caption{Top: Differential yield (averaged over azimuthal angles $\phi_{\Pt}$ and $\phi_{\kt}$) for production of quark anti-quark pair in $\gamma^*_{\mathrm{L}} +  p/ \mathrm{Au}$ scattering. Bottom: Ratio of the differential yield in the (I)TMD to the CGC.}
\label{Fig_N0L}
\end{minipage}
\begin{minipage}{1.0\textwidth}
    \centering
\includegraphics[width = 5.7in]{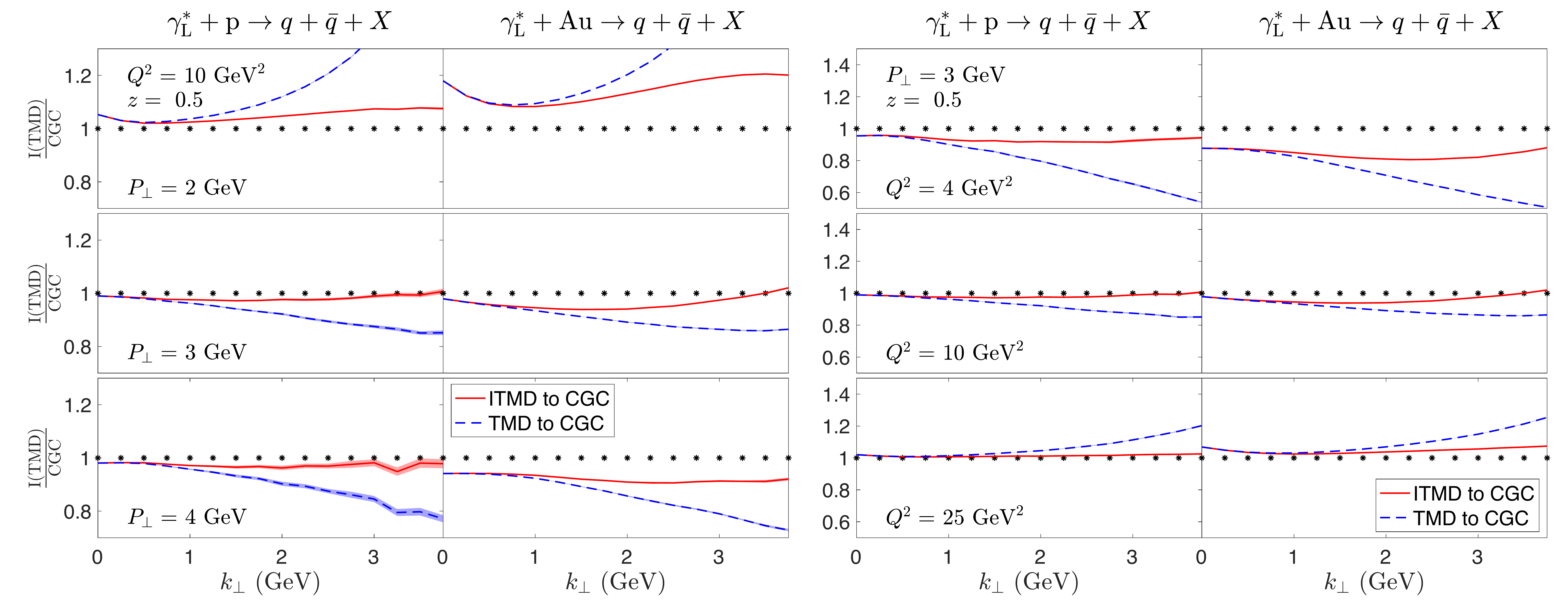}
\caption{Ratio of the averaged angle differential yield (I)TMD to CGC at different values of $Q$ (left), and at different values of $P_\perp$ (right). }
\label{Fig_ratio_N0L}
\end{minipage}
\end{figure}

We now move to the differential yield $N^{\rm{L}}_0(P_\perp,k_\perp)$ for the longitudinally polarized virtual photon case, shown in Fig.\,\ref{Fig_N0L} as a function of momentum imbalance $k_\perp$, and at a fixed virtuality $Q^2=10 \ \rm{GeV}^2$ and relative momentum $P_\perp = 2 \ \rm{GeV}$. The results depicted are qualitatively similar to those in the transversely polarized case, but with some notable differences. We observe deviations between the CGC and the TMD at large momentum imbalance $k_\perp$ because of the importance of kinematic power corrections. Once these contributions are resummed in the ITMD the agreement with the CGC results is improved. We note that the agreement between CGC and ITMD at large momentum imbalance is not as good as in the transversely polarized case, which might be caused by the bias for larger dipole size contributions in the light-cone wave-function of longitudinally polarized photons\footnote{In contrast to fully inclusive DIS, in dijet production the longitudinal momentum fractions $z_{1,2}$ are kept fixed. At the chosen values of $z_1=z_2=0.5$ here (and at fixed virtuality $Q^2$), one can verify that the light-cone wave-function for a transverse photon $\sim K_1(\varepsilon r_\perp)$ grows more quickly than that of a longitudinal photon $\sim K_0(\varepsilon r_\perp)$ in the limit $r_\perp \rightarrow 0$. 
}. This reveals that genuine saturation corrections also appear away from the back-to-back limit in the longitudinally polarized case.

Ratios of ITMD and TMD to GCC for other choices of $P_\perp$ and $Q^2$ are shown in Fig.\,\ref{Fig_ratio_N0L}. We find both enhancement and suppression of the quark anti-quark production in the CGC relative to the ITMD, depending on the virtuality $Q$ and the relative momenta $P_\perp$. Overall we observe that the inclusion of kinematic power corrections improves the agreement with the CGC, and that increasing $Q$ or $P_\perp$ reduces genuine higher twist contributions as one should expect. The apparent better agreement in Fig.\,\ref{Fig_ratio_N0L} at $P_\perp = 3 \ \mathrm{GeV}$ and $Q^2 = 10 \ \mathrm{GeV}^2$ is coincidental as the CGC is transitioning from enhancement to suppression as $Q^2$ grows.

\begin{figure}[h!]
\includegraphics[width = 3.0in]{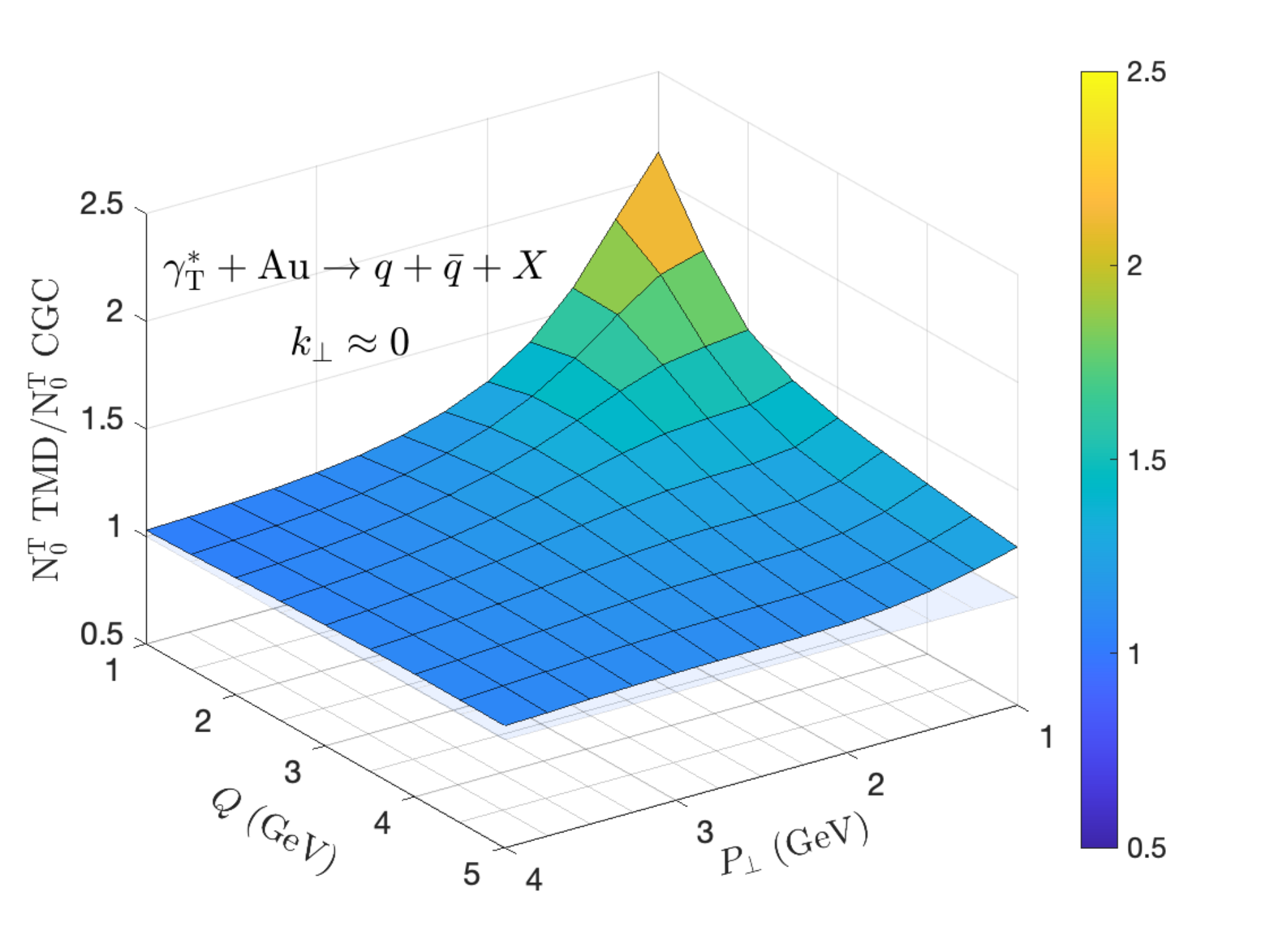}
\includegraphics[width = 3.0in]{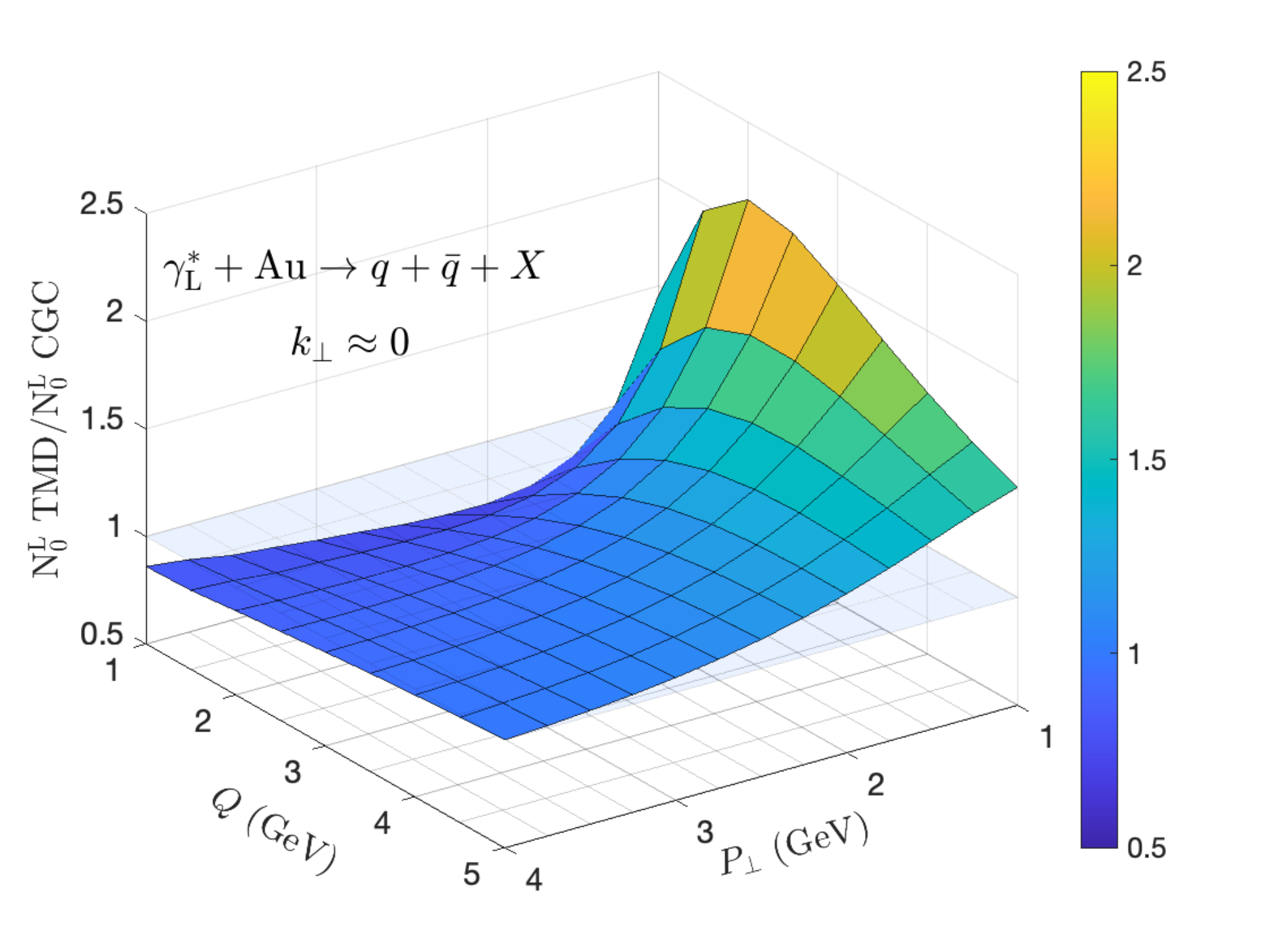}
\caption{Ratio of TMD to CGC for near back-to-back kinematics $k_\perp \approx 0$ in $\gamma^*_{\lambda} + \mathrm{Au} \rightarrow q + \bar{q} + X$, as a function of virtuality $Q$ and relative momentum $P_\perp$. Left: transversely polarized case. Right: longitudinally polarized case.}
\label{Fig_BB}
\end{figure}

In order to single out the effect of genuine higher twists, we study the ratio of TMD to CGC near the back-to-back configuration $k_\perp \approx 0$, where kinematic power corrections vanish (see Sec\,\ref{sec:inclusive_dijets_ght}). In Fig.\,\ref{Fig_BB} we show this ratio in nuclear DIS, and observe that if either the relative momentum $P_\perp$ or the virtuality $Q$ are large enough this ratio goes to unity, and thus the genuine twists are suppressed. However, significant differences are observed at low to moderate values of $Q$ and $P_\perp$. They amount to more than a factor of 2 difference between the CGC and the TMD framework. In particular, a suppression of the CGC relative to the TMD prediction is observed at all back-to-back kinematics in the case of transverse polarization, while both suppression and enhancement are seen for the longitudinally polarized case depending on the virtuality and relative momenta.  

The genuine saturation contribution to dijet production in the back-to-back limit $k_\perp \to 0$ (and in the light-cone gauge) is the result of multiple scattering, such that the total transverse momentum transferred to the $q\bar{q}$ pair is zero (since it compensates between several scatterings). The TMD and ITMD calculations miss this contribution as their result is proportional to the number of gluons with zero transverse momentum

\subsection{Elliptic Anisotropy}
\label{sec:numerical_results_dijets_elliptic}

In this section we study the elliptic anisotropy in the angle $\phi$ separately for transversely and longitudinally polarized photons. This quantity can be accessed by measuring the distribution of the momentum imbalance $\kt$ with respect to the relative momentum $\Pt$.

In the TMD framework, the magnitude of the elliptic anisotropy is proportional to the ratio of polarized to unpolarized Weizsäcker-Williams TMD (see Eqs.\,\eqref{eq:v2LTMD} and \eqref{eq:v2TTMD}):
\begin{align}
    |v_{2,\mathrm{L/T}}| \propto  \frac{x h^0}{x G^0} \,.
\end{align}

The study of $v_2$ as a function of the dijet momentum imbalance $k_\perp$ is a promising observable to extract the behavior of the linearly polarized gluons inside protons and nuclei. In particular, it is expected that $xh^0/xG^0 \rightarrow 0$ in the regime of small momentum imbalance ($k_\perp \ll Q_s$), while the gluons are completely linearly polarized $xh^0/xG^0 \rightarrow 1$ for the perturbative limit $k_\perp \gtrsim Q_s$ \cite{Dominguez:2011br,Metz:2011wb,Dumitru:2015gaa}. 

\begin{figure}[h!]
\centering
\includegraphics[width = 2.9in]{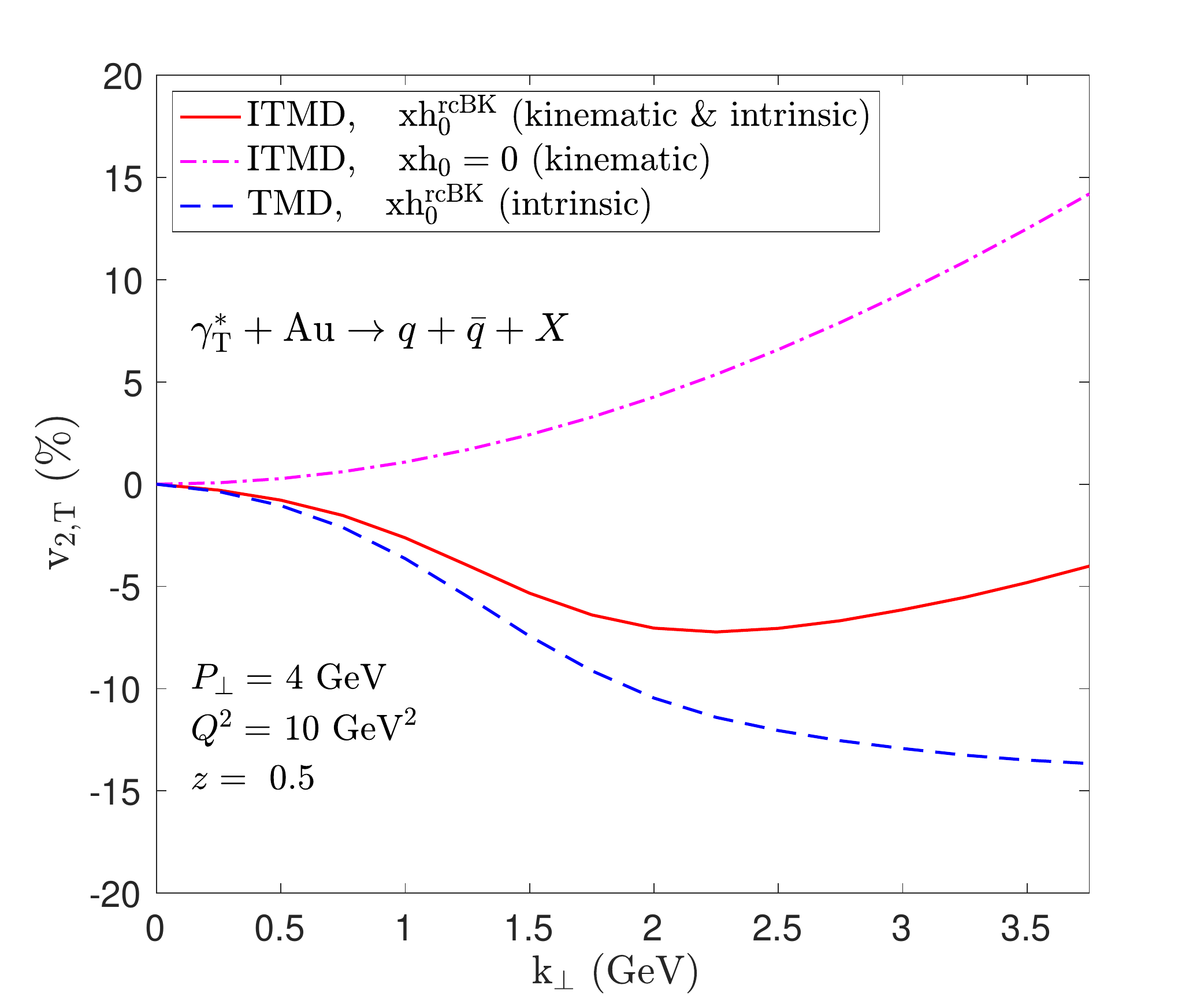}
\includegraphics[width = 2.9in]{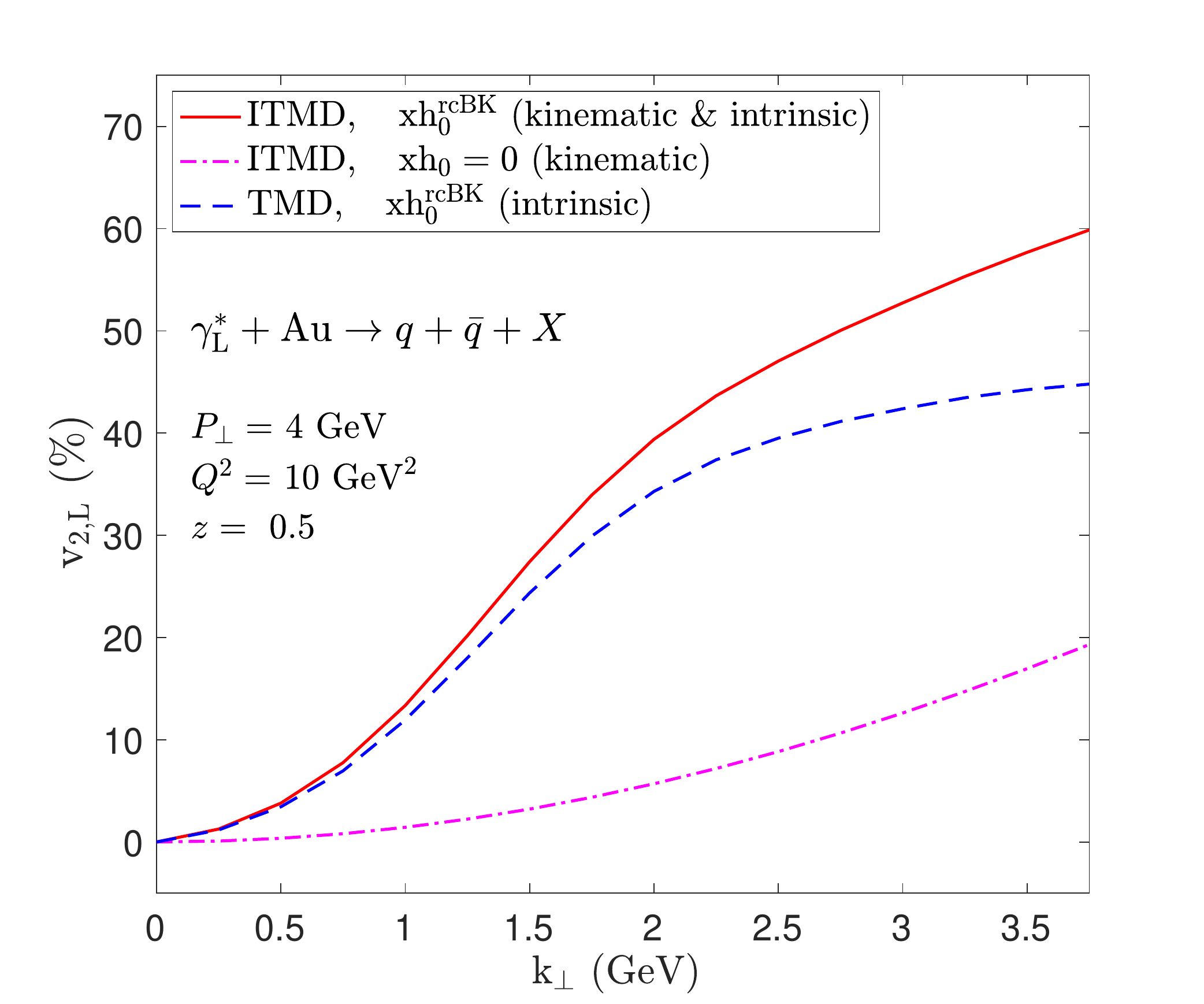}
\caption{Elliptic anisotropy in $\gamma^*_{\lambda} + \mathrm{Au} \rightarrow q + \bar{q} + X$ for transversely polarized photons (left), and longitudinally polarized photons (right). We show results in the TMD (dashed) and ITMD (solid) framework. To illustrate the effect purely from kinematic twists, we also show the result for the ITMD in which we turn off the linearly polarized WW gluon TMD (dashed dotted).}
\label{Fig_TMD_ITMD_on_off}
\end{figure}

\begin{figure}[h!]
\centering
\includegraphics[width = 4.50in]{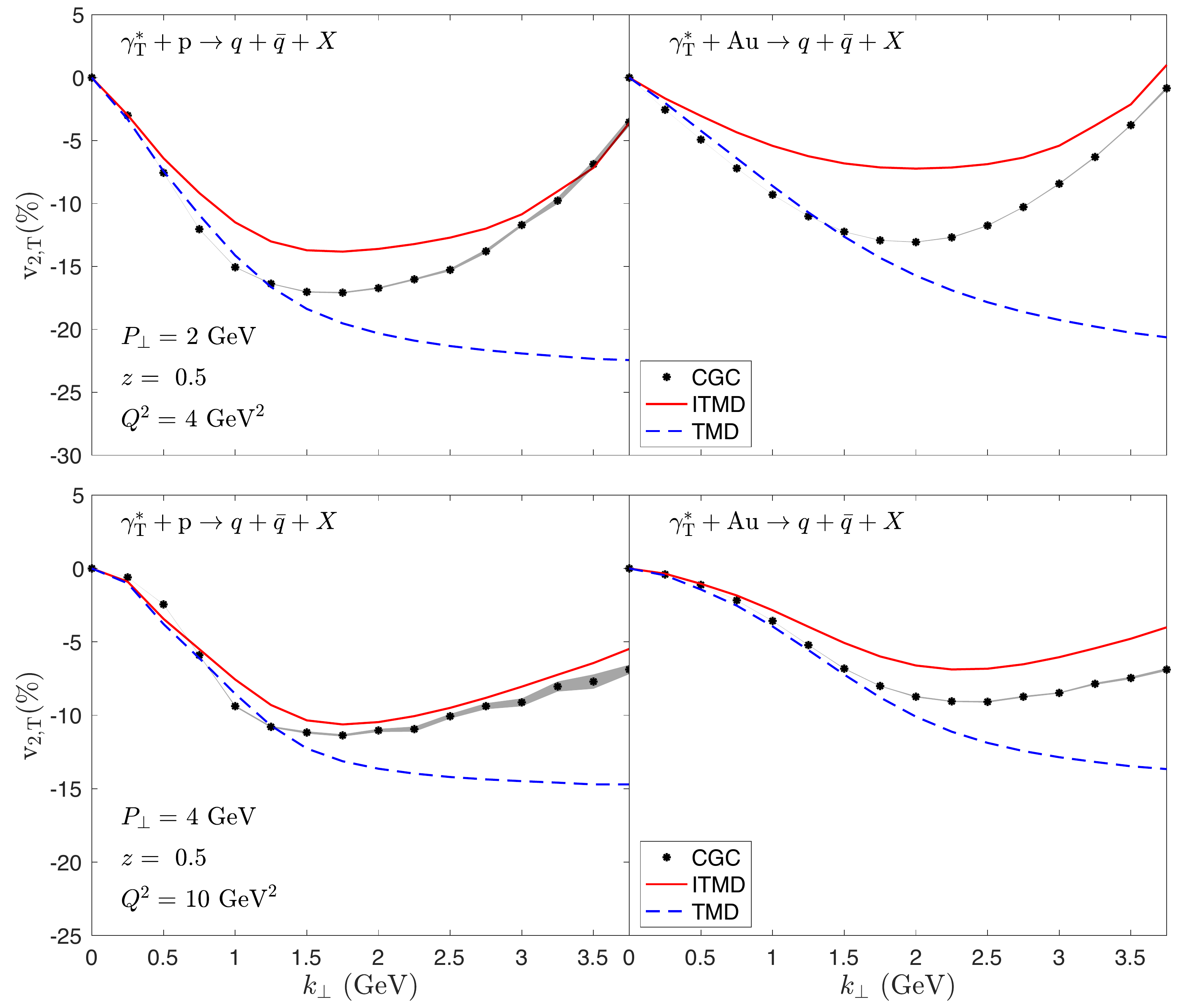}
\caption{Elliptic anisotropy $v_2$ in the angle $\phi$ for production of quark anti-quark pairs in $\gamma^*_{\mathrm{T}} +  p$ (left) and  $\gamma^*_{\mathrm{T}} + \mathrm{Au}$ (right) scattering. The grey band shows the numerical uncertainty of the CGC calculation.}
\label{Fig_ratio_v2T}
\end{figure}
However, it is important to note that kinematic power corrections in the ITMD hard factor also induce correlations between the $\Pt$ and $\kt$ (see Eqs\,\eqref{eq:ITMD_hardL_analytic} and \eqref{eq:ITMD_hardT_analytic}). These azimuthal correlations also couple to the linearly polarized WW. Consequently, distinguishing contributions purely from kinematic twists and those solely from the linearly polarized WW is not possible. However, we can  compare the full ITMD predictions to the case where we set the  linearly polarized WW gluon distribution to zero by hand, which will show the contribution from kinematic twists only. The comparison is shown in Fig.\,\ref{Fig_TMD_ITMD_on_off} as a function of the momentum imbalance $k_\perp$. For both polarizations we find that the kinematic power corrections result in positive contributions to $v_{2}$, which grow with the dijet momentum imbalance. On the other hand, in the TMD curve (which does not include kinematic power corrections) the elliptic anisotropy is sourced only by the linearly polarized gluons; in this case the sign of $v_{2}$ depends on the polarization of the photon (see Eqs.\,\eqref{eq:v2LTMD} and \eqref{eq:v2TTMD}). 
This additional $v_2$ from kinematic twists should be taken into account when extracting the linearly polarized gluon distribution in azimuthal dijet measurements \cite{Dumitru:2018kuw}. In addition to kinematic corrections, genuine higher twists contribute to azimuthal correlations through higher body operators (see Sec.\,\ref{sec:inclusive_dijets_ght}). Our goal in this section is to quantitatively study the different contributions to $v_2$ by comparing CGC, ITMD and TMD formalisms.

In Fig.\,\ref{Fig_ratio_v2T} we present our results for the elliptic anisotropy as a function of the momentum imbalance $k_\perp$ when the virtual photon in DIS is transversely polarized. We consider proton and gold targets and different $P_\perp$ and $Q^2$. We observe sizeable deviations between the TMD and CGC results when $P_\perp \sim k_\perp$, signaling the appearance of azimuthal correlations due to kinematic power corrections and genuine higher twists. In particular, in the CGC we observe minima in $v_{2,\mathrm{T}}$, which are absent in the TMD limit, and the anisotropies are considerably reduced at large $k_\perp$ (as argued previously the contribution to $v_{2,\mathrm{T}}$ from the intrinsic correlations due to linearly polarized gluons and those from kinematic twists come with different signs, see Fig.\ref{Fig_TMD_ITMD_on_off}).

In order to distinguish the relative importance of kinematic and genuine saturation corrections, we compare the CGC to the ITMD framework. We observe that the deviations in $v_{2\mathrm{T}}$ are enhanced for the nuclear DIS and for the lower scales in $P_\perp$ and $Q$. Interestingly, our results also indicate that the contributions of kinematic and genuine saturation contributions appear to come with different signs. In particular, we observe that at low momentum imbalance $k_\perp$ this results in an apparent better agreement between the CGC and its TMD limit despite the exclusion of kinematic power corrections.

\begin{figure}[h!]
\centering
\includegraphics[width = 4.50in]{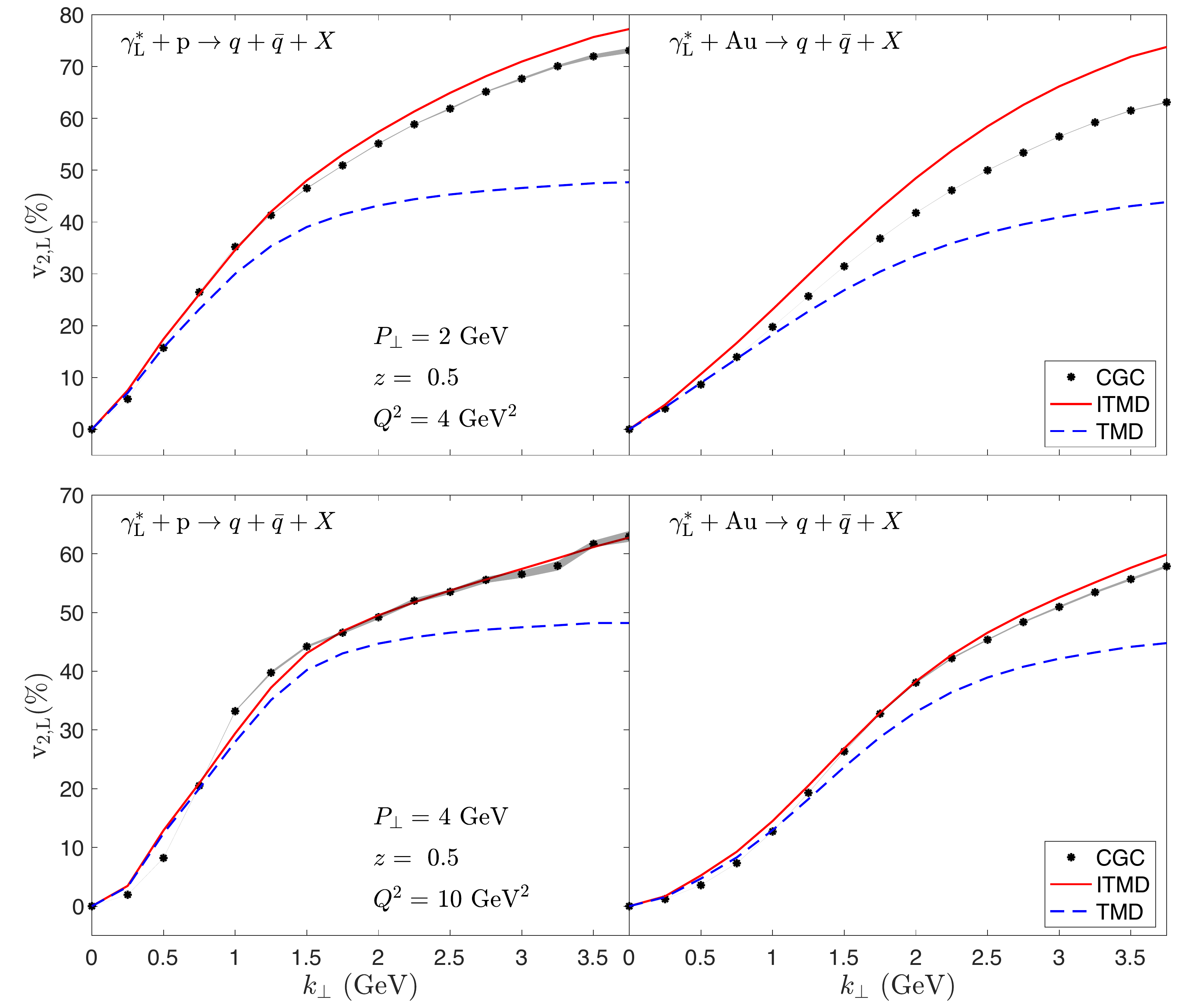}
\caption{Elliptic anisotropy $v_2$ in the angle $\phi$ for production of quark anti-quark pair in $\gamma^*_{\mathrm{L}} +  p/ \mathrm{Au}$ scattering. The grey band shows the numerical uncertainty of the CGC calculation.}
\label{Fig_ratio_v2L}
\end{figure}
Our results for the elliptic anisotropy for longitudinally polarized photons are shown in Fig.\,\ref{Fig_ratio_v2L}. In this case we observe that kinematic power corrections in the ITMD increase the azimuthal modulations compared to the TMD limit, which results in very large $v_{2,\mathrm{L}}$ at large momentum imbalance. In the CGC, this enhancement is partially tamed by genuine saturation corrections (most pronounced at low values of $P_\perp$ and $Q^2$ and for nuclear DIS).

Note that in the TMD framework  $v_{2,\mathrm{L}}$ cannot exceed $1/2$, which is because of the bound in Eq.\,\eqref{eq:TMD_bound}. On the other hand, in ITMD and CGC the anisotropies exceed $1/2$ due to kinematic ($k_\perp/Q_\perp$) and genuine ($Q_s/Q_\perp$) power corrections. Since the cross-section must remain positive, this indicates the presence of higher azimuthal correlations, which render the cross-section finite as we will show in the next section.

\subsection{Quadrangular Anisotropy}

\label{sec:numerical_results_dijets_quadrangular}

In this section we study the quadrangular anisotropy $v_4$ in the azimuthal angle between $\Pt$ and $\kt$. We note that this contribution is completely absent in the TMD framework. Contributions to the quadrangular anisotropy have been computed for the first time in \cite{Dumitru:2016jku} by considering the first correction to the leading power of the dipole size $r_\perp$ when taking the TMD limit. Such higher powers of the dipole size can account for $k_\perp/Q_\perp$ and $Q_s/Q_\perp$ corrections without distinction unless one is careful with the separation between kinematic and genuine higher twists. This distinction for $v_4$ is the purpose of this section, where our results will be presented in the ITMD and CGC frameworks.

\begin{figure}[h!]
\includegraphics[width = 2.9in]{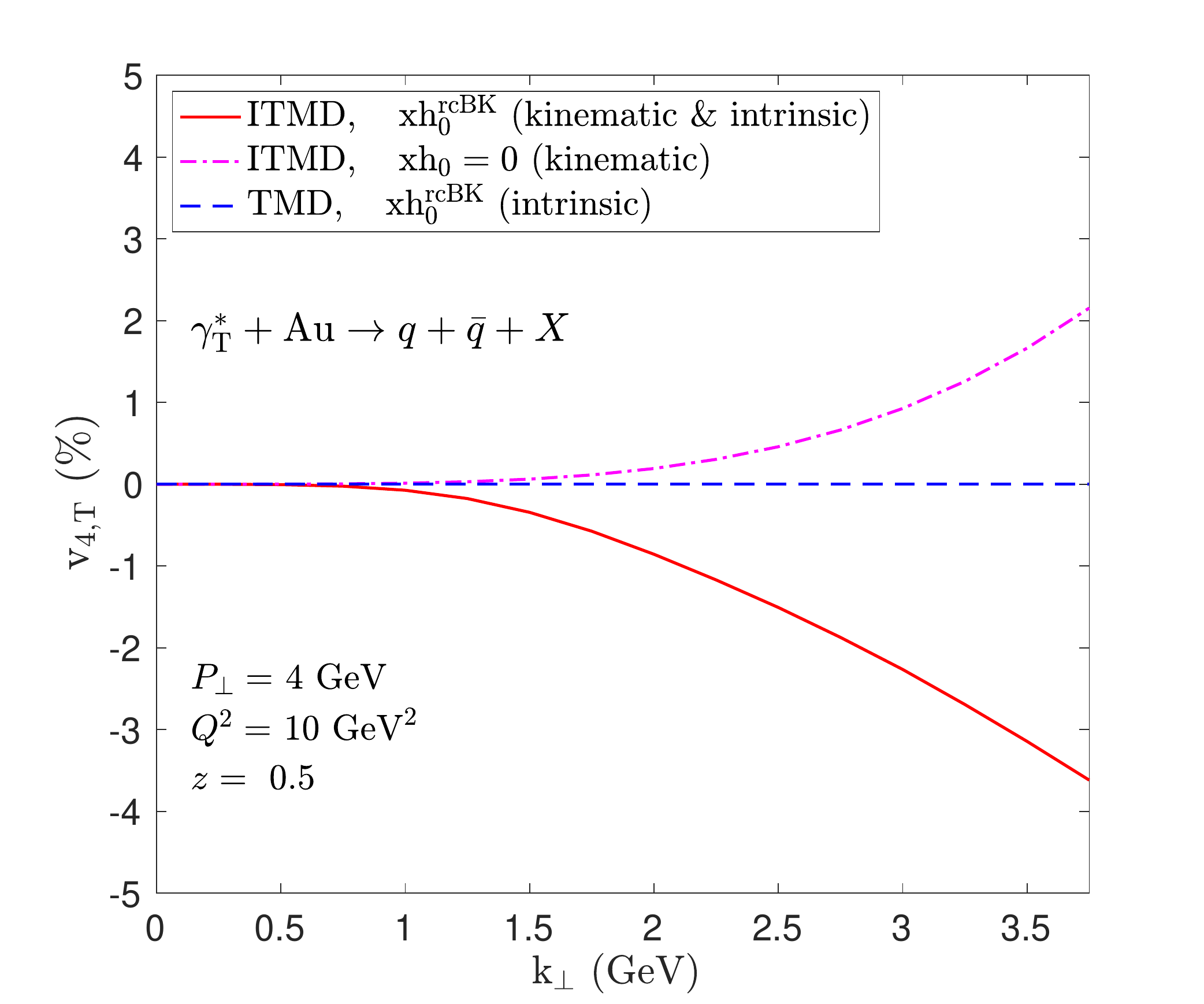}
\includegraphics[width = 2.9in]{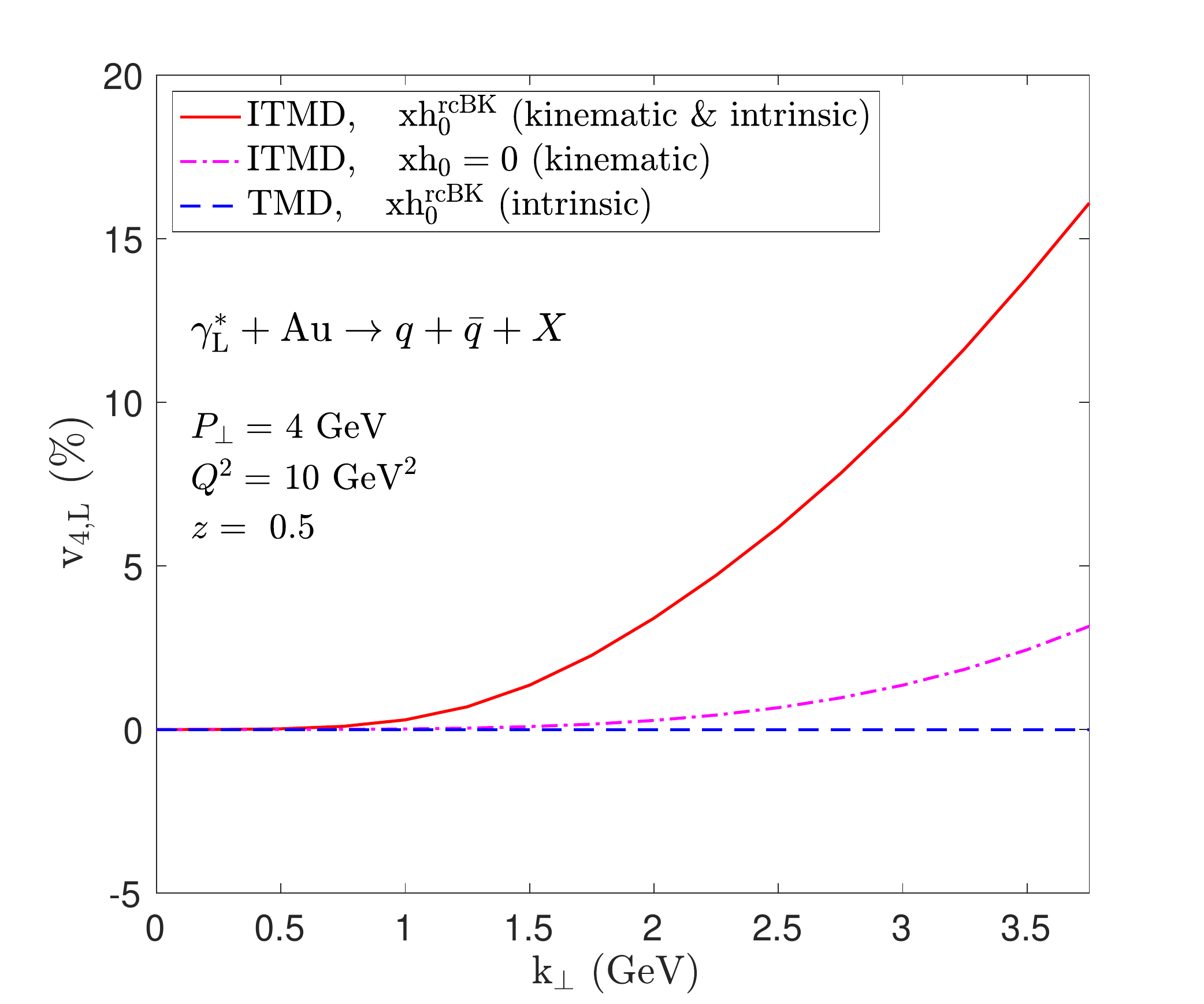}
\caption{Quadrangular anisotropy in $\gamma^*_{\lambda} + \mathrm{Au} \rightarrow q + \bar{q} + X$: for transversely polarized photons (left), and longitudinally polarized photons (right). We show results in the TMD and ITMD framework. To illustrate the effect purely from kinematic twists, we also show the result for the ITMD in which we turn off the linearly polarized WW gluon TMD (dashed dotted).}
\label{Fig_TMD_ITMD_on_off_v4}
\end{figure}
A quadrangular anisotropy can be generated by correlations between $\Pt$ and $\kt$ in the hard factor in the ITMD (see Eqs.\eqref{eq:ITMD_hardL_analytic} and \eqref{eq:ITMD_hardT_analytic}) as well as correlations in the genuine higher twists embodied in Eq.\,\eqref{eq:M2-b2b}. Fig.\,\ref{Fig_TMD_ITMD_on_off_v4} demonstrates in scattering on Au targets that while the linearly polarized WW gluon distribution produces only an elliptic anisotropy in the TMD framework, in the ITMD framework the linearly polarized WW gluon distribution also has a contribution to $v_4$, as it couples to higher modes. To illustrate this, observe that using Eqs.\,\eqref{eq:ww_decomposition} and \eqref{eq:ITMD_xsec}, we can write the differential cross-section for the ITMD as
\begin{align}
    \der \sigma \sim \Hcal^G(\Pt,\kt) xG^0(x,k_\perp) + \Hcal^h(\Pt,\kt) xh^0(x,k_\perp) \,,
\end{align}
where we introduced the hard factors:
\begin{align}
    \Hcal^G(\Pt,\kt) &= \delta^{ij} \Hcal^{ij}_{\mathrm{ITMD}}(\Pt,\kt)\,,
    \end{align}
    \begin{align}
    \Hcal^h(\Pt,\kt) &= \left(2\kt^i\kt^j/k_\perp^2-\delta^{ij} \right) \Hcal^{ij}_{\mathrm{ITMD}}(\Pt,\kt) \,.
\end{align}
One could then expand $\Hcal^G$ and $\Hcal^h$ in modes in $\phi$:
\begin{align}
    \Hcal^G(\Pt,\kt) &= \Hcal^G_0(P_\perp,k_\perp)+ 2\Hcal^G_2(P_\perp,k_\perp) \cos(2\phi) + 2\Hcal^G_4(P_\perp,k_\perp) \cos(4\phi) + ... \,, \\ \Hcal^h(\Pt,\kt) &= \Hcal^h_0(P_\perp,k_\perp)+ 2\Hcal^h_2(P_\perp,k_\perp) \cos(2\phi) + 2\Hcal^h_4(P_\perp,k_\perp) \cos(4\phi) + ... \,.
\end{align}
Then we can obtain expressions for the elliptic and quadrangular anisotropies:
\begin{align}
    v_{2} &= \frac{xG^0(x,k_\perp)\Hcal^G_2(P_\perp,k_\perp)+ xh^0(x,k_\perp)\Hcal^h_2(P_\perp,k_\perp)}{xG^0(x,k_\perp)\Hcal^G_0(P_\perp,k_\perp)+ xh^0(x,k_\perp)\Hcal^h_0(P_\perp,k_\perp)} \, \\
    v_{4} &= \frac{xG^0(x,k_\perp)\Hcal^G_4(P_\perp,k_\perp)+ xh^0(x,k_\perp)\Hcal^h_4(P_\perp,k_\perp)}{xG^0(x,k_\perp)\Hcal^G_0(P_\perp,k_\perp)+ xh^0(x,k_\perp)\Hcal^h_0(P_\perp,k_\perp)} \,.
\end{align}

This implies that when $\Hcal^h_4(P_\perp,k_\perp)$ is non-zero, $xh^0(x,k_\perp)$ can contribute to $v_4$.  While analytic expressions for these modes in the ITMD are difficult to obtain, the numerical results in Fig.\,\ref{Fig_TMD_ITMD_on_off_v4} suggest that $\Hcal^h_4(P_\perp,k_\perp)$ is larger than $\Hcal^G_4(P_\perp,k_\perp)$.

Turning off the linearly polarized WW gluon TMD by hand in the ITMD framework reveals the $v_4$ purely from kinematic twists. Its sign is positive for both transverse and longitudinal photon polarizations, while the full ITMD result has opposite signs for the two polarizations. 

\begin{figure}[h!]
\centering
\includegraphics[width = 4.50in]{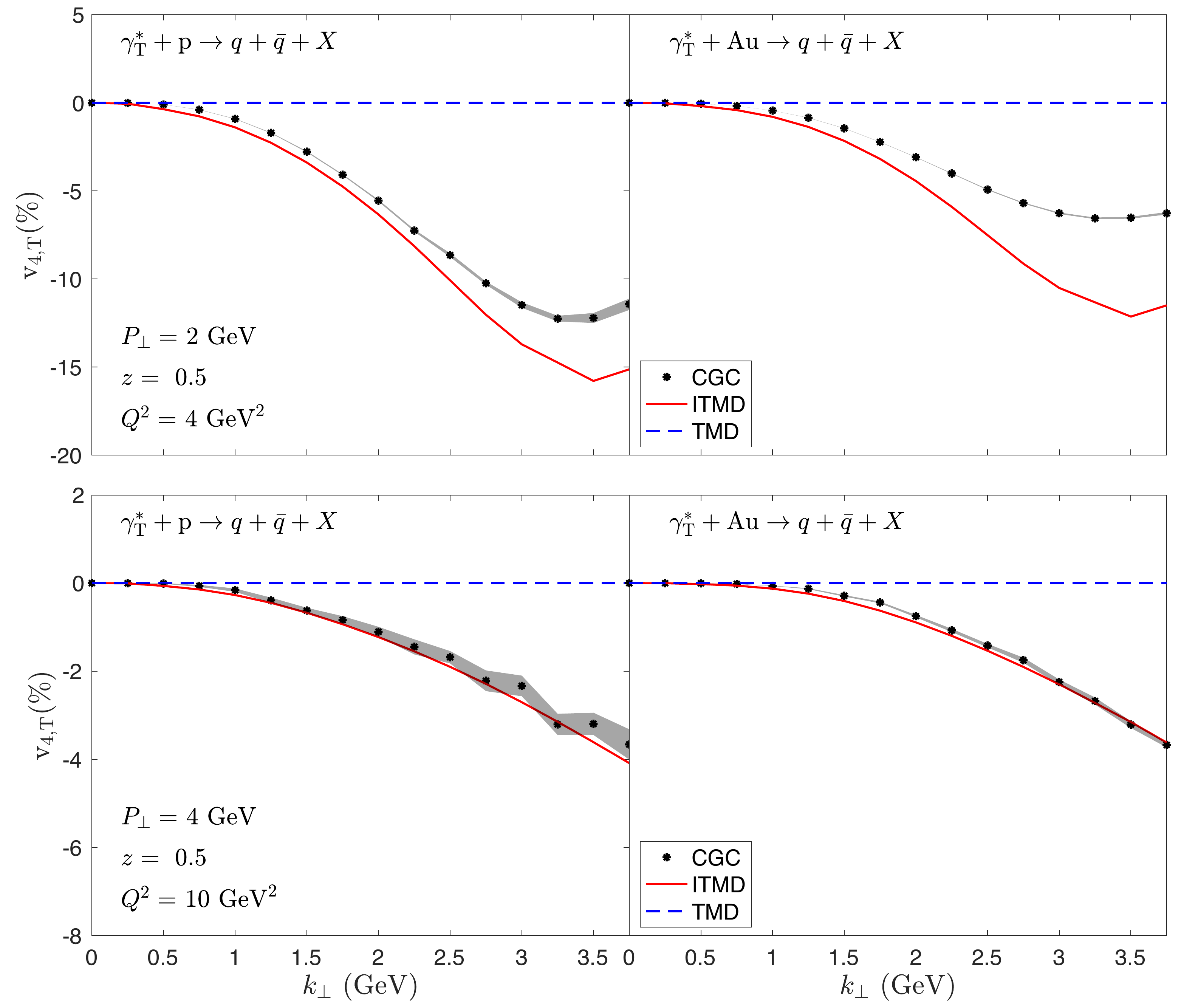}
\caption{Quadrangular anisotropy $v_4$ in the angle $\phi$ for production of quark anti-quark pairs in $\gamma^*_{\mathrm{T}} + p$ (left) and $\gamma^*_{\mathrm{T}} + \mathrm{Au}$ (right) scattering. The grey band shows the numerical uncertainty of the CGC calculation.}
\label{Fig_ratio_v4T}
\end{figure}

We present the CGC and ITMD results for the quadrangular anisotropy as a function of momentum imbalance $k_\perp$ at two different values of $P_\perp$ and $Q^2$ in Fig.\,\ref{Fig_ratio_v4T} for the case in which the virtual photon is transversely polarized. First, we note that the quadrangular anisotropy $v_{4,\rm{T}}$ is negative (similar to $v_{2,\rm{T}}$) and on the order of a few percent at low momentum imbalance $( k_\perp \lesssim P_\perp)$. In the lower panels (higher $P_\perp$ and $Q$) we observe that the results between the ITMD and CGC schemes closely agree with each other, showing that the quadrangular anisotropy is mostly driven by kinematic power corrections. The genuine higher twists tend to suppress $v_{4,\rm{T}}$, this effect can be seen more clearly in the upper right plot corresponding to nuclear DIS and at low values of $P_\perp$ and $Q$.

\begin{figure}[h!]
\centering
\includegraphics[width = 4.50in]{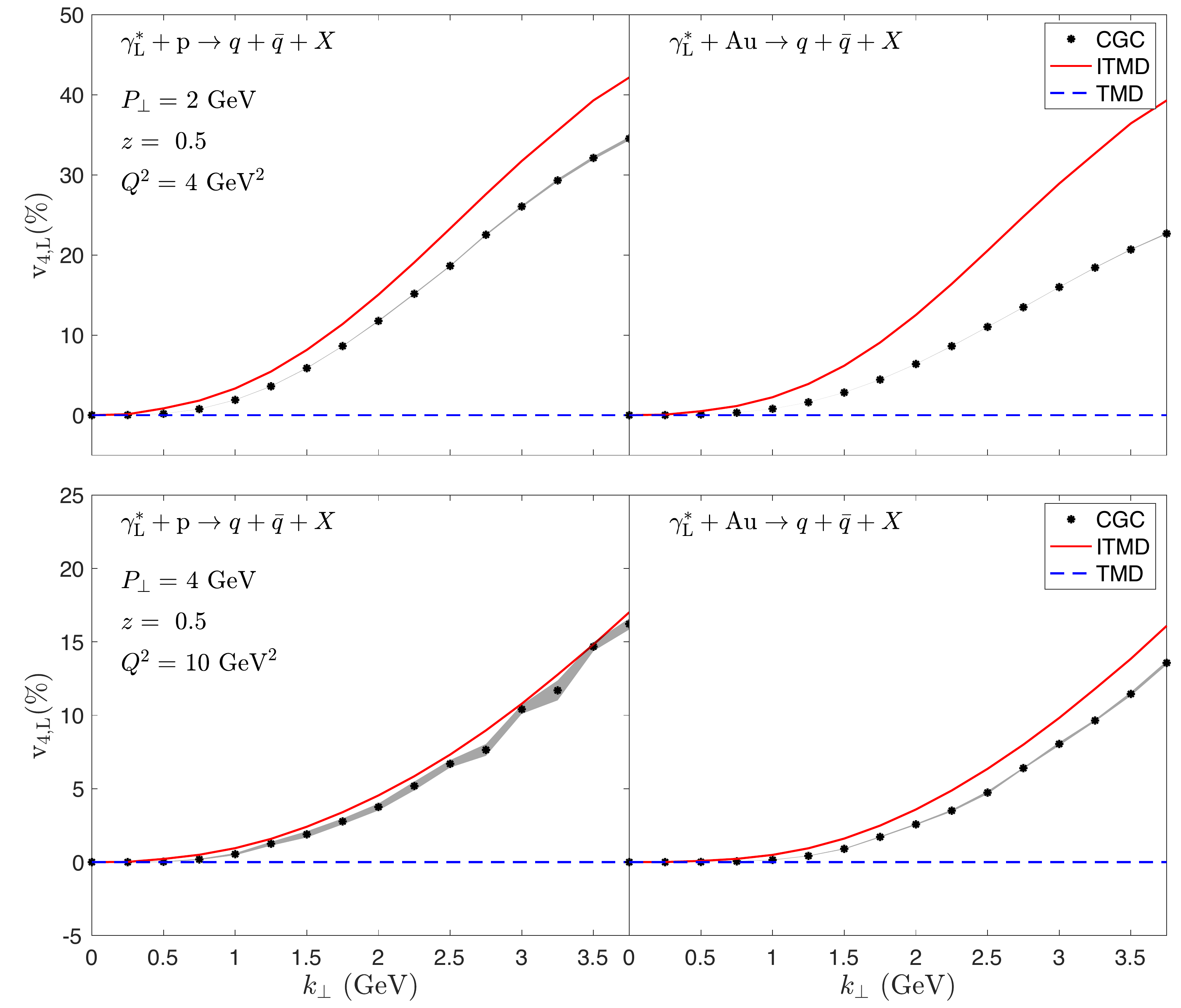}
\caption{Quadrangular anisotropy $v_4$ in the angle $\phi$ for production of quark anti-quark pairs in $\gamma^*_{\mathrm{T}} + p$ (left) and $\gamma^*_{\mathrm{T}} + \mathrm{Au}$ (right) scattering. The grey band shows the numerical uncertainty of the CGC calculation.}
\label{Fig_ratio_v4L}
\end{figure}

Our results to the quadrangular anisotropy when the photon is longitudinally polarized are shown in Fig.\,\ref{Fig_ratio_v4L}. The behavior of $v_{4,\rm{L}}$ is similar to that of $v_{2,\rm{L}}$: It is positive and significantly larger in magnitude than in the transversely polarized case. We observe that the quadrangular anisotropy is mostly sourced by kinematic power corrections; except for nuclear DIS and at small $Q$ and $P_\perp$ (upper right panel). As in the transversely polarized case, the effect of the genuine higher twists is to suppress the quadrangular anisotropy.

We close this section by pointing out that our results for $v_{4,\mathrm{L}}$ and $v_{4,\mathrm{T}}$ as a function of momentum imbalance $k_\perp$ and at $P_\perp = 4 \ \mathrm{GeV}$ and $Q^2 = 10 \ \mathrm{GeV}^2$ have the same sign and similar magnitudes to those estimated in \cite{Dumitru:2015gaa}. However, for this choice of kinematics the ITMD seems to be sufficient to describe the quadrangular anisotropy, especially in $\gamma^*+p$ scattering, and thus one only needs the WW gluon TMD (supplemented with ITMD hard factors) and no higher order derivative operators of Wilson lines.

\section{Conclusions}
\label{sec:conclusions_dijets}

The uncovering of the small-$x$ structure in protons and nuclei is one of the major goals of the Electron-Ion Collider. The study of azimuthal dihadron and dijet anisotropies can shed light both on the emergence of gluon saturation as well as the nature of the Weizsäcker-Williams gluon  TMD at small $x$ (both linearly polarized and unpolarized). In this manuscript, we explicitly computed the production of quark anti-quark dijets at leading order in the ITMD scheme and the CGC EFT, extending the results in \cite{Mantysaari:2019hkq}. 
We briefly reviewed the origin of kinematic twists and genuine higher twist corrections by expressing the product of Wilson lines as a transverse gauge link at $x^-=-\infty$ \cite{Boussarie:2020vzf}, generalizing the small dipole size expansion in \cite{Dominguez:2011wm}. The genuine higher twists (genuine saturation effects) enter with powers of $Q_s/Q_\perp$, and these are present in addition to the saturation contributions $Q_s/k_\perp$ in the WW gluon TMD at small $x$. Meanwhile, the ITMD resums kinematic power corrections to all orders in $k_\perp/Q_\perp$.

We found that for kinematics where either $P_\perp$ (dijet relative momentum) or $Q$ (virtuality of the DIS photon) is sufficiently larger than the saturation scale $Q_s$, the ITMD framework provides a good approximation to the CGC quark anti-quark differential cross-section and its anisotropies in the azimuthal angle $\phi$ between $\Pt$ and $\kt$. Compared to the standard TMD framework, we observe that the resummation of kinematic power corrections in the ITMD hard factors result in additional azimuthal correlations between $\Pt$ and $\kt$; in particular, we observe a non-zero quadrangular anisotropy. Thus, we expect that a proper extraction of the linearly polarized WW gluon TMD $xh_0(x, k_\perp)$ from dijet azimuthal asymmetries \cite{Dumitru:2018kuw} will require the implementation of the ITMD hard factors that have been computed analytically in this manuscript. An advantage of the ITMD framework is that it only depends on the WW gluon TMD in contrast to the CGC, which requires the computation of the quadrupole. This drastically simplifies the complexity of the computation of the differential cross-section for a wide set of kinematics, and makes it suitable for coupling to Monte-Carlo based hadronization and fragmentation routines. Efforts in this direction can be found in \cite{vanHameren:2021sqc} where projections for the EIC have been made using the ITMD$^{*}$\footnote{The ITMD$^{*}$ framework is equivalent to the ITMD framework in the absence of linearly polarized gluons, see Appendix\,\ref{sec:computing_J_hardfactor_ITMD_dijets}.} framework coupled to the event generator in  \cite{vanHameren:2016kkz}.

On the other hand, genuine higher twists effects which are absent in the ITMD framework, are most significant for scattering off a heavy target, as expected due to the enhancement of the nuclear saturation scale $Q_s^2 \sim A^{1/3}$. The genuine higher twist contributions suppress both the differential yield (near the back-to-back configuration) as well as the anisotropies (elliptic and quadrangular) in the angle $\phi$ for realistic EIC kinematics\footnote{Except for the angle averaged differential yield for the longitudinally polarized photon, where we observe an enhancement of the CGC relative to the TMD when the dijet momenta are large and the virtualities are small (see Fig.\,\ref{Fig_BB}).}. These genuine saturation effects may be accessed in the back-to-back measurement of low transverse momentum dihadrons and at low photon virtualities, and could be an important experimental tool for the gluon saturation searches at the EIC \cite{Zheng:2014vka}. For non back-to-back configurations, it might be possible to probe genuine saturation effects when the virtual photon is longitudinally polarized as Fig.\,\ref{Fig_ratio_N0L} showed deviations between CGC and ITMD across a wide range in $k_\perp$. In Fig.\,\ref{Fig_BB} we observed that in $e+\mathrm{Au}$ collisions the back-to-back peak for quark anti-quark production in the CGC is suppressed by a factor of 2 relative to the TMD framework for $P_\perp, Q \sim 1 \ \mathrm{GeV}$, whereas when either $P_\perp$ or $Q \sim 4 \ \mathrm{GeV}$ both frameworks agree to a less than a few percent difference.

To go towards more phenomenological applications, we plan to incorporate in our analysis the effect of Sudakov resummation \cite{Mueller:2012uf,Mueller:2013wwa} and final state soft gluon radiation \cite{Hatta:2020bgy,Hatta:2021jcd}, which have been shown to significantly impact the measurement of azimuthal dihadron and dijet anisotropies at the EIC \cite{vanHameren:2021sqc,Zheng:2014vka,Zhao:2021kae}. Furthermore, recent progress towards next-to-leading order computations for DIS at small-$x$ \cite{Lappi:2016oup,Lappi:2016fmu,Boussarie:2016ogo,Ducloue:2017ftk,Beuf:2017bpd,Hanninen:2017ddy,Roy:2019hwr,Beuf:2021qqa,Mantysaari:2021ryb,Xiao:2017yya,Hentschinski:2021lsh} will allow us to extend our computation to higher accuracy (for a recent computation of NLO contributions for dijet production in the CGC see \cite{Caucal:2021ent}). The inclusion of parton showers, hadronization, full jet reconstruction, and appropriate background processes will be necessary to fully assess the impact of our study on future EIC phenomenology. Finally, it would be interesting to compare the results of the suppression of the back-to-back peak due to gluon saturation with those caused by the momentum broadening due to cold nuclear matter energy loss and coherent power corrections \cite{Xing:2012ii} (see also \cite{Bergabo:2021woe} for a recent computation in the CGC).

\section*{Acknowledgements}

We thank Niklas Mueller for his collaboration at the early stages of this project. We are very grateful to Tolga Altinoluk, Xiaoxuan Chu, Piotr Kotko, Cyrille Marquet, and Raju Venugopalan for reading this manuscript and their valuable comments. We are also thankful to the anonymous referee for her/his insightful suggestions. H.M. is supported by the Academy of Finland project 314764, and by the EU Horizon 2020 research and innovation programme, STRONG-2020 project (grant agreement No 824093). F.S. and B.P.S. are supported under DOE Contract No.~DE-SC0012704. F.S is also supported by the joint Brookhaven National Laboratory-Stony Brook University Center for Frontiers in Nuclear Science (CFNS). This research used resources of the National Energy Research Scientific Computing Center, which is supported by the Office of Science of the U.S. Department of Energy under Contract No. DE-AC02-05CH11231.

\newpage

\appendix

\section{Wilson lines and transverse gauge links}
\label{sec:Wilson_and_gaugelink}

The goal of this section is to briefly summarize the derivation of Eq.\,\eqref{eq:dipole_transversegaugelink} which relates the product of two Wilson lines to a transverse gauge link at $x^-=-\infty$:
\begin{align}
    V(\xt)V^\dagger(\yt) = \mathcal{P} \exp\left[-ig \int_{\yt}^{\xt} \der \zt^i  \Ats^i(\zt) \right] \,.
    \label{eq:dipole_transversegaugelink_2}
\end{align}
As discussed in Sec.\,\ref{sec:field_stregnth_WilsonLines} this relation is helpful to establish the relation between CGC and TMD amplitudes rigorously since it allows for a systematic expansion in powers of $g\Ats$. More details and generalizations of the presented derivations can be found in \cite{Altinoluk:2019fui,Altinoluk:2019wyu,Boussarie:2020vzf}. 

To begin we recall that the gauge transformation (introduced in Eq.\,\eqref{eq:gauge_trans_A+_Ai}) going from $\partial_\mu A^\mu =0$ Lorenz gauge to $\tilde{A}^+ =0$ light-cone gauge is given by
\begin{align}
    \Omega_{\xt}(x^-) = \Pcal \exp( ig \int_{x^-}^{+\infty} \der z^- A^{+}  (z^-,\vect{x})  ) \,.
    \label{eq:gauge_trans_A+_Ai_2}
\end{align}
It follows by comparing Eq.\,\eqref{eq:gauge_trans_A+_Ai_2} with the definition of the light-like Wilson line in Eq.\,\eqref{eq:WilsonLine} that
\begin{align}
    V(\xt) = \Omega_{\xt}(-\infty) \,.
\end{align}

Similarly, the product of two Wilson lines can be expressed as the product of two gauge rotations at the $x^- = -\infty$ boundary
\begin{align}
    V(\xt)V^\dagger(\yt) = \Omega_{\xt}(-\infty) \Omega^{-1}_{\yt}(-\infty) \,.
    \label{eq:dipole_gaugerotations}
\end{align}
In order to connect this result with that in Eq.\,\eqref{eq:dipole_transversegaugelink_2} we shall express the product of these gauge rotations as a transverse gauge link. This follows by first noting
\begin{align}
    \Omega^{-1}_{\xt}(-\infty)  &= \Omega^{-1}_{\yt}(-\infty)+ \int_{\yt}^{\xt} \der \zt^i \partial_i \Omega^{-1}_{\zt}(-\infty) \,, \nonumber \\
    &= \Omega^{-1}_{\yt}(-\infty) +ig \int_{\yt}^{\xt} \der \zt^i \Omega^{-1}_{\zt}(-\infty) \Ats^i(\zt) \,,
    \label{eq:gaugerotation_transport}
\end{align}
where in the second equality we related the derivative of the gauge rotation to the gauge field in $\tilde{A}^+ =0$ gauge by $ ig \Ats^i(\zt) = \Omega_{\zt}(-\infty) \partial_i \Omega^{-1}_{\zt}(-\infty)$ (see Eq.\,\eqref{eq:transversegauge_gaugerotation}). From Eq.\,\eqref{eq:gaugerotation_transport} then we find the recursive relation
\begin{align}
    \Omega_{\xt}(-\infty)\Omega^{-1}_{\yt}(-\infty) = \mathbb{1} -ig \int_{\yt}^{\xt} \der \zt^i \Omega_{\xt}(-\infty) \Omega^{-1}_{\zt}(-\infty) \Ats^i(\zt) \,,
    \label{eq:transversegaugelink_recursive1}
\end{align}
and in a similar fashion one can show
\begin{align}
    \Omega_{\xt}(-\infty)\Omega^{-1}_{\yt}(-\infty) = \mathbb{1} -ig \int_{\yt}^{\xt} \der \zt^i  \Ats^i(\zt) \Omega_{\zt}(-\infty) \Omega^{-1}_{\yt}(-\infty) \,.
    \label{eq:transversegaugelink_recursive2}
\end{align}
These two recursive relations are satisfied by
\begin{align}
    \Omega_{\xt}(-\infty)\Omega^{-1}_{\yt}(-\infty) =\mathcal{P} \exp\left[-ig \int_{\yt}^{\xt} \der \zt^i  \Ats^i(\zt) \right] \,,
\end{align}
combining this expression with Eq.\,\eqref{eq:dipole_gaugerotations} results in the desired relation in Eq.\,\eqref{eq:dipole_transversegaugelink_2}. 

The integral is independent of the choice of path connecting the transverse points $\yt$ and $\xt$ due to the non-Abelian Stokes' theorem and the fact that the transverse components of the field strength are zero $\tilde{F}^{ij} =0$ (at a given $x^-$, $\Ats^i(x^-,\xt)$ is a pure gauge field in two dimensions), more precisely $\Pcal \exp\left(\oint_{\mathcal{C}} \der \zt^i \Ats^{i}(\zt) \right) = \mathbb{1}$ for any closed path $\mathcal{C}$ contained in the transverse plane\footnote{The non-Abelian Stokes' theorem reads $ \Pcal \exp(\oint_{\mathcal{C}} \der x_{\mu} A^{\mu}) = \Pcal \exp(\int_{\mathcal{S}} \der \sigma_{\mu\nu} U F^{\mu\nu} U^\dagger)$ where $\der \sigma_{\mu\nu}$ is the surface measure on $\mathcal{S}$ and $U$ denotes a Wilson line connecting the point
$x \in \mathcal{S}$ enclosed by the surface measure to an arbitrary base point $O$ on $\mathcal{C}$ (see \cite{fishbane1981stokes}). }.

It follows from Eq.\,\eqref{eq:transversegaugelink_recursive1} that the product of two Wilson lines satisfy an identical recursive relation
\begin{align}
    V(\xt)V^\dagger(\yt) = \mathbb{1}-ig \int_{\yt}^{\xt} \der \zt^i  \Ats^i(\zt)V(\zt)V^\dagger(\yt) \,.
\end{align}
After the application of the recursive relation in Eq.\,\eqref{eq:transversegaugelink_recursive2} we find
\begin{align}
    \mathbb{1}-V(\xt)V^\dagger(\yt) =  \left[\mathbb{1}-V(\xt)V^\dagger(\yt) \right]_{\rm{ITMD}} + \left[\mathbb{1}-V(\xt)V^\dagger(\yt) \right]_{\rm{g.h.t.}} \,,
\end{align}
where we define
\begin{align}
    \left[\mathbb{1}-V(\xt)V^\dagger(\yt) \right]_{\rm{ITMD}} &=ig \int_{\yt}^{\xt} \der \zt^i  \Ats^i(\zt) \label{eq:dipole_ITMD_expansion} \,, \\
    \left[\mathbb{1}-V(\xt)V^\dagger(\yt) \right]_{\rm{g.h.t.}} &= g^2 \int_{\yt}^{\xt} \!\!\!\! \der \ztone^i  \int_{\yt}^{\ztone} \!\!\!\! \der \zttwo^j  \Ats^i(\ztone) V(\ztone) V^\dagger(\zttwo) \Ats^j(\zttwo) \,.
    \label{eq:dipole_ght_expansion}
\end{align}
We would like to obtain a tractable expression for Eq.\,\eqref{eq:dipole_ITMD_expansion} which will be the starting point for our study of the production of quark and anti-quark pairs in the ITMD framework.

Since the choice of path connecting $\yt$ and $\xt$ in Eq.\,\eqref{eq:dipole_ITMD_expansion} is arbitrary, we choose a straight line path
\begin{align}
    \zt(s) = \yt + s \rt,\quad s \in [0,1] \,,
\end{align}
where $\rt =\xt -\yt$. Then we have
\begin{align}
    \left[\mathbb{1}-V(\xt)V^\dagger(\yt) \right]_{\rm{ITMD}} = i \rt^i \int_{0}^{1} \der s \ g \Ats^i(\yt + s\rt) \,.
\end{align}
The integral over $s$ is easily done in momentum space using the identity
\begin{align}
    \Ats^i(\zt) = \int \frac{\der^2 \lt}{(2\pi)^2} e^{i \lt\cdot \zt} \int \der^2 \vt e^{-i \lt \cdot \vt} \Ats^i(\vt) \,,
\end{align}
and the simple integral over the phase
\begin{align}
    \int \der s \ e^{i \lt \cdot (\yt + s\rt)}=\left( \frac{e^{i\lt \cdot \xt}-e^{i\lt \cdot \yt}}{i \lt \cdot \rt}\right)\,.
\end{align}
We find
\begin{align}
    \left[\mathbb{1}-V(\xt)V^\dagger(\yt) \right]_{\rm{ITMD}}
    & = i \rt^i  \int \der^2 \vt \int \frac{\der^2 \lt}{(2\pi)^2}  e^{-i \lt \cdot \vt} \ g \Ats^i(\vt) \left( \frac{e^{i\lt \cdot \xt}-e^{i\lt \cdot \yt}}{i \lt \cdot \rt}\right) \,.
\end{align}
Despite the fact that this expression might look more formidable than Eq.\,\eqref{eq:dipole_ITMD_expansion}, we will show in Sec.\,\ref{sec:inclusive_dijets_iTMD} that the resultant differential cross-section for quark and anti-quark dijet production has a relatively simple analytic expression.

Following the same procedure and with some algebra we find a similar expression for Eq.\,\eqref{eq:dipole_ght_expansion}
\begin{align}
    \left[\mathbb{1}-V(\xt)V^{\dagger}(\yt)\right]&_{\mathrm{g.h.t}} = \int\frac{{\rm d}^{2}\ltone}{(2\pi)^{2}}\frac{{\rm d}^{2}\lttwo}{(2\pi)^{2}}\int{\rm d}^{2}\vtone{\rm d}^{2}\vttwo{\rm e}^{-i(\ltone\cdot\vtone)-i(\lttwo\cdot\vttwo)}\frac{\rt^{i}\rt^{j}}{i(\lttwo\cdot\rt)}\nonumber \\
    & \times\left(\frac{{\rm e}^{i(\ltone+\lttwo)\cdot\boldsymbol{x}_{\perp}}-{\rm e}^{i(\ltone+\lttwo)\cdot\boldsymbol{y}_{\perp}}}{i(\ltone+\lttwo)\cdot\rt}-{\rm e}^{i(\lttwo\cdot\boldsymbol{y}_{\perp})}\frac{{\rm e}^{i(\ltone\cdot\boldsymbol{x}_{\perp})}-{\rm e}^{i(\ltone\cdot\boldsymbol{y}_{\perp})}}{i(\ltone\cdot\rt)}\right)\nonumber \\
    & \times g^2 A^{i}(\vtone)V(\vtone)V^{\dagger}(\vttwo)A^{j}(\vttwo)\,.
\end{align}

\section{Useful integrals}
\label{sec:Appendix_usefulinregrals}
We list some useful transverse integrals:
\begin{align}
    \int \frac{\der^2 \rt}{2\pi} e^{-i \Pt \cdot \rt} K_0(\varepsilon r_\perp) &= \frac{1}{P_\perp^2 + \varepsilon^2} \,,
    \label{eq:integral_1}\\
    \int \frac{\der^2 \rt}{2\pi} e^{-i \Pt \cdot \rt} \frac{i \varepsilon \rt^j}{r_\perp} K_1(\varepsilon r_\perp) &= \frac{\Pt^j}{P_\perp^2 + \varepsilon^2} \,,
    \label{eq:integral_5}\\
    \int \frac{\der^2 \lt}{2\pi} e^{i \lt \cdot \rt} \frac{1}{\lt^2 + \varepsilon^2} &= K_0(\varepsilon r_\perp) \,,
    \label{eq:integral_3}\\
    \int \frac{\der^2 \lt}{2\pi} e^{i \lt \cdot \rt} \frac{\lt^j}{\lt^2 + \varepsilon^2} &= \frac{i \varepsilon \rt^j}{r_\perp} K_1(\varepsilon r_\perp) \,,
    \label{eq:integral_4}\\
    \int \frac{\der^2 \rt}{2\pi} e^{-i \Pt \cdot \rt} i \rt^j K_0(\varepsilon r_\perp) &= \frac{2 \Pt^j }{(P_\perp^2 + \varepsilon^2)^2} \,,
    \label{eq:integral_2}\\
    \int \frac{\der^2 \rt}{2\pi} e^{-i \Pt \cdot \rt} \rt^k \frac{ \varepsilon \rt^j}{r_\perp} K_1(\varepsilon r_\perp) &= \frac{1}{P^2 + \varepsilon^2}\left(\delta^{jk} - \frac{2\Pt^j\Pt^k}{P_\perp^2 + \varepsilon^2} \right) \,.
    \label{eq:integral_6}
\end{align}

\section{Explicit representation of Dirac spinors}
\label{sec:explicit_spinor_rep}

We work in the Dirac basis for gamma matrices
\begin{align}
    \gamma^0 = \begin{bmatrix}
    \mathbb{1} & 0 \\
    0 & -\mathbb{1}
    \end{bmatrix} \ \ \ \ \ , \ \  \gamma^i= \begin{bmatrix}
    0 & \sigma^i \\
    -\sigma^i & 0
    \end{bmatrix} \ \ \ , \ \sigma^1 = \begin{bmatrix}
    0 & 1 \\
    1 & 0 
    \end{bmatrix} \ \ \ , \ \sigma^2 = \begin{bmatrix}
    0 & -i \\
    i & 0 
    \end{bmatrix} \ \ \ , \ \sigma^3 = \begin{bmatrix}
    1 & 0 \\
    0 & -1 
    \end{bmatrix} \,.
\end{align}
where $\mathbb{1}$ is the two-by-two identity matrix.

The helicity operator $h$ is defined as 
\begin{align}
    h = \frac{2 \vec{k}\cdot\vec{S}}{|\vec{k}|} \,, \quad \vec{S} = \frac{1}{2}\begin{bmatrix} \vec{\sigma} & 0 \\
0 & \vec{\sigma} 
\end{bmatrix} \,.
\end{align}
where $\vec{k} = (\kt,k^3)$ is the three momentum.
    
It is straightforward to check that the (massless) Dirac equation
\begin{align}
    \slashed{k} u(k) = 0 \,,
\end{align}
has the following solutions\footnote{In the massless case, the spinors corresponding to particle and anti-particle are the same, but correspond to opposite helicities.}
\begin{align}
    u_+(k) =v_{-}(k) =  \frac{1}{2^{1/4}}\begin{bmatrix}
    \sqrt{k^+} e^{-i\phi_k} \\
    \sqrt{k^-} \\
    \sqrt{k^+} e^{-i\phi_k} \\
    \sqrt{k^-}
    \end{bmatrix} \,, \ \ \ \, \ \ u_-(k) =v_{+}(k)  = \frac{1}{2^{1/4}}\begin{bmatrix}
    \sqrt{k^-}  \\
    -\sqrt{k^+} e^{i\phi_k} \\
    -\sqrt{k^-}  \\
    \sqrt{k^+} e^{i\phi_k}
    \end{bmatrix} \,,
\end{align}
where the subscripts $\pm$ denote the helicities\footnote{Note that in this manuscript, $p^-$ is the large component of the spinor momenta; thus $p^3 < 0$.}, $\phi_k$ is the azimuthal angle of $\kt$, and the normalization is chosen so that
\begin{align}
    \slashed{k} = \sum_{\sigma} u_\sigma(k) \bar{u}_\sigma(k) \,,
\end{align}
where the barred spinors are defined as usual by $\bar{u} = u^\dagger \gamma^0$.

The following identities hold:
\begin{align}
    \bar{u}_{\sigma}(k_1) \gamma^- v_{\sigma'}(k_2) &= 2 \sqrt{k_1^- k_2^-} \delta_{\sigma,- \sigma'} \label{eq:spinor_gamma_1} \,, \\
    \bar{u}_{\sigma}(k_1) \gamma^i \gamma^j \gamma^- v_{\sigma'}(k_2) &= 2 \sqrt{k_1^- k_2^-} \left( -\delta^{ij} + i \sigma \epsilon^{ij} \right) \delta_{\sigma,- \sigma'}
    \label{eq:spinor_gamma_2} \,.
\end{align}

\section{Computing the perturbative factors $\Ncal^{\lambda}$}
\label{sec:computing_N_CGC_dijets}

To compute $\Ncal^\lambda_{\sigma\sigma'} (\rt)$ we first perform the $l^-$ and $l^+$ integrations. The former is done immediately using the delta function $\delta(k_1^- - l^-)$, while the latter is performed via contour integration using Cauchy's theorem, we find
\begin{align}
    \Ncal^\lambda_{\sigma\sigma'} (\rt) = \int \frac{\der^2 \lt}{(2\pi)} \frac{N^{\lambda}_{\sigma\sigma'}(l) e^{i \lt \cdot \rt}}{z_1 z_2 Q^2 + \lt^2}
    \label{eq:dijet-LO-pert_2} \,.
\end{align}
To compute the transverse integration, we must first work out the Dirac algebra. We find
\begin{align}
    N^{\lambda =0}_{\sigma\sigma'} &= 2 (z_1 z_2)^{3/2} Q \delta_{\sigma,-\sigma'} \label{eq:dijet-LO-dirac_L} \,, \\
    N^{\lambda =\pm 1}_{\sigma\sigma'} &= (z_1 z_2)^{1/2} (\lt \cdot \et^{\lambda }) \left[(z_1-z_2) -\lambda \sigma \right] \delta_{\sigma,-\sigma'}
    \label{eq:dijet-LO-dirac_T} \,,
\end{align}
where we used Eqs.\,\eqref{eq:spinor_gamma_1} and \eqref{eq:spinor_gamma_2}, and $ \epsilon^{ij} \et^{\lambda,i} = - i \lambda \et^{\lambda,j}$. This last identity of the polarization vector holds for circularly polarized states $\et^{\pm 1} = \frac{1}{\sqrt{2}}(1,\pm i)$ .

Eqs.\,\eqref{eq:dijet-LO-pert_L} and \eqref{eq:dijet-LO-pert_T} then follow by inserting Eqs.\,\eqref{eq:dijet-LO-dirac_L} and \eqref{eq:dijet-LO-dirac_T} into Eq.\,\eqref{eq:dijet-LO-pert_2} and performing the transverse integrations with the aid of identities in Appendix\,\ref{sec:Appendix_usefulinregrals}.

\section{Computing the ITMD hard factors $\mathcal{J}^\lambda$}
\label{sec:computing_J_hardfactor_ITMD_dijets}

To compute the ITMD hard factors in Eqs.\,\eqref{eq:ITMD_hardL_analytic} and \eqref{eq:ITMD_hardT_analytic} we insert Eqs.\,\eqref{eq:dijet-LO-pert_L} and \eqref{eq:dijet-LO-pert_T} into Eq.\,\eqref{eq:ITMD_J} and find
\begin{align}
    \Jcal^{\lambda=0,i}_{\sigma,\sigma'}(\Pt,\kt) &= 2 (z_1 z_2)^{3/2} Q \delta_{\sigma,-\sigma'} J_\mathrm{L}^{i}(\Pt+z_{1}\kt,\kt) \,, \\
    \Jcal^{\lambda=\pm 1,i}_{\sigma,\sigma'}(\Pt,\kt) &=  (z_1 z_2)^{1/2} \left[(z_2 -z_1) + \sigma \lambda \right] \delta_{\sigma,-\sigma'} \et^{\lambda = \pm 1,j} J_\mathrm{T}^{ij}(\Pt+z_{1}\kt,\kt) \,,
\end{align}
where 
\begin{align}
    J_\mathrm{L}^{i}(\pt,\kt) & \equiv\!\int\!\frac{{\rm d}^{2} \rt}{2\pi}\,{\rm e}^{-i(\pt\cdot\rt)} i \rt^{i}K_{0}(\varepsilon|\rt|)\frac{{\rm e}^{i(\kt\cdot\rt)}-1}{i(\kt\cdot\rt)} \,,\label{eq:I1-def}
\end{align}
and
\begin{align}
    J_\mathrm{T}^{ij}(\pt,\kt) & \equiv\!\int\!\frac{{\rm d}^{2}\rt}{2\pi}\,{\rm e}^{-i(\pt\cdot\rt)}\rt^{i}\frac{\varepsilon\rt^{j}}{|\rt|}K_{1}(\varepsilon|\rt|)\frac{{\rm e}^{i(\kt\cdot\rt)}-1}{i(\kt\cdot\rt)} \,.\label{eq:I2-def} 
\end{align}
The first steps to perform these integrals explicitly is to integrate the $\rt^i$ factors by parts into derivatives w.r.t. $\pt^i$, and to use the integral form
\begin{equation}
\frac{{\rm e}^{i(\kt\cdot\rt)}-1}{i(\kt\cdot\rt)}=\int_{0}^{1}\!{\rm d}\alpha \, {\rm e}^{i\alpha(\kt\cdot\rt)}\label{eq:string-length}.
\end{equation}
Using Eq.~(\ref{eq:integral_1}) for $J^i_\mathrm{L}$ and Eq.~(\ref{eq:integral_5}) for $J_T^{ij}$ then yields:
\begin{equation}
    J_\mathrm{L}^{i}(\pt,\kt)= - \frac{\partial}{\partial\pt^{i}}\int_{0}^{1}{\rm d}\alpha\frac{1}{(\pt-\alpha\kt)^{2}+\varepsilon^{2}} \,, \label{eq:I1-step1}
\end{equation}
and
    \begin{equation}
    J_\mathrm{T}^{ij}(\pt,\kt)=\frac{\partial}{\partial\pt^{i}}\int_{0}^{1}{\rm d}\alpha\frac{\pt^{j}-\alpha\kt^{j}}{(\pt-\alpha\kt)^{2}+\varepsilon^{2}}\label{eq:I2-step2} \,.
\end{equation}
Let us now address the $\alpha$ integral. Introducing
\begin{equation}
A^{2}=\sqrt{\kt^{2}(\pt^{2}+\varepsilon^{2})-(\pt\cdot\kt)^{2}} \,,
\end{equation}
noting that
\begin{align}
\frac{1}{(\boldsymbol{p}_{\perp}-\alpha\boldsymbol{k}_{\perp})^{2}+\varepsilon^{2}} & =\frac{1}{\boldsymbol{k}_{\perp}^{2}\left[\left(\alpha-\frac{(\boldsymbol{p}_{\perp}\cdot\boldsymbol{k}_{\perp})}{\boldsymbol{k}_{\perp}^{2}}\right)^{2}+\frac{A^{4}}{\boldsymbol{k}_{\perp}^{4}}\right]}\label{eq:frac} \,,
\end{align}
and using the standard integrals
\begin{equation}
\int_{\alpha_{0}}^{\alpha_{1}}\frac{{\rm d}\alpha}{\alpha^{2}+a^{2}}=\frac{1}{a}\left[\arctan\left(\frac{\alpha_{1}}{a}\right)-\arctan\left(\frac{\alpha_{0}}{a}\right)\right],\label{eq:arctan-int}
\end{equation}
and
\begin{align}
\int_{\alpha_{0}}^{\alpha_{1}}{\rm d}\alpha\frac{\alpha}{\alpha^{2}+a^{2}} & =\frac{1}{2}\ln\left(\frac{\alpha_{1}^{2}+a^{2}}{\alpha_{0}^{2}+a^{2}}\right)\label{eq:log-int} \,,
\end{align}
we have:
   \begin{align}
    J_{\mathrm{L}}^{i}(\boldsymbol{p}_{\perp},\boldsymbol{k}_{\perp}) & =-\frac{\partial}{\partial\boldsymbol{p}_{\perp}^{i}}\frac{1}{A^{2}}\left[\arctan\left(\frac{(\boldsymbol{k}_{\perp}-\boldsymbol{p}_{\perp})\cdot\boldsymbol{k}_{\perp}}{A^{2}}\right)+\arctan\left(\frac{\boldsymbol{p}_{\perp}\cdot\boldsymbol{k}_{\perp}}{A^{2}}\right)\right] \label{eq:I1-before-der} \,,
\end{align}
and
\begin{align}
    J_{\mathrm{T}}^{ij}(\boldsymbol{p}_{\perp},\boldsymbol{k}_{\perp}) & =-\frac{1}{\boldsymbol{k}_{\perp}^{2}}\frac{\partial}{\partial\boldsymbol{p}_{\perp}^{i}}\left[\frac{\boldsymbol{k}_{\perp}^{j}}{2}\ln\left(\frac{(\boldsymbol{k}_{\perp}-\boldsymbol{p}_{\perp})^{2}+\varepsilon^{2}}{\boldsymbol{p}_{\perp}^{2}+\varepsilon^{2}}\right)\right.\label{eq:I2-before-der}\\
    & \left.-\frac{\boldsymbol{k}_{\perp}^{2}\boldsymbol{p}_{\perp}^{j}-(\boldsymbol{p}_{\perp}\cdot\boldsymbol{k}_{\perp})\boldsymbol{k}_{\perp}^{j}}{A^{2}}\left\{ \arctan\left(\frac{(\boldsymbol{k}_{\perp}-\boldsymbol{p}_{\perp})\cdot\boldsymbol{k}_{\perp}}{A^{2}}\right)+\arctan\left(\frac{\boldsymbol{p}_{\perp}\cdot\boldsymbol{k}_{\perp}}{A^{2}}\right)\right\} \right] \,. \nonumber 
\end{align}
With the help of the relation
\begin{equation}
\frac{\partial}{\partial\pt^{i}}A^{4}=2\left[\kt^{2}\pt^{i}-\kt^{i}(\pt\cdot\kt)\right],\label{eq:drho}
\end{equation}
and the following additional relation for $J_\mathrm{T}^{ij}$:
\begin{equation}
\kt^{2}\pt^{i}\pt^{j}+\pt^{2}\kt^{i}\kt^{j}=\left[\pt^{2}\kt^{2}-(\pt\cdot\kt)^{2}\right]\delta^{ij}+(\pt\cdot\kt)(\pt^{i}\kt^{j}+\pt^{j}\kt^{i})\label{eq:tensors} \,.
\end{equation}
Along with tedious but straightforward algebra, we find eventually:
    \begin{align}
    J_\mathrm{L}^{i}(\ktone,\kt) & =\frac{ \left[\kt^{2}\Pt^{i}-\left(\Pt\cdot\kt\right)\kt^{i}\right]}{\mathcal{X}^{3}} \left[ \arctan\left(\frac{\kt \cdot \ktone}{\mathcal{X}} \right) + \arctan\left(\frac{\kt \cdot \kttwo}{\mathcal{X}} \right)\right] \nonumber \\
    & - \frac{\left[\kt^{2}\Pt^{i}-\left(\Pt\cdot\kt\right)\kt^{i}\right]}{\mathcal{X}^2}\left(\frac{-\ktone\cdot\kttwo + \varepsilon^2}{(\ktone^{2}+\varepsilon^{2})(\kttwo^{2}+\varepsilon^{2})}\right)\nonumber \\
    & + \frac{\ktone^{i}-\kttwo^{i}}{(\ktone^{2}+\varepsilon^{2})(\kttwo^{2}+\varepsilon^{2})} \,,\label{eq:I1-final}
\end{align}
and
    \begin{align}
    J_\mathrm{T}^{ij}(\ktone,\kt) & =\frac{\varepsilon^{2}(\delta^{ij}\kt^{2}-\kt^{i}\kt^{j})}{\mathcal{X}^{3}}\left[ \arctan\left(\frac{\kt \cdot \ktone}{\mathcal{X}} \right) + \arctan\left(\frac{\kt \cdot \kttwo}{\mathcal{X}} \right)\right] \nonumber \\
    & +\frac{1}{\kt^{2}}\frac{\varepsilon^{2}(\delta^{ij}\kt^{2}-\kt^{i}\kt^{j})\kt^{k}}{\mathcal{X}^2}\left(\frac{\ktone^{k}}{\ktone^{2}+\varepsilon^{2}}+\frac{\kttwo^{k}}{\kttwo^{2}+\varepsilon^{2}}\right)\nonumber \\
    & +\frac{1}{\kt^{2}}(\kt^{i}\delta^{jk}+\kt^{j}\delta^{ik}-\kt^{k}\delta^{ij})\left(\frac{\ktone^{k}}{\ktone^{2}+\varepsilon^{2}}+\frac{\kttwo^{k}}{\kttwo^{2}+\varepsilon^{2}}\right) \,, \label{eq:I2-final}
\end{align}
where
\begin{align}
    \mathcal{X}^2 &= \Pt^{2}\kt^{2}-(\Pt\cdot\kt)^{2}+\varepsilon^{2}\kt^{2} \,.
\end{align}
Therefore, we find the following expressions for the ITMD hard factors in the amplitude of quark anti-quark production:
\begin{align}
    \Jcal^{\lambda=0,i}_{\sigma,\sigma'}(\Pt,\kt) &= 2 (z_1 z_2)^{3/2} Q \delta_{\sigma,-\sigma'} \left \{ \frac{\ktone^{i}-\kttwo^{i}}{(\ktone^{2}+\varepsilon^{2})(\kttwo^{2}+\varepsilon^{2})} \right. \nonumber \\
    &+\frac{ \left[\kt^{2}\Pt^{i}-\left(\Pt\cdot\kt\right)\kt^{i}\right]}{\mathcal{X}^{3}} \left[ \arctan\left(\frac{\kt \cdot \ktone}{\mathcal{X}} \right) + \arctan\left(\frac{\kt \cdot \kttwo}{\mathcal{X}} \right)\right] \nonumber \\
    & \left. - \frac{\left[\kt^{2}\Pt^{i}-\left(\Pt\cdot\kt\right)\kt^{i}\right]}{\mathcal{X}^2}\left(\frac{-\ktone\cdot\kttwo + \varepsilon^2}{(\ktone^{2}+\varepsilon^{2})(\kttwo^{2}+\varepsilon^{2})}\right) \right \} \,,
\end{align}
and
\begin{align}
    \Jcal^{\lambda=\pm 1,i}_{\sigma,\sigma'}(\Pt,\kt) &= (z_1 z_2)^{1/2} \left[(z_2 -z_1) + \sigma \lambda \right] \delta_{\sigma,-\sigma'} \et^{\lambda = \pm 1,j} \nonumber \\
    & \times \left\{ \frac{\varepsilon^{2}(\delta^{ij}\kt^{2}-\kt^{i}\kt^{j})}{\mathcal{X}^{3}}\left[ \arctan\left(\frac{\kt \cdot \ktone}{\mathcal{X}} \right) + \arctan\left(\frac{\kt \cdot \kttwo}{\mathcal{X}} \right)\right] \right. \nonumber \\
    & +\frac{1}{\kt^{2}}\frac{\varepsilon^{2}(\delta^{ij}\kt^{2}-\kt^{i}\kt^{j})\kt^{k}}{\mathcal{X}^2}\left(\frac{\ktone^{k}}{\ktone^{2}+\varepsilon^{2}}+\frac{\kttwo^{k}}{\kttwo^{2}+\varepsilon^{2}}\right)\nonumber \\
    & \left. +\frac{1}{\kt^{2}}(\kt^{i}\delta^{jk}+\kt^{j}\delta^{ik}-\kt^{k}\delta^{ij})\left(\frac{\ktone^{k}}{\ktone^{2}+\varepsilon^{2}}+\frac{\kttwo^{k}}{\kttwo^{2}+\varepsilon^{2}}\right) \right\} \,.
\end{align}

It is illustrative to consider two interesting limits:  the so-called ITMD$^{*}$ limit \cite{Altinoluk:2021ygv}, and photo-production limit $Q^2 \rightarrow 0$.

The hard factors in the ITMD$^{*}$ scheme can be obtained in a diagrammatic approach with off-shell gluons \cite{Kotko:2015ura}, and they can be obtained from our results by the following projection \cite{Altinoluk:2021ygv}: 
\begin{align}
    \Hcal_{\rm{ITMD}^{*}}^{ij,\lambda}(\Pt,\kt) = \Hcal_{\rm{ITMD}}^{ij,\lambda}(\Pt,\kt) \frac{\kt^{i}\kt^{j}}{k_\perp^2} \,.
\end{align}
Therefore, in the ITMD$^{*}$ limit it is enough to consider the projections:
\begin{align}
    \Jcal^{\lambda=0,i}_{\sigma,\sigma'}(\Pt,\kt) \frac{\kt^{i}}{k_\perp} & = 2  (z_1z_2)^{3/2} Q \delta_{\sigma,-\sigma'} \left(\frac{1}{\kttwo^{2}+\varepsilon^{2}}-\frac{1}{\ktone^{2}+\varepsilon^{2}}\right) \frac{1}{k_\perp} \label{eq:I1-final-cont} \,, \\
    \Jcal^{\lambda=\pm1 ,i}_{\sigma,\sigma'}(\Pt,\kt) \frac{\kt^{i}}{k_\perp} & = (z_1z_2)^{1/2} \left[  (z_2-z_1) + \sigma \lambda \right] \delta_{\sigma,-\sigma'} \left(\frac{\kttwo \cdot \et^{\lambda}}{\kttwo^{2}+\varepsilon^{2}} + \frac{\ktone \cdot \et^{\lambda}}{\ktone^{2}+\varepsilon^{2}}\right) \frac{1}{k_\perp} \,.\label{eq:I2-final-cont} 
\end{align}
Alternatively, the projections in Eqs.\,\eqref{eq:I1-final-cont} and \eqref{eq:I2-final-cont} could have been easily obtained by first projecting Eq.\,\eqref{eq:ITMD_J} with $\kt^i$, resulting in:
\begin{align}
    \kt^j \Jcal_{\sigma\sigma'}^{\lambda,j}(\Pt,\kt)  = \int \frac{\der^2 \rt}{2\pi}  \left(e^{i\kttwo \cdot\rt}-e^{-i\ktone \cdot\rt}\right) \ \Ncal_{\sigma\sigma^{\prime}}^{\lambda}(\rt) \,.
    \label{eq:iTMD_J2}
\end{align}
We refer the reader to \cite{Altinoluk:2021ygv} where the differences between the ITMD$^*$ and ITMD schemes have been numerically studied for the electro-production of heavy quarks. 

Finally, in the photo-production limit, we find
\begin{align}
    \lim_{Q^2 \rightarrow 0} \Jcal_{\sigma\sigma'}^{\lambda =\pm 1,i}(\Pt,\kt) &= (z_1z_2)^{1/2} \left[  (z_2-z_1) + \sigma \lambda \right] \delta_{\sigma,-\sigma'} \et^{\lambda=\pm 1,j} \frac{c^{ijk}}{\kt^{2}} \left(\frac{\ktone^{k}}{\ktone^{2}}+\frac{\kttwo^{k}}{\kttwo^{2}}\right)  \,,
\end{align}
where
\begin{align}
    c^{ijk} = \kt^{i}\delta^{jk}+\kt^{j}\delta^{ik}-\kt^{k}\delta^{ij} \,,
\end{align}
which is compatible with the result from~\cite{Altinoluk:2019fui}.

\section{Operator definition of the WW gluon TMD}
\label{sec:Appendix_WW}

In this appendix we briefly review the relation between the operator definition of the Weizsäcker-Williams distribution and the definition in Eq.\,\eqref{eq:WW_gluon_1}, which was fleshed out for the first time in \cite{Dominguez:2011wm}. In what follows let us go back to the general case where we have not fixed the gauge condition for the gauge field $A^{\mu,a}$.

The operator definition of a generic gluon TMD is constructed from the bilocal operator field strength tensors appropriately dressed by gauge links for gauge invariance. Let us define the finite path Wilson lines along the light-cone direction as
\begin{align}
    [b_2^-,b_1^-]_{\bt} = \Pcal \exp \left[ig\int_{b_1^-}^{b_2^-} A^{+,a}(z^-,\bt) t^a dz^- \right] \,,
\end{align}
and that along the transverse direction (where the path is a straight line) as
\begin{align}
    [\bttwo,\btone]_{b^-} = \Pcal \exp \left[-ig\int_{\btone}^{\bttwo} \At^{i,a}(b^-,\zt) t^a d \zt^i \right] \,.
\end{align}
The Weizsäcker-Williams distribution is defined as
\begin{align}
    xG^{ij} (x,\kt) = \frac{4}{\left \langle P | P \right \rangle} \int & \frac{\der b^- \der b'^- \der^2 \bt   \der^2 \bt'}{(2\pi)^3} e^{-i xP^+ (b^--b'^-)-i \kt \cdot (\bt-\bt') } \nonumber \\
    & \left \langle P \Big| \Tr\left[ F^{i+}(b) U^{[+]\dagger}_{b',b} F^{j+}(b') U^{[+]}_{b',b}\right] \Big| P \right \rangle 
    \label{eq:WW_gluon_2} \,,
\end{align}
which involves the future pointing staple shaped infinite gauge link:
\begin{align}
    U^{[+]}_{b_2,b_1} = [b_2^-,+\infty]_{\bttwo} [\bttwo,\btone]_{+\infty} [+\infty,b_1^-]_{\btone} \,.
\end{align}
To make the connection of the operator definition in Eq.\,\eqref{eq:WW_gluon_2} with that given in Eq.\,\eqref{eq:WW_gluon_1}, let us now work in the Lorenz gauge $\partial_\mu A^\mu =0$ and in the small $x$ limit, in which case $\At^i$ is suppressed. We note that the (infinite) light-like Wilson line defined in Eq.\,\eqref{eq:WilsonLine} can be expressed as
\begin{align}
    V(\bt) = [\infty,-\infty]_{\bt} \,,
\end{align}
and the derivative $\partial^i$ acting on the Wilson line is given by
\begin{align}
    \partial^i V^\dagger(\bt) = -ig \int_{-\infty}^{\infty} \der b^-  [-\infty,b^-]_{\bt} F^{i+}(b^-,\bt) [b^-,+\infty]_{\bt} \,,
\end{align}
where we used $A^\mu = \delta^{\mu+} A^+$ and thus $F^{i+}(b^-,\bt) =\partial^i A^+(b^-,\bt)$ in light-cone gauge.

Thus we find
\begin{align}
    \Acalts^i(\kt) = \int \der b^-  \der^2 \bt  e^{-i \kt \cdot \bt } [+\infty,b^-]_{\bt} F^{i+}(b^-,\bt)[b^-,+\infty]_{\bt} \,,
\end{align}
where we used $\Ats^i(\bt) = \frac{i}{g} V(\bt) \partial^i V^\dagger(\bt)$.

Then it follows from Eq.\,\eqref{eq:WW_gluon_1}:
\begin{align}
    xG^{ij} (x,\kt) = \frac{4}{(2\pi)^3} \int & \der b^- \der b'^- \der^2 \bt   \der^2 \bt  e^{-i \kt \cdot (\bt-\bt') } \nonumber \\
    & \left\langle \Tr\left[ F^{i+}(b^-,\bt) U^{[+]\dagger} F^{j+}(b'^-,\bt') U^{[+]}\right] \right \rangle_x \,,
\end{align}
where we used $[\bt',\bt]_{+\infty} = \mathbb{1}$ in $\partial_\mu A^\mu =0$ gauge with the subgauge condition that transverse fields cancel at $+\infty$, which is allowed for both components in the small $x$ limit.

The equivalence to the operator definition in Eq.\,\eqref{eq:WW_gluon_2} follows (strictly speaking) in the limit $x \rightarrow 0$, and by noting the relation between the CGC average and the operator expectation value:
\begin{align}
    \left\langle \mathcal{O} \right \rangle_x \leftrightarrow \frac{\left\langle P | \mathcal{O} | P \right \rangle}{\left\langle P | P \right \rangle} \,.
\end{align}

\bibliographystyle{JHEP-2modlong.bst}
\bibliography{refs}

\end{document}